\documentclass{article}

\usepackage{amssymb,amsthm,latexsym,amsmath,graphics,setspace}
\usepackage{epsfig,epsf,amsfonts}
\usepackage{color,multicol,colordvi,graphicx,multirow}
\usepackage{fancyhdr}
\usepackage{authblk}
\newtheorem{proposition}{Proposition}

\numberwithin{equation}{section}

\newcommand{\paren}[1]{\left(#1\right)}
\newcommand{\jump}[1]{\left[#1\right]}
\newcommand{\D}[2]{\frac{d#1}{d#2}}
\newcommand{\DD}[3]{\frac{d^{#1}{#2}}{d{#3}^{#1}}}
\newcommand{\PD}[2]{\frac{\partial#1}{\partial#2}}
\newcommand{\PDD}[3]{\frac{\partial^{#1}{#2}}{\partial{#3}^{#1}}}

\newcommand{\at}[2]{\left. #1 \right|_{#2}}

\newcommand{\bm}[1]{\boldsymbol{#1}}
\newcommand{\mb}[1]{\bm{#1}}
\newcommand{\abs}[1]{\left\lvert #1 \right\rvert}

\newcommand{\dual}[2]{\left\langle #1,#2 \right\rangle}
\newcommand{\wh}[1]{\widehat{#1}}
\newcommand{\wt}[1]{\widetilde{#1}}
\newcommand{\mc}[1]{\mathcal{#1}}
\newcommand{\citep}[1]{\cite{#1}}
\newcommand{\citet}[1]{\cite{#1}}

\author[*]{Yoichiro Mori}
\author[**]{Yuan-Nan Young}
\affil[*]{School of Mathematics, University of Minnesota, Minneapolis MN, 55455, USA}
\affil[**]{Department of Mathematical Sciences, New Jersey Institute of Technology, Newark NJ, 07102, USA}
\title{From Electrodiffusion Theory to the Electrohydrodynamics of Leaky Dielectrics 
through the Weak Electrolyte Limit}

\begin{document}

\maketitle

\begin{abstract}
The Taylor-Melcher (TM) model is the standard model for describing the dynamics of
poorly conducting leaky dielectric fluids under an electric field. 
The TM model treats
the fluids as Ohmic conductors, without modeling the underlying ion dynamics. On the other hand,
electrodiffusion models, which have been successful in describing electrokinetic phenomena, 
incorporate ionic concentration dynamics.
Mathematical reconciliation of the electrodiffusion picture and the TM model has been a major issue
for electrohydrodynamic theory. Here, we derive the TM model from an electrodiffusion model
in which we explicitly model the electrochemistry of ion dissociation.
We introduce salt dissociation reaction terms in the bulk electrodiffusion equations and
take the limit in which the salt dissociation is weak; the assumption of weak dissociation
corresponds to the fact that the TM model describes poor conductors.
Together with the assumption that the Debye length is small,
we derive the TM model with or without the
surface charge convection term depending upon the scaling of relevant dimensionless parameters.
An important quantity that emerges is the Galvani potential (GP), the jump in voltage across
the liquid-liquid interface between the two leaky dielectric media; 
the GP arises as a natural consequence of the interfacial boundary conditions
for the ionic concentrations, and is absent under certain parametric conditions.
When the GP is absent, we recover the TM model. Our analysis also reveals 
the structure of the Debye layer at the liquid-liquid interface, which suggests 
how interfacial singularities may arise under strong imposed electric fields.
In the presence of a non-zero GP,
our model predicts that the liquid droplet will drift under an imposed electric field,
the velocity of which is computed explicitly to leading order. 
\end{abstract}

\section{Introduction}\label{sect:intro}

\subsection{Background}

The Taylor-Melcher (TM) model was first proposed by Taylor to describe the deformation of an oil droplet 
immersed in another poorly conducting medium under the influence of a DC electric field \citep{Taylor1966_PRSLa}. 
The TM model and its variants have since been widely used to model electrohydrodynamic
phenomena of poorly conducting (or leaky dielectric) fluids, ranging from electrodeformation, 
ink-jet printing, droplet fabrication in microfluidics and oil separation
\citep{MelcherTaylor1969_AnnuRevFluidMech,Saville1997_AnnuRevFluidMech}.

The TM model treats the two leaky dielectric fluids as electrically neutral media of
constant conductivity and dielectric constant. Under an imposed electric field, bulk currents 
carry electric charge to the interface of the two leaky dielectric fluids, 
leading to an interfacial accumulation
of monopolar charge (which we shall call the electric monopolar layer, or EML). 
The stress generated by this interfacial charge (EML) generates fluid flow,
and this fluid flow, in turn, results in interfacial charge convection.
In his original analysis, Taylor neglected charge convection \citep{Taylor1966_PRSLa}, which leads to the decoupling of the electrostatic 
and fluid equations. This made it possible to obtain explicit solutions under the assumption of small droplet deformation.
Although this approximation has since been commonly used \citep{MelcherTaylor1969_AnnuRevFluidMech,Saville1997_AnnuRevFluidMech},
some authors have argued that surface charge convection is important in explaining electrohydrodynamic 
phenomena in certain situations, especially under strong electric fields {
\citep{feng1996computational,xu2006settling,roberts2009ac,roberts2010electrohydrodynamic,salipante2010electrohydrodynamics,salipante2013electrohydrodynamic,he2013electrorotation,Lanauze2015_JFM,hu2015hybrid,vlahovska2016electrohydrodynamic,Das2017_JFM,das2017electrohydrodynamics,sengupta2017role}}.

Electrical currents in leaky dielectrics are carried by ions. The TM model, however, treats 
electrical current as Ohmic, without modeling the underlying ionic concentration dynamics. 
Equations of ionic electrodiffusion and advection, sometimes referred to as Poisson-Nernst-Planck (PNP) models, 
have been widely used to model the dynamics of electrolyte solutions \citep{rubinstein1990electro}. PNP models have been particularly successful
in describing electrokinetic phenomena, in which electrical double layers (EDL) at material interfaces play a key role \citep{delgado2001interfacial,squires2004induced,bazant2009towards,bruus2007theoretical,chang2010electrokinetically}. 
A more complete description of electrohydrodynamic phenomena of poorly conducting media 
should thus be based on the PNP equations of ionic transport,  
and the TM model should be derived as a suitable limit of such a model.
The absence of such a PNP model for leaky dielectrics has resulted in separate developments of EML and EDL theories; a derivation of the TM model 
from a suitable PNP model promises to unify our understanding of EML and EDL phenomena \citep{bazant2015electrokinetics}.
A need for such a model is also highlighted by the presence of electrohydrodynamic phenomena 
that cannot be explained by the TM model, including the drift of droplets under DC electric fields \citep{Taylor1966_PRSLa,vizika1992electrohydrodynamic,Saville1997_AnnuRevFluidMech}
and the formation of singularities under strong fields {\citep{Mora2007_AnnuRevFluidMech,brosseau2017streaming,sengupta2017role}}.

There have been several prior attempts to derive the TM model from a PNP model 
\citep{baygents1990circulation,zholkovskij2002electrokinetic,Schnitzer2015_JFM}.
In \citet{zholkovskij2002electrokinetic}, the authors consider the limit of weak electric field and small Debye length.
In a recent study, \citet{Schnitzer2015_JFM} perform an asymptotic analysis 
based on an earlier attempt by \citet{baygents1990circulation}, 
arriving at the TM model in the limit of small Debye length and large electrical field strength. 
There are several limitations in the above studies.
All analyses are limited to near spherical interfacial geometry, 
and do not produce the surface charge convection term in the TM model.
The study of \citet{zholkovskij2002electrokinetic} is limited to binary electrolytes of equal diffusivity.
In \citet{Schnitzer2015_JFM}, the product of diffusivity and viscosity for each ion is assumed constant across the two solvents.

{One of the main results of our paper is a derivation of the TM model with surface charge convection as a limit of a suitable PNP model
for arbitrary interfacial geometry without parametric assumptions on the diffusivity of ions or the viscosity of the fluids.
Furthermore, when the {\em Galvani potential} is present at the interface at rest (see below), we show that the droplet will undergo electromigration 
to leading order. We now give an overview of our results emphasizing the physical picture.}

{\subsection{Overview of Results and Comparison with Previous Studies}}
{
\subsubsection{The Electrodiffusion Model, the Weak Electrolyte Assumption and the Charge Diffusion Model}\label{intro_ediffmodel}
The most important feature of our study is that we consider a {\em weak} electrolyte solution.
Consider the salt dissociation reaction:
\begin{equation}\label{dissociation}
{\rm S} \rightleftharpoons {\rm C}^++{\rm A}^-
\end{equation}
where S is the salt and C$^+$ and A$^-$ are the cation and anion respectively.
In a weak electrolyte, most of the salt does not dissociate into their constitutive ions. 
That is to say, if $c_*=a_*$ and $s_* $ are the typical concentrations of the cation/anions and the salt respectively, 
we have:
\begin{equation}\label{alphadef0}
\frac{c_*}{s_*}\equiv \alpha \ll 1.
\end{equation}
This weak electrolyte assumption corresponds to the fact that we are interested in {\em poorly} conducting media.}

{In Section \ref{sec:modified_saville_model}, we present our electrodiffusion model.
Let $\Omega_{\rm i, e}$ be the regions occupied by the interior and exterior leaky dielectrics respectively and 
let $\Gamma$ be the interface between the two media (see Figure \ref{modified_saville_fig}).
We write down the electrodiffusion-advection equations for the solute species S, C$^+$ and A$^-$ with 
dissociation reaction terms to be satisfied inside $\Omega_{\rm i}$ and $\Omega_{\rm e}$ (Eq. \eqref{ceqn0}-\eqref{fluideqn}).
The unknown functions are the concentrations of the cation $c$, anion $a$ and solute $s$ (Eq. \eqref{ceqn0}-\eqref{seqn0}), 
as well as the electrostatic potential $\phi$ (Eq. \eqref{poisson}), velocity field $\bm{u}$ and the pressure $p$ (Eq. \eqref{fluideqn}).
These equations, satisfied in the bulk, are essentially the same as those presented in 
\citet{Saville1997_AnnuRevFluidMech}; we shall thus refer to this as the 
{\em modified Saville model}. The main difference is that 
we prescribe the requisite interfacial conditions 
for the ionic concentrations at $\Gamma$ whereas \citet{Saville1997_AnnuRevFluidMech} does not. 
As we shall see, this difference is crucial; in particular, these interfacial conditions lead naturally to the Galvani potential, 
which plays a central role in our analysis (see Section \ref{EDLGP}).
Another important difference is
our identification of the small parameter $\alpha$ (see \eqref{alphadef0}) that captures the weak electrolyte limit.}

{
In the modified Saville model, 
we assume that the interface  $\Gamma$ is a sharp (mathematical) interface that carries no chemical density, and hence no charge density.
Since the interface $\Gamma$ carries no charge, both the electrostatic potential and the electric flux density 
must be continuous across the interface $\Gamma$ (Eq. \eqref{poissonbc}):
\begin{equation}\label{poissonbc0}
\at{\phi}{\Gamma_{\rm i}}=\at{\phi}{\Gamma_{\rm e}}, \; 
\at{\epsilon_{\rm i}\PD{\phi}{\bm{n}}}{\Gamma_{\rm i}}=\epsilon_{\rm e}\at{\PD{\phi}{\bm{n}}}{\Gamma_{\rm e}}
\end{equation}
where $\epsilon_{\rm i,e}$ is the interior/exterior dielectric constant and 
$\at{\cdot}{\Gamma_{\rm i,e}}$ denotes the value of the quantity in question evaluated at the interior
or exterior face of the interface $\Gamma$ respectively, and $\bm{n}$ is the outward normal on $\Gamma$.}

{Our assumption above that $\Gamma$ carries no charge is motivated by simplicity, and is in contrast to \citet{Saville1997_AnnuRevFluidMech,Schnitzer2015_JFM}. Note that, in the modified Saville model, the Debye layer 
is fully resolved; assuming that $\Gamma$ carries surface charge, then, amounts to claiming 
that there is a concentrated charge density at the liquid-liquid interface (thinner than the Debye layer) much like the Stern layer 
for solid-liquid interfaces \citep{bazant2009towards}. It is not clear if such a charge density is of significant magnitude even if present.  
From a modeling perspective, it is easy to include such a charge density, at the expense of making
the model more complex. In fact, one of the interesting features of our derivation in Section \ref{leaky} is that the 
surface charge density in the Taylor-Melcher model appears naturally in a suitable limit (to be discussed 
shortly) even though we do not explicitly have a surface charge density to start with.}

{ For the ionic concentrations $c$ and $a$, we first 
impose the usual flux continuity conditions across $\Gamma$.
We further assume that the electrochemical potentials of $c$ and $a$ are equal across $\Gamma$:
\begin{equation}\label{electrochemcont}
\begin{split}
E_{\rm C, i}+RT\ln (\at{c}{\Gamma_{\rm i}})+F\at{\phi}{\Gamma_{\rm i}}&
=E_{\rm C,e}+RT\ln (\at{c}{\Gamma_{\rm e}})+F\at{\phi}{\Gamma_{\rm e}},\\
E_{\rm A, i}+RT\ln (\at{a}{\Gamma_{\rm i}})
-F\at{\phi}{\Gamma_{\rm i}}
&=E_{\rm A,e}+RT\ln (\at{a}{\Gamma_{\rm e}})-F\at{\phi}{\Gamma_{\rm e}},
\end{split}
\end{equation}
where $RT$ is the ideal gas constant times the absolute temperature and $F$ is the Faraday constant.
The constants $E_{\rm \cdot,\cdot}$ reflect the differences in solvation energy of the cation/anions 
across in the interior and exterior leaky dielectric media. 
Using \eqref{poissonbc0}, we may write the above as follows:
\begin{equation}\label{lXE}
\at{c}{\Gamma_{\rm i}}=l_{\rm C}\at{c}{\Gamma_{\rm e}}, \; \at{a}{\Gamma_{\rm i}}=l_{\rm A}\at{a}{\Gamma_{\rm e}}, \;\;
l_{\rm X}=\exp(-(E_{\rm X,i}-E_{\rm X,e})/RT),\; {\rm X}={\rm C,A}. 
\end{equation}
The constants $l_{\rm C}$ and $l_{\rm A}$ are the concentration ratios across the interface $\Gamma$,
and are known as {\em partition coefficients} \citep{hung1980electrochemical}.
The above boundary conditions are the same as those used in \citet{zholkovskij2002electrokinetic}.}

{We identify two small parameters in the modified Saville model, the ratio $\alpha$ in \eqref{alphadef0} and the 
and the ratio $\delta$ between the Debye length $r_{\rm D}$ and droplet size $L$:
\begin{equation}\label{deltadefn}
\delta=\frac{r_{\rm D}}{L}, \; r_{\rm D}=\sqrt{\frac{\epsilon_* RT/F}{Fc_*}}
\end{equation}
where $\epsilon_*$ is the representative dielectric constant and
$F$ is the Faraday constant. The assumption that $\delta$ is small is well-accepted, 
and is used in all previous derivations of the Taylor-Melcher model \citep{baygents1990circulation,zholkovskij2002electrokinetic,Schnitzer2015_JFM}.
As discussed above, the smallness of $\alpha$ stems from our weak electrolyte assumption.
We first take the limit $\alpha\to 0$ and subsequently take the limit $\delta \to 0$.
This is most natural if 
\begin{equation}\label{scaling_intro}
\alpha \ll \delta \ll 1.
\end{equation}
}

{In Section \ref{CD}, we take the limit $\alpha\to 0$. Under suitable scaling, $\alpha \ll 1$ 
implies that the dissociation reaction \eqref{dissociation} is so fast that it is effectively at equilibrium:
\begin{equation}
\frac{ca}{s}= K_{\rm eq},
\end{equation}
where $K_{\rm eq}$ is the equilibrium constant for this reaction.
Since we may assume that the salt concentration $s$ is constant in space and time (with potentially 
different values in $\Omega_{\rm i}$ and $\Omega_{\rm e}$), 
the above relation allows us to eliminate both $c$ and $a$ in favor of the charge density $q=c-a$. 
The resulting model is the {\em charge diffusion model}, whose unknown functions are $q, \phi, \bm{u}$ and $p$.
The charge $q$ satisfies a nonlinear drift-diffusion advection equation.}

{
\subsubsection{The Electric Double Layer and the Galvani Potential}\label{EDLGP}
We subsequently take the limit $\delta=r_{\rm D}/L\to 0$ in the charge diffusion model. 
In the bulk, away from the interface $\Gamma$, we obtain electroneutrality
($q=0$) and Ohm's law for electric current conduction.
A spatially constant Ohmic conductivity results naturally from the weak electrolyte limit and 
electroneutrality. This is in contrast to \citet{Schnitzer2015_JFM}, in which spatially constant Ohmic conductivity 
results from strong advection due to a large imposed electric field.}

\begin{figure}
\begin{center}
\includegraphics[width=0.85\textwidth]{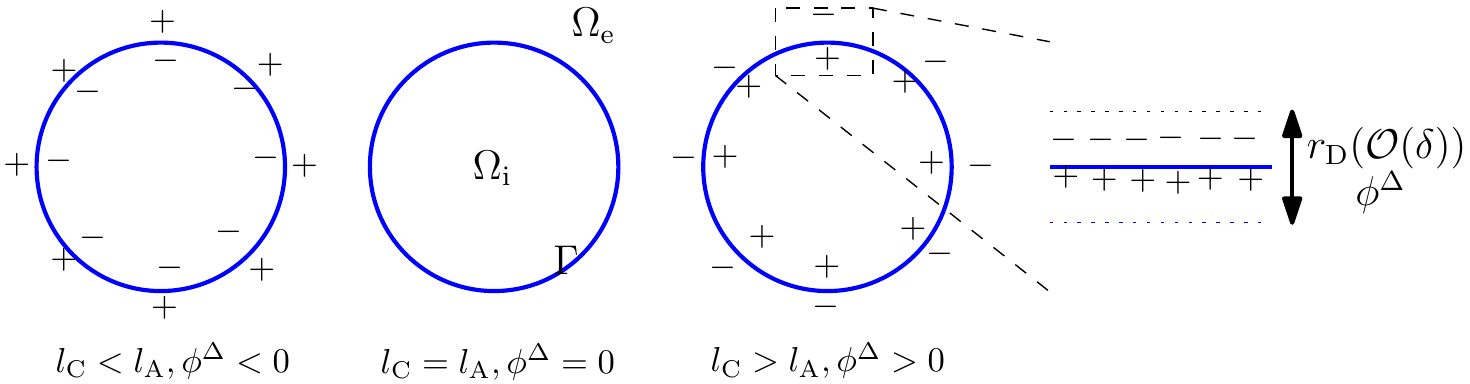}
\end{center}
\caption{\label{galvani_EDL_fig} {When $l_{\rm C}\neq l_{\rm A}$, an electric double layer develops 
across the interface $\Gamma$ even when the system is at rest, 
with a resulting voltage jump (the Galvani potential) of 
$\phi^\Delta=\frac{RT}{2F}\ln(l_{\rm C}/l_{\rm A})$.}}
\end{figure}

\begin{figure}
\begin{center}
\includegraphics[width=0.5\textwidth]{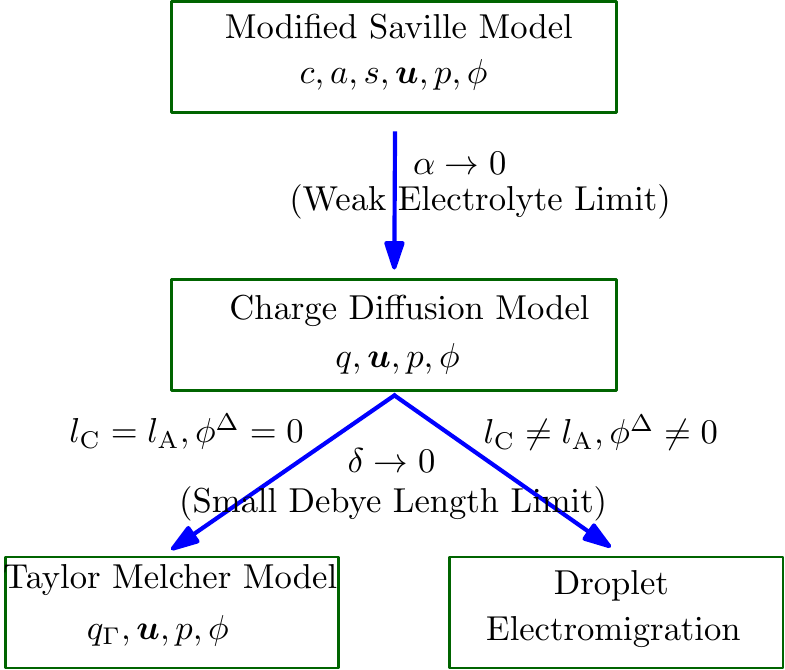}
\end{center}\caption{\label{model_hierarchy_fig} {A schematic showing the inter-relation between the different models 
and asymptotic limits. When $\alpha\to 0$ leads to the charge diffusion model (Section \ref{CD}). 
We subsequently take the limit $\delta \to 0$. Here, the limiting behavior is fundamentally different depending on 
whether $l_{\rm C}=l_{\rm A}$ or $l_{\rm C}\neq l_{\rm A}$. When $l_{\rm C}=l_{\rm A}$, we obtain the Taylor-Melcher 
model (Section \ref{leaky}) whereas when $l_{\rm C}\neq l_{\rm A}$ we obtain droplet electromigration under an imposed electric field
(Section \ref{DL}).
}}
\end{figure}

{At the interface $\Gamma$, the limit $\delta \to 0$
results in a boundary layer of thickness $r_{\rm D}$.
The properties of this Debye layer depends critically on the ratio of the partition coefficients $l_{\rm C}/l_{\rm A}$.
When $l_{\rm C}/l_{\rm A}\neq 1$, an EDL with a voltage jump $\phi^\Delta =\frac{RT}{2F}\ln(l_{\rm C}/l_{\rm A})$
develops across the Debye layer, even in the absence of an imposed electric field (see Figure \ref{galvani_EDL_fig}). 
We may identify $\phi^\Delta$ as the {\em Galvani potential}, 
whose presence is a well-documented feature of liquid-liquid interfaces \citep{girault1989electrochemistry,reymond2000electrochemistry}
(much like the $\zeta$ potential of liquid-solid interfaces).
The cases $\phi^\Delta=0$ and $\phi^\Delta\neq 0$ lead to fundamentally different 
behaviors. The case $\phi^\Delta=0$, treated in Section \ref{leaky}, leads to the Taylor-Melcher 
model, whereas the case $\phi^\Delta\neq 0$, treated in Section \ref{DL}, leads to droplet electromigration 
(see Figure \ref{model_hierarchy_fig} for schematic).
}

\begin{figure}
\begin{center}
\includegraphics[width=0.6\textwidth]{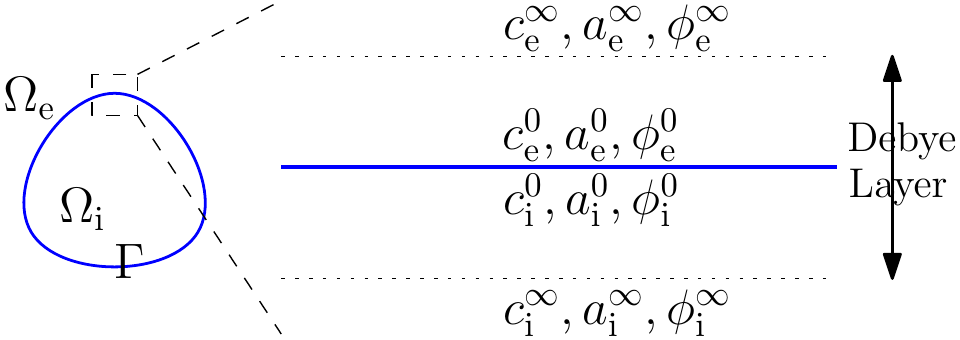}
\end{center}
\caption{\label{galvani_potential_fig}{Schematic diagram illustrating the heuristic derivation of the Galvani potential. 
The Galvani potential $\phi^\Delta=\phi_{\rm i}^\infty-\phi_{\rm e}^\infty$ 
is the the difference in the voltage across the Debye layer.}}
\end{figure}

{We now include a heuristic calculation as to how the Galvani potential $\phi^\Delta$ arises.
Consider a patch of the interface $\Gamma$. This interface is sandwiched by Debye layers on both sides 
of the interface, as shown in Figure \ref{galvani_potential_fig}. Let $c_{\rm i}^0$ be the cation concentration 
on the interior face of $\Gamma$ and $c_{\rm i}^\infty$ be the cation concentration just outside the Debye layer 
(in the outer layer). Adopt a similar notation for the the exterior concentrations $c_{\rm e}^{0,\infty}$ as well 
as the anion concentrations $a_{\rm i,e}^{0,\infty}$. We also introduce the notation $\phi_{\rm i,e}^0$ for the 
voltage at the interior/exterior face of $\Gamma$. The Galvani potential is the difference:
\begin{equation}
\phi^\Delta=\phi_{\rm i}^\infty-\phi_{\rm e}^\infty.
\end{equation}
Suppose the system is in equilibrium. Then, there is no chemical flux across the Debye 
layer, and thus, the chemical potential across the Debye layer must be equal. For the 
cation concentration, we must thus have:
\begin{equation}
\begin{split}
&E_{\rm C,i}+RT\ln c_{\rm i}^\infty+F\phi_{\rm i}^\infty
=E_{\rm C, i}+RT\ln c_{\rm i}^0+F\phi_{\rm i}^0\\
=&E_{\rm C,e}+RT\ln c_{\rm e}^0+F\phi_{\rm e}^0
=E_{\rm C,e}+RT\ln c_{\rm e}^\infty+F\phi_{\rm e}^\infty.
\end{split}
\end{equation}
The second equality follows from \eqref{electrochemcont}. Therefore, 
\begin{equation}\label{phiDeltac}
\phi^\Delta=\frac{RT}{F}\ln\paren{\frac{c_{\rm e}^\infty}{c_{\rm i}^\infty}}
-\frac{1}{F}(E_{\rm C,i}-E_{\rm C,e}).
\end{equation}
A similar calculation for the anion concentration yields:
\begin{equation}\label{phiDeltaa}
\phi^\Delta=-\frac{RT}{F}\ln\paren{\frac{a_{\rm e}^\infty}{a_{\rm i}^\infty}}
+\frac{1}{F}(E_{\rm A,i}-E_{\rm A,e})
\end{equation}
In the limit as $\delta \to 0$, the bulk (outer layer) is electroneutral, and therefore, $a_{\rm i}^\infty=c_{\rm i}^\infty$
and $a_{\rm e}^\infty=c_{\rm e}^\infty$. Thus, combining \eqref{phiDeltac} and \eqref{phiDeltaa}, we have:
\begin{equation}\label{phiDeltaE}
\phi^\Delta=\frac{1}{2F}\paren{-(E_{\rm C,i}-E_{\rm C,e})+(E_{\rm A,i}-E_{\rm A,e})}=\frac{RT}{2F}\ln\paren{\frac{l_{\rm C}}{l_{\rm A}}}
\end{equation}
where we used \eqref{lXE} in the last equality.}
{As can be seen from \eqref{phiDeltaE}, 
the EDL and its attendant Galvani potential 
arise as a consequence of the mismatch in the solvation energies (or partition coefficient) between the cation and anion.
The presence of such a potential jump when $l_{\rm C}\neq l_{\rm A}$ is also noted in \citet{zholkovskij2002electrokinetic}.}

{
\subsubsection{Derivation of the Leaky Dielectric Model when $l_{\rm C}=l_{\rm A}$, and Implications for Singularity Formation}\label{LD_intro}}

In Section \ref{leaky}, we consider the limit $\delta\to 0$ in the charge diffusion model
when $l_{\rm C}=l_{\rm A}$, in which case the Galvani potential $\phi^\Delta=0$. { In this case, 
there is no interfacial layer at rest when there is no flow. It is only with flow that an interfacial charge layer of Debye 
layer thickness emerges at $\Gamma$.}
Assume that the surface tension $\gamma$ scales 
like $\delta^2$ and a suitably defined P\'eclet number $Pe$ scales like $\delta^{-2}$ as $\delta \to 0$. 
{This particular scaling is chosen so that the electrohydrodynamic time, the Maxwell-Wagner charge relaxation time
and the capillary time scale are all of the same order \citep{salipante2010electrohydrodynamics}.}
A boundary layer analysis then yields the TM model with charge convection in the limit 
as $\delta\to 0$ for arbitrary interfacial geometry. An interesting feature of our derivation is that, unlike the 
PNP models of \citet{Saville1997_AnnuRevFluidMech,Schnitzer2015_JFM}, the modified Saville model, 
and hence the charge diffusion model does not have any built-in 
surface charges; the surface charge density $q_\Gamma$ in the TM model emerges naturally from the bulk charge $q$
of the charge diffusion model as $\delta \to 0$. Variants of the TM model are obtained when $Pe$
is scaled differently with respect to $\delta$.
When $Pe$ is smaller than $\mc{O}(\delta^{-2})$, we recover the TM model without surface charge convection to leading order.
It should be pointed out, however, that $Pe=\mc{O}(\delta^{-2}), \gamma=\mc{O}(\delta^2)$ is the thermodynamically 
canonical scaling, {the precise meaning of which is discussed in Appendix \ref{app:energy}}. 

{In addition to the recovery of the TM model, our analysis yields a set of equations governing the charge distribution inside
the interfacial Debye layer, which we study in Section \ref{bndry_layer}. These equations show that the interfacial charge density 
profile depends strongly on the properties of the local flow field. Let $\bm{u}_\parallel$ be the tangential 
component of the flow field and $\nabla_\Gamma \cdot \bm{u}_\parallel$ be its surface divergence. 
Consider a point $\bm{x}_0\in \Gamma$ at which point the flow is stagnant ($\bm{u}=0$),  
and suppose $\nabla_\Gamma \cdot \bm{u}_\parallel(\bm{x}_0)>0$.
Then, the Debye layer charge density decays 
supra-exponentially at $\bm{x}_0$ with distance from the interface $\Gamma$. On the other hand, 
if $\nabla_\Gamma\cdot \bm{u}_\parallel(\bm{x}_0)<0$, the Debye layer charge density decays only algebraically 
with distance from the interface, resulting in a thicker charge layer. In fact, when $\nabla_\Gamma \cdot \bm{u}_\parallel(\bm{x}_0)\leq -1/\tau$, where $\tau$ is the charge relaxation time scale of the bulk medium, 
the boundary layer assumption breaks down at $\bm{x}_0$.
}

\begin{figure}
\begin{center}
\includegraphics[width=0.7\textwidth]{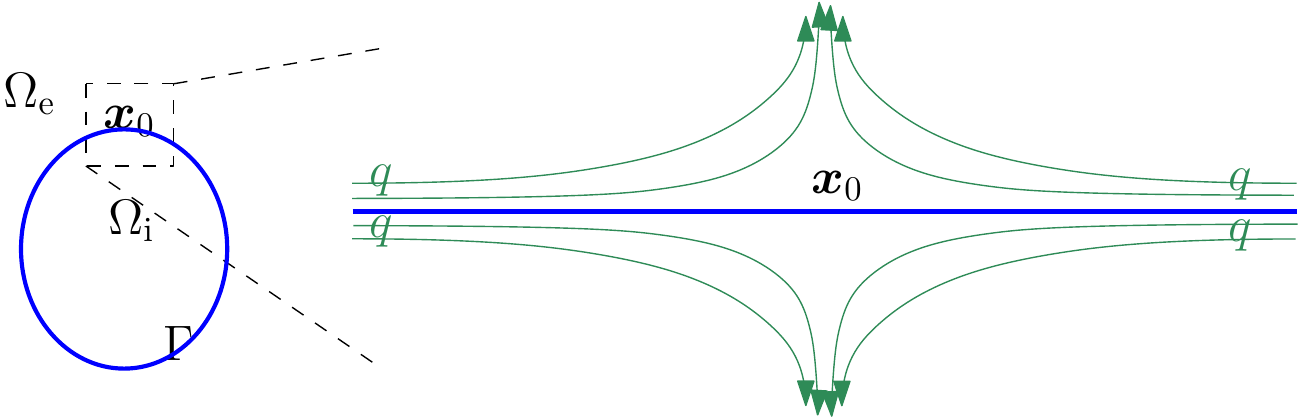}
\end{center}
\caption{\label{charge_accumulation_fig} {Suppose $\Gamma$ is stationary and 
$\bm{x}_0\in \Gamma$ is a stagnation point of the flow field and $\nabla_\Gamma \cdot \bm{u}_\parallel<0$(see discussion for details). 
Then, the surface charge $q$ will be carried to the point $\bm{x}_0$ via the flow field (green arrows)
and accumulate at $\bm{x}_0$, potentially overwhelming charge relaxation.}}
\end{figure}

{We now give a heuristic explanation as to why such a breakdown may take place
using only the TM model. Suppose the surface $\Gamma$ is stationary (but the fluid velocity, 
of course, is non-zero) and, at a point $\bm{x}_0\in \Gamma$, the flow is stagnant. 
Then, at $\bm{x_0}$, we have the equation:
\begin{equation}
\PD{q_\Gamma}{t}
=-(\nabla_\Gamma \cdot \bm{u}_\parallel)q_\Gamma-\jump{\wh{\sigma}\PD{\phi}{\mb{n}}},\; \jump{\epsilon \PD{\phi}{\bm{n}}}=q_\Gamma,
\end{equation}
where we have followed the notation of Section \ref{leaky} (see \eqref{phiqgamma} and \eqref{LD_qGammaeqn}).
Here, $\wh{\sigma}$ is the bulk conductivity. The above suggests that:
\begin{equation}
\PD{q_\Gamma}{t}
=-(\nabla_\Gamma \cdot \bm{u}_\parallel)q_\Gamma-\jump{\wh{\sigma}\PD{\phi}{\mb{n}}}
\sim -\paren{\nabla\cdot \bm{u}_\parallel+\frac{1}{\tau}}q_\Gamma, \; \tau=\frac{\epsilon}{\wh{\sigma}}.
\end{equation}
The relation $\sim$ is meant to indicate that this is only approximate; indeed, $\sim$ can be replaced by an equality 
only if the value of $\tau=\epsilon/\wh{\sigma}$ is equal in $\Omega_{\rm i}$ and $\Omega_{\rm e}$.
Nonetheless, if $\nabla \cdot \bm{u}_\parallel+1/\tau<0$, the surface charge $q_\Gamma$ is expected to grow exponentially,
leading to unbounded accumulation of surface charge. As we shall see in Section \ref{bndry_layer}, the precise 
condition for boundary layer breakdown is:
\begin{equation}
\nabla_\Gamma \cdot \bm{u}_\parallel +\frac{1}{\tau_{\rm max}}\leq 0,
\end{equation}
where $\tau_{\rm max}$ is the larger of the value of $\tau$ in $\Omega_{\rm i}$ or $\Omega_{\rm e}$ (see \eqref{taumax}).
The physical picture that emerges is that, when surface charge convection is strong enough, 
it may overwhelm bulk charge relaxation leading to boundary layer breakdown (see Figure \ref{charge_accumulation_fig}).}

{This suggests the following scenario for the formation of interfacial geometric singularities,
such as the Taylor cone \citep{Mora2007_AnnuRevFluidMech} or the recently reported equatorial streaming \citep{brosseau2017streaming}.
Stronger electric fields may generate strong charge convection at the interface, which will result in the thickening 
of the Debye layer at locations where $\nabla_\Gamma \cdot \bm{u}_\parallel<0$. At a certain field strength, 
the Debye layer charge distribution broadens to the extent that the boundary layer assumption fails, 
at which point the TM model will no longer be valid. At this point, the interfacial stress 
balance also fails, indicating the emergence of geometric singularities. 
Boundary layer matching cannot be achieved if surface charge accumulation due to the flow field overwhelms conductive charge dissipation; 
at this point, stress balance is also violated. 
When the prolate deformation is favored, strong electric fields should lead to charge accumulation at the
``poles" where the surface divergence is maximally negative, leading to a breakdown of the interfacial charge layer
and eventually to a Taylor cone. The recent paper \citet{sengupta2017role} indicates that run-away surface 
charge accumulation may indeed be the mechanism for singularity formation in prolate drops.
On the other hand, if oblate deformation is favored, strong electric fields may lead to charge accumulation 
at the ``equator", leading to equatorial streaming \citep{brosseau2017streaming}.}

The foregoing derivation assumed that the imposed voltage is on the order of the thermal voltage. 
In Section \ref{large_voltage}, we extend our derivation to the case when the imposed electric field is large.
Following  \citet{Saville1997_AnnuRevFluidMech,Schnitzer2015_JFM}, let $\beta$ be the ratio between the imposed voltage (characteristic 
electric field times droplet size) and the thermal voltage. Then, under the assumption:
\begin{equation}\label{strong_scaling_intro}
1\ll \beta\ll \frac{1}{\delta}
\end{equation}
we can obtain the TM model, in much the same way as before. 
An important difference, however, is that there are now two boundary layers (Figure \ref{two_layers}).
The charge layer widens to $r_{\rm E}=\sqrt{\beta}r_{\rm D}$ inside of which is an inner-inner layer of width $r_{\rm D}$.

{\subsubsection{Comparison with Previous Derivations of the Taylor-Melcher Model}}

{Before proceeding further with an overview of our results,} 
we compare our derivation of the TM model outlined above with those of \citet{baygents1990circulation,zholkovskij2002electrokinetic,Schnitzer2015_JFM}. 

In all of these studies, the authors consider a standard PNP model for {\em strong} binary electrolytes
in which ions are completely dissociated, in contrast to our derivation in which we consider weak electrolytes.
Indeed, the derivation in these studies do {\em not} seem to rely on whether the ionic medium is a poor or good conductor. 
In this sense, these derivations may be addressing the validity of the TM model outside of the leaky dielectric regime.

These studies derive the stationary TM model without surface charge convection for near spherical geometries.
Here, we derive the dynamic TM model with surface charge convection for arbitrary geometry.
{As such, previous studies provide little insight into geometric singularity formation, which, many argue, 
relies on surface charge convection \citep{salipante2010electrohydrodynamics,Lanauze2015_JFM,hu2015hybrid,Das2017_JFM,sengupta2017role}.}

The derivation of bulk Ohmic conduction from an electrodiffusion model, a key ingredient in any derivation of the TM model,
is of interest beyond leaky dielectrics and has been discussed by many authors
\citep{Saville1997_AnnuRevFluidMech,squires2004induced,chen2005convective,chen2011electrohydrodynamic}. 
To the best of our knowledge, all previous derivations of Ohmic conduction starts with a strong electrolyte model.
In \citet{zholkovskij2002electrokinetic,squires2004induced}, it is assumed that the diffusion coefficients of 
the positive and negative ions are the same. In this case, a partial decoupling of the equations for $q=c-a$ and $c+a$ leads
to Ohmic conduction. When the diffusion coefficients are not equal, one typically assumes
that the electric field is strong so that drift due to the electric field overwhelms diffusion 
\citep{Saville1997_AnnuRevFluidMech,chen2005convective,chen2011electrohydrodynamic,Schnitzer2015_JFM}.
In \citet{chen2005convective,chen2011electrohydrodynamic}, the conductivity is a function of position 
that satisfies an advection-diffusion equation. In \citet{Schnitzer2015_JFM}, the conductivity is spatially constant; the constant ionic 
concentrations in the far field get swept into the region of interest by strong fields. 

Our derivation of Ohmic conduction, in contrast, relies on the weak electrolyte assumption.
The concentration of ions is governed to leading order by local ion dissociation reactions, which leads to
Ohmic conduction with spatially constant conductivity for any diffusion coefficient.

\citet{Schnitzer2015_JFM} make the interesting suggestion that conductivity may not be a bulk material 
property but a surface property. In our case, as we shall discuss in Section \ref{CD}, 
conductivity is a bulk material property if the dissociating neutral species 
in reaction \eqref{dissociation} is the solvent itself.

A key parametric assumption made in \citet{Schnitzer2015_JFM}, as emphasized in \citet{bazant2015electrokinetics}, 
is that, for each ion, the product of the diffusion coefficient and viscosity is constant across solvents. 
This allows the authors to perform a matched asymptotic calculation that leads to the TM model.
This assumption, sometimes known as Walden's rule, may be justified by the Stokes-Einstein relation for viscosity and diffusivity; 
as such it depends on the approximation that the effective radii of ions do not change from solvent to solvent. 
Significant deviations from Walden's rule are well-documented \citep{steel1958individual}.
In our derivation, no such assumptions on the diffusion coefficients or the viscosities are needed.

In Section \ref{large_voltage}, we show that our derivation of the TM model is valid for large voltage satisfying \eqref{strong_scaling_intro}.
This scaling is precisely the parametric ordering assumed in \citet{baygents1990circulation,Schnitzer2015_JFM}.
In fact, our derivation of the TM model is valid from thermal voltage up to imposed voltages satisfying $\beta \ll \delta^{-1}$.
This is in contrast to \citet{baygents1990circulation,Schnitzer2015_JFM} where $\beta\gg 1$ is needed (which stems in part from 
their need to obtain Ohm's law, as discussed above). Our derivation of the TM model is thus valid over a wider range of imposed voltages,
which seems to be supported by the absence of experimental 
reports indicating a break down of the TM model at low voltages.
{We also point out that, as discussed at the end of Section \ref{LD_intro}, 
our analysis indicates the presence of two interfacial layers 
for large imposed electric fields. This is in contrast to 
\citet{Schnitzer2015_JFM} who argue that the Debye layer is the only interfacial layer even at large voltages.}

{
\subsubsection{Electromigration under Imposed Electric Field when $l_{\rm C}\neq l_{\rm A}$}}
In Section \ref{DL}, we consider the case when $l_{\rm C}\neq l_{\rm A}$.
{In this case, we have an EDL across the interface $\Gamma$ (see Figure \ref{galvani_EDL_fig}).
We compute the asymptotic limit as $\delta\to 0$ assuming the scaling $\gamma=\mc{O}(\delta)$ and $Pe=\mc{O}(\delta^0)$. 
The important conclusion here is that we obtain droplet electromigration under an imposed electric field, 
which may explain the experimental reports of droplet electromigration in leaky dielectrics \citep{Taylor1966_PRSLa,vizika1992electrohydrodynamic}.}

\begin{figure}
\begin{center}
\includegraphics[width=0.25\textwidth]{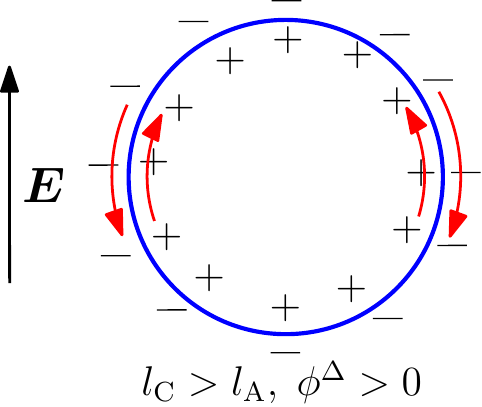}
\end{center}
\caption{\label{velocity_slip_fig} {In the presence of a non-zero Galvani potential, an imposed electric field 
will exert opposite forces on the two sides of the interface, producing a velocity slip across the Debye layer. 
The explicit expression for this is given in Eq. \eqref{vi0slip}.}}
\end{figure}

{
In Section \ref{subsec:lc_neq_la},
we find that there is an initial time layer during which the shape of the interface 
quickly approaches a sphere; this is due in part to assumption that surface tension is strong 
($\gamma=\mc{O}(\delta)$ in Section \ref{DL}
whereas $\gamma=\mc{O}(\delta^2)$ in Section \ref{leaky}).
Dynamics after this initial layer is governed by the jump conditions for the velocity, stress and voltage
across the interface $\Gamma$, which are obtained in Sections \ref{subsec:lc_neq_la} and \ref{perturb_sphere} 
via matched asymptotic calculations across the 
electric double layer. Of particular interest is the interfacial condition
for the tangential velocity across the leaky dielectric interface (Eq. \eqref{vi0slip}).
There is a tangential velocity slip, which may be interpreted as the liquid-liquid generalization of 
the Smoluchowski slip velocity relation for solid-liquid interfaces (see Figure \ref{velocity_slip_fig}).
}

{In Section \ref{sph_drop}, these interface conditions are used to 
derive an explicit formula for the drift velocity and compute the flow field around the leaky dielectric droplet under 
an imposed electric field.
An interesting result of our analysis is that the deformation of the leaky 
dielectric sphere does not depend, to leading order, on the partition coefficient ratio $l_{\rm C}/l_{\rm A}$. Thus, the Taylor criterion 
for prolate/oblate deformation of a sphere applies even to the case when $l_{\rm C}\neq l_{\rm A}$.
One of the most important predictions of the TM model can thus be obtained 
even in a regime in which the TM model is not valid. This suggest that it may be misleading to use the verification of the Taylor deformation 
criterion as evidence of the validity of the TM model, especially in the presence of electromigration.
}

{We also find that the sign of $\phi^\Delta$ does not necessarily dictate the direction of electromigration.
If $\phi^\Delta>0$, for example, the droplet is positively charged with respect to the outside liquid (see Figure \ref{galvani_EDL_fig}).
It might be natural to expect that the droplet will move in the direction of the electric field, but we find that it can migrate 
in either direction depending on material parameters.}

{Our calculation here is reminiscent of those in 
\citet{booth1951cataphoresis,baygents1991electrophoresis,pascall2011electrokinetics}, where the authors 
compute electromigration velocities of droplets under different assumptions on the nature of the droplet and of surrounding fluid.
In particular, the results in \citet{baygents1991electrophoresis} show, similarly to our calculation, that the
migration velocity of conducting droplets need not be in the direction expected by the sign of $\phi^\Delta$.}

{\subsubsection{Concluding Discussion and Appendices}}

The picture that emerges from our analysis in Sections \ref{leaky} and \ref{DL} is that EML phenomena 
dominate in the absence of the Galvani potential whereas EDL phenomena appear in its presence. 
In our model, electrophoretic motion of leaky dielectrics is an EDL phenomenon;
the identification of leaky dielectric electrohydrodynamics with EML phenomena may thus be misleading.
{In Section \ref{discussion} we discuss the implications and questions that arise from our study.
In particular, we estimate the magnitude of the dimensionless parameters and
discuss whether the TM model with or without surface charge convection corresponds to typical leaky dielectric systems. 
Appendix \ref{app:energy} discusses the thermodynamic thread 
that runs through the modified Saville, charge diffusion and Taylor-Melcher models.}
The other appendices contain technical material used in the main text.

\section{{Governing Equations: Modified Saville Model}}\label{sec:modified_saville_model}

\subsection{Setup}
\begin{figure}
\begin{center}
\includegraphics[width=0.35\textwidth]{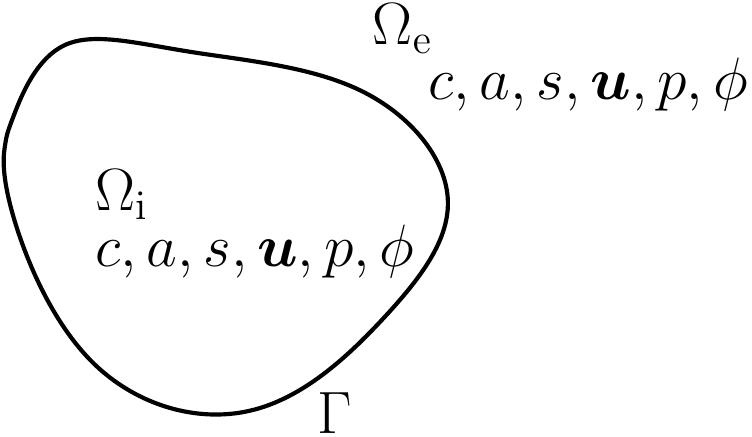}
\end{center}
\caption{\label{modified_saville_fig} Setup for the modified Saville model. 
The interior/exterior medium is denoted by $\Omega_{\rm i}$/$\Omega_{\rm e}$. 
The interface is denoted by $\Gamma$.
The unknown functions in $\Omega_{\rm i,e}$ are {the cation/anion/salt concentrations $c,a,s$, velocity and pressure fields $\bm{u},p$ and 
and the electrostatic potential $\phi$.}}
\end{figure}

Figure~\ref{modified_saville_fig} shows the configuration considered here:
a Newtonian viscous fluid in the interior domain $\Omega_{\rm i}$ is separated from
another Newtonian viscous fluid in the exterior domain $\Omega_{\rm e}$ by
an interface denoted by $\Gamma$. We let the whole region be $\Omega=\Omega_{\rm i}\cup \Omega_{\rm e}\cup \Gamma$. 
The region $\Omega$ may either be a bounded domain (so that there is a boundary $\partial \Omega$) or the whole of $\mathbb{R}^3$.
Unlike the analysis in \citet{Schnitzer2015_JFM}, here the fluid interface $\Gamma$ is not restricted to a spherical surface, and
is allowed to be time-dependent (and hence a time-dependent $\Omega_{\rm i}$).
We shall sometimes use the notation $\Gamma_t$ 
to make this time dependence explicit.
{In the following, for convenience of the reader, we may reiterate the definition of some variables that were introduced in 
the Section \ref{intro_ediffmodel}. }

We consider a salt (S) which may dissociate into cations (C$^+$) and anions (A$^-$)
in a solvent, as in reaction \eqref{dissociation}. 
As discussed in Section \ref{sect:intro}, we consider the weak-electrolyte case, in which most of the 
salt does not dissociate into its constituent ions. This may happen, for example, if NaCl is dissolved in a non-polar solvent.
When the above dissociation reaction is at equilibrium in the non-polar solvent, 
there should be considerably more S than C$^+$ or A$^-$. 

In the bulk ($\Omega_{\rm i} \cup \Omega_{\rm e}$) the concentrations of C$^+$, A$^-$ and S are denoted by $c$, $a$ and $s$ respectively.
The equations satisfied by these variables in $\Omega_{\rm i}\cup \Omega_{\rm e}$ are:
\begin{align}\label{ceqn0}
\PD{c}{t}+\nabla \cdot (\mb{u}c)&=\nabla \cdot \paren{D_{\rm C}\paren{\nabla c+c\frac{F}{RT}\nabla \phi}}+k_+s-k_-ca,\\
\label{aeqn0}
\PD{a}{t}+\nabla \cdot (\mb{u}a)&=\nabla \cdot \paren{D_{\rm A}\paren{\nabla a-a\frac{F}{RT}\nabla \phi}}+k_+s-k_-ca,\\
\label{seqn0}
\PD{s}{t}+\nabla \cdot (\mb{u}s)&=\nabla \cdot \paren{D_{\rm S}\nabla s}-k_+s+k_-ca.
\end{align}
Here, $\mb{u}$ is the solvent velocity, $D_{\rm C,A,S}$ are the diffusion coefficients,
$F$ is the Faraday constant, $RT$ is the ideal gas constant times absolute temperature,
$\phi$ is the electrostatic potential and $k_{\pm}$ are the rates of forward and backward reaction in \eqref{dissociation}.
The diffusion coefficients $D_{\rm C,A,S}$ and the reaction rate constants $k_{\pm}$
may differ in $\Omega_{\rm i}$ and $\Omega_{\rm e}$, but we assume they are constant within each region.
We will sometimes use the notation $D_{\rm C,i}$ or $D_{\rm C,e}$ to refer to the diffusion constant 
in regions $\Omega_{\rm i}$ and $\Omega_{\rm e}$ respectively. Analogous expressions will be used for $D_{\rm A,S}$
and for $k_{\pm}$.
We need an equation for the fluid velocity $\mb{u}$ as well as the electrostatic potential $\phi$. 
For the electrostatic potential, we have:
\begin{equation}
-\nabla \cdot (\epsilon \nabla \phi)=F(c-a),\label{poisson}
\end{equation}
where $\epsilon$ is the dielectric constant. For the fluid velocity, we assume Stokes flow:
\begin{equation}\label{fluideqn}
\mu \Delta \mb{u}-\nabla p=F(c-a)\nabla \phi, \; \nabla \cdot \mb{u}=0,
\end{equation}
where $\mu$ is the viscosity and $p$ is the pressure.  The dielectric constant $\epsilon$ and the viscosity $\mu$
may take different but spatially constant values in $\Omega_{\rm i}$ and $\Omega_{\rm e}$, and we use the notation 
$\epsilon_k, \mu_k, k={\rm i,e}$ to denote the constants in the two regions.
{We have chosen to ignore the inertial term as is customary in most treatments of the electrohydrodynamics
of leaky dielectrics; \citet{Saville1997_AnnuRevFluidMech} estimates the Reynolds number to be on the order of $10^{-4}$
even under high imposed electric fields.}
Note that the force balance equation in \eqref{fluideqn} can also be written as:
\begin{align}
&\nabla \cdot(\Sigma(\mb{u},p)+\Sigma_{\rm M}(\phi))=0,\\
\Sigma(\mb{u},p)&=2\mu\nabla_{S}\mb{u}-pI,\;  2\nabla_S\mb{u}=(\nabla\mb{u}+(\nabla{\mb{u}})^T),\\
\Sigma_{\rm M}(\phi)&=\epsilon\paren{\nabla \phi\otimes \nabla \phi-\frac{1}{2}\abs{\nabla\phi}^2 I}.
\end{align}
The tensor $\Sigma_{\rm M}$ is known as the Maxwell stress tensor.

{The above equations, posed in the bulk $\Omega_{\rm i,e}$, are the same as those proposed in \citet{Saville1997_AnnuRevFluidMech}.
However, in \citet{Saville1997_AnnuRevFluidMech}, the specification for interfacial conditions for the ionic concentrations is incomplete.
The author proceeds using heuristic arguments to simplify and ignore certain terms to arrive at the Taylor Melcher model.
In contrast, here, we shall completely specify the interface conditions at $\Gamma$, 
and will use the resulting system of equations as the starting point
for all subsequent discussion. 
Given this, we shall refer to our model as the {\em modified Saville} model.}
At $\Gamma$, 
the boundary conditions for the electrostatic potential are
\begin{equation}\label{poissonbc}
\jump{\phi}=0, \; \jump{\epsilon \PD{\phi}{\mb{n}}}=0,
\end{equation}
where $\mb{n}$ is the normal on $\Gamma$ (pointing from region $\Omega_{\rm i}$ to $\Omega_{\rm e}$)
and $\jump{w}$ is the jump in the value of $w$ across the interface $\Gamma$:
\begin{equation}
\jump{w}\equiv \at{w}{\Gamma_{\rm i}}-\at{w}{\Gamma_{\rm e}}
\end{equation}
with $w_{\Gamma_{\rm i,e}}$ denoting the value of {$w$} evaluated at the $\Omega_{\rm i,e}$ face of $\Gamma$.
For the fluid equations, we have the following interface conditions.
\begin{equation}
\jump{\mb{u}}=0, \; \jump{(\Sigma(\mb{u},p)+\Sigma_{\rm M}(\phi))\mb{n}}=-\gamma_* \kappa\mb{n},
\end{equation}
where $\gamma_*$ is the surface tension coefficient and $\kappa$ is the sum of the principal curvatures of the surface $\Gamma$.
The above boundary condition on the stress, with the help of \eqref{poissonbc}, can also be recast as:
\begin{equation}\label{stressbc}
\jump{\Sigma(\mb{u},p)\mb{n}}=-\jump{\frac{\epsilon}{2}\paren{\paren{\PD{\phi}{\mb{n}}}^2-\abs{\nabla_\Gamma\phi}^2}}\mb{n}
-\gamma_*\kappa \mb{n},
\end{equation}
where $\nabla_\Gamma$ denotes the surface gradient operator on the interface $\Gamma$.
Finally, we impose the kinematic (no-slip) condition; the interface moves with the local fluid velocity. 

On both sides of the interface $\Gamma$, we set{
\begin{equation}\label{concbc}
\begin{split}
\jump{-D_{\rm C}\paren{\nabla c+c\frac{F}{RT}\nabla \phi}\cdot \mb{n}}=0,& \;l_{\rm C}\at{c}{\Gamma_{\rm  e}}=\at{c}{\Gamma_{\rm i}},\\
\jump{-D_{\rm A}\paren{\nabla a-a\frac{F}{RT}\nabla \phi}\cdot \mb{n}}=0,& \;l_{\rm A}\at{a}{\Gamma_{\rm  e}}=\at{a}{\Gamma_{\rm i}},\\
\jump{-D_{\rm S}\nabla s\cdot \mb{n}}=0,& \;l_{\rm S}\at{s}{\Gamma_{\rm  e}}=\at{s}{\Gamma_{\rm i}},
\end{split}
\end{equation}
where $l_{\rm C,A,S}$ are positive constants known as partition coefficients}.
These boundary conditions are the same as those imposed in \citet{zholkovskij2002electrokinetic}, 
{and were missing in the electrodiffusion model proposed in \citet{Saville1997_AnnuRevFluidMech}}.
{We refer the reader back to the discussion surrounding equation \eqref{lXE} for a thermodynamic interpretation of the
partition coefficients $l_{\rm C,A,S}$.}
The partition coefficients $l_{\rm C,A,S}$ must satisfy the following relations:
\begin{equation}\label{th_restriction}
\frac{l_{\rm C}l_{\rm A}}{l_{\rm S}}=\frac{K_{{\rm eq},{\rm i}}}{K_{{\rm eq},\rm e}}, \mbox{ with }\; K_{{\rm eq},i}=\frac{k_{+,i}}{k_{-,i}}, \; i={\rm i,e},
\end{equation}
where $k_{\pm,i}, i={\rm i,e}$ are the rate constants in $\Omega_{i},i={\rm i,e}$ respectively. 
{The above is needed for thermodynamic consistency; the requirement that 
the cation, the anion and the salt each have a defined energy level leads to the above restriction, and guarantees that 
the system of equations, as a whole, satisfies a free energy identity. The details of this are discussed in Appendix \ref{app:energy}.}

If $\Omega$ is bounded, we must impose boundary conditions 
on the outer boundary $\partial \Omega$. Here, 
the boundary conditions are
\begin{equation}
c=c_*, \; a=a_*, \; s=s_*, \; \mb{u}=0, \; \phi=\phi_{\rm b}.
\end{equation}
Assuming electroneutral boundary conditions, 
the concentrations $c_*, a_*$ and $s_*$ satisfy the following equations:
\begin{equation}\label{c0eq}
k_{+,{\rm e}}s_*-k_{-,{\rm e}}c_*a_*=0, \; c_*=a_*,
\end{equation}
where $k_{\pm,{\rm e}}$ denotes the reaction rate constants in $\Omega_{\rm e}$.
We may interpret $\phi_{\rm b}$ as the externally imposed voltage, and may allow $\phi_{\rm b}$
to depend on time.

If $\Omega=\mathbb{R}^3$, we may set:
\begin{equation}\label{concbcatinfty}
\lim_{\abs{\mb{x}}\to \infty}c=c_*, \; \lim_{\abs{\mb{x}}\to \infty}a=a_*, \lim_{\abs{\mb{x}}\to \infty}s=s_*
\end{equation}
where $c_*, a_*$ and $s_*$ are to satisfy \eqref{c0eq}. For the voltage and velocity field, we may set
\begin{equation}\label{phibcatinfty}
\lim_{\abs{\mb{x}}\to \infty}(\phi-\phi_{\rm b})=0, \; \lim_{\abs{\mb{x}}\to \infty}\mb{u}=0, 
\end{equation}
where $\phi_{\rm b}(\bm{x},t)$ is a given function that prescribes the behavior of the voltage at infinity.

\subsection{Non-dimensionalization}\label{non_dimensionalization}
We now non-dimensionalize the above equations. Let the quantities $\wh{\cdot}$ denote the dimensionless quantities.
We set:
\begin{equation}\label{dless_scaling}
\begin{split}
c&=c_*\wh{c}, \; a=c_*\wh{a}, \; s=s_*\wh{s},\;\phi=\frac{RT}{F}\wh{\phi},\; \phi_{\rm b}=\frac{RT}{F}\wh{\phi_{\rm b}},\\
D_{\rm C,A,S}&=D_*\wh{D}_{\rm C,A,S}, \; \epsilon=\epsilon_*\wh{\epsilon}, \; \mu=\mu_*\wh{\mu},\;
\mb{x}=L\wh{\mb{x}}, \; t=t_0\wh{t}=\frac{L}{u_*}\wh{t},\; \kappa=\frac{\wh{\kappa}}{L},\\
\mb{u}&=u_*\wh{\mb{u}}=\frac{\epsilon_*(RT/F)^2}{\mu_*L}\wh{\mb{u}},\; 
p=\frac{\epsilon_*(RT/F)^2}{L^2}\wh{p},
\end{split}
\end{equation}
where $D_*$, $\epsilon_*$, $\mu_*$ and $L$ are the characteristic diffusion coefficient, dielectric constant, 
viscosity and length respectively. {We have taken the characteristic voltage to be the thermal voltage $RT/F$
rather than the characteristic magnitude of the externally imposed voltage. The latter scaling will be discussed 
in Section \ref{large_voltage}.}
Using the above non-dimensionalization, we have, in $\Omega_{\rm i}\cup \Omega_{\rm e}$:
\begin{align}
\label{dlessc}
Pe\paren{\PD{c}{t}+\nabla \cdot (\mb{u}c)}&
=\nabla \cdot \paren{D_{\rm C}\paren{\nabla c+c\nabla \phi}}+{\frac{k_{\rm e} k}{\alpha}\paren{s-\frac{ca}{K}}},\\
\label{dlessa}
Pe\paren{\PD{a}{t}+\nabla \cdot (\mb{u}a)}&=\nabla \cdot \paren{D_{\rm A}\paren{\nabla a-a\nabla \phi}}+
{\frac{k_{\rm e}k}{\alpha}\paren{s-\frac{ca}{K}}},\\
\label{dlesss}
Pe\paren{\PD{s}{t}+\nabla \cdot (\mb{u}s)}&=\nabla \cdot \paren{D_{\rm S}\nabla s}-{k_{\rm e} k\paren{s-\frac{ca}{K}}},\\
\label{dlesscapoisson}
-\delta^2\nabla \cdot (\epsilon \nabla \phi)&=(c-a),\\
\label{dlesscastokes}
\delta^2(\mu \Delta \mb{u}-\nabla p)&=(c-a)\nabla \phi, \; \nabla \cdot \mb{u}=0,
\end{align}
where we have dropped $\wh{\cdot}$ for notational simplicity. 
$Pe$ is the P\'eclet number 
\begin{equation}\label{dlessconstants}
Pe=\frac{u_*L}{D_*}, \; {k_{\rm e}=\frac{k_{+,{\rm e}}L^2}{D_*}},\; \alpha=\frac{c_*}{s_*}=\frac{k_{+,{\rm e}}}{k_{-,{\rm e}}c_*},\;
\delta=\frac{r_{\rm D}}{L}, \; r_{\rm D}=\sqrt{\frac{\epsilon_*RT/F}{Fc_*}},
\end{equation}
and
\begin{equation}
{
k=\begin{cases}
k_{+,{\rm i}}/k_{+,{\rm e}} & \text{ in } \Omega_{\rm i},\\
1 & \text{ in } \Omega_{\rm e},
\end{cases},}\quad
K=\begin{cases}
K_{\rm i}\equiv K_{{\rm eq},{\rm i}}/K_{{\rm eq},{\rm e}} = \frac{l_{\rm C}l_{\rm A}}{l_{\rm S}} & \text{ in } \Omega_{\rm i},\\
1 & \text{ in } \Omega_{\rm e}.
\end{cases}
\end{equation}
Note that the last equality comes from \eqref{c0eq}, and the expression for $K_{\rm i}$ is 
derived from
the thermodynamic restrictions \eqref{th_restriction}.
%
The dimensionless boundary conditions at the interface are 
\begin{align}\label{dlessconcic}
\jump{-D_{\rm C}\paren{\nabla c+c\nabla \phi}\cdot \mb{n}}& =0, \;l_{\rm C}\at{c}{\Gamma_{\rm e}}=\at{c}{\Gamma_{\rm i}},\\
\jump{-D_{\rm A}\paren{\nabla a-a\nabla \phi}\cdot \mb{n}}& =0, \;l_{\rm A}\at{a}{\Gamma_{\rm e}}=\at{a}{\Gamma_{\rm i}},\\
\jump{-D_{\rm S}\nabla s\cdot \mb{n}}& =0, \;l_{\rm S}\at{s}{\Gamma_{\rm e}}=\at{s}{\Gamma_{\rm i}},\\
\label{dlesspoissonbc}
\jump{\epsilon\nabla\phi\cdot\mb{n}} &=0,\;\jump{\phi}=0, \\
\label{dlstressbc}
\delta^2\jump{\Sigma(\mb{u},p)\mb{n}}&=-\delta^2\jump{\frac{\epsilon}{2}\paren{\paren{\PD{\phi}{\mb{n}}}^2-\abs{\nabla_\Gamma\phi}^2}}\mb{n}-\gamma \kappa\mb{n}, \\
\label{dlnoslip}
\jump{\mb{u}}&=0,
\end{align}
with the dimensionless surface tension coefficient $\gamma=\frac{\gamma_*}{c_*RTL}$.
The dimensionless boundary conditions at $\partial \Omega$ are:
\begin{equation}\label{dlessconcbc}
c=a=s=1,\; \phi=\phi_{\rm b},\;\mb{u}=0.
\end{equation}
When $\Omega=\mathbb{R}^3$, the above must be replaced by appropriate limits as $\abs{\bm{x}}\to \infty$ as in \eqref{concbcatinfty} and \eqref{phibcatinfty}.

\section{Charge Diffusion Model}\label{CD}

We first make the assumption that $\alpha\ll 1$ and perform asymptotic calculations  to 
reduce the above full electrokinetic model to a
 {\em charge diffusion model}, where a single equation for the charge density $q=c-a$ replaces the equations 
for $c, a$ and $s$. This charge diffusion model will then be reduced 
further in the Sections \ref{leaky} and \ref{DL} by assuming that $\delta$ is small, to derive the leaky 
dielectric model and possible corrections. The validity of the parametric assumption $\alpha\ll \delta\ll 1$ as in \eqref{scaling_intro} will 
be discussed in Section \ref{mod_large}.
{We point out that this limiting procedure is different from the one in \citet{Saville1997_AnnuRevFluidMech}, 
in which the limit $k_{\rm e}\gg 1$ is considered. 
}

%
Subtracting \eqref{dlessa} from \eqref{dlessc} we obtain the following equation for the charge density $q=c-a$:
\begin{equation}\label{dlessq}
Pe\paren{\PD{q}{t}+\nabla \cdot (\mb{u}q)}=\nabla \cdot (D_{\rm C}\nabla c-D_{\rm A}\nabla a+(D_{\rm C}c+D_{\rm A}a)\nabla \phi).
\end{equation}
Expanding variables in powers of $\alpha \ll 1$
\begin{equation}
\label{alphaexp}
s=s_{(0)}+\alpha s_{(1)}+\mc{O}(\alpha^2), \; c=c_{(0)}+\alpha c_{(1)} +\mc{O}(\alpha^2), \; a=a_{(0)}+\alpha a_{(1)}+\mc{O}(\alpha^2),
\end{equation}
%
%
 the leading order equation obtained from \eqref{dlessq} is simply:
\begin{equation}\label{dlessq0}
Pe\paren{\PD{q_{(0)}}{t}+\nabla \cdot (\mb{u}_{(0)}q_{(0)})}=\nabla \cdot (D_{\rm C}\nabla c_{(0)}-D_{\rm A}\nabla a_{(0)}+(D_{\rm C}c_{(0)}+D_{\rm A}a_{(0)})\nabla \phi_{(0)})
\end{equation}
From equation \eqref{dlessc} we see that
\begin{equation}\label{equil}
s_{(0)}-\frac{c_{(0)}a_{(0)}}{K}=0.
\end{equation}
We thus conclude from \eqref{dlesss} that, to leading order in $\alpha$, 
\begin{equation}
Pe\paren{\PD{s_{(0)}}{t}+\nabla \cdot(\mb{u}_{(0)}s_{(0)})}=\nabla \cdot(D_{\rm S}\nabla s_{(0)})
\end{equation}
with boundary conditions:
\begin{equation}
s_{(0)}=1 \text{ on } \partial \Omega \text{ and } \jump{D_{\rm S}\nabla s_{(0)}}=0, \; 
\at{s_{(0)}}{\Gamma_{\rm i}}=l_{\rm S}\at{s_{(0)}}{\Gamma_{\rm e}} \text{ on } \Gamma.
\end{equation}
We see that, regardless of $\mb{u}_{(0)}$, $s_{(0)}$ approaches the steady state:
\begin{equation}\label{s0eqn}
s_{(0)}=s_{\rm st}=
\begin{cases}
l_{\rm S} &\text{ in } \Omega_{\rm i},\\
1 &\text{ in } \Omega_{\rm e}.
\end{cases}
\end{equation}
We shall thus assume that $s=s_{\rm st}$ at all times. Using this with \eqref{equil}, we 
may eliminate $c_{(0)}, a_{(0)}$ in favor of an equation for $q_{(0)}=c_{(0)}-a_{(0)}$ only.
We have:
\begin{equation}\label{c0a0q}
c_{(0)}=\frac{1}{2}\paren{q_{(0)}+\sqrt{4S+q_{(0)}^2}},\; a_{(0)}=\frac{1}{2}\paren{-q_{(0)}+\sqrt{4S+q_{(0)}^2}}
\end{equation}
where
\begin{equation}\label{Sexp}
S=Ks_{\rm st}=
\begin{cases}
S_{\rm i}=K_{\rm i}l_{\rm S} &\text{ in } \Omega_{\rm i},\\
1 &\text{ in } \Omega_{\rm e}.
\end{cases}
\end{equation}
We may now substitute \eqref{c0a0q} into \eqref{dlessq0} to obtain the following equation for $q_{(0)}$:
\begin{align}\label{qeqn}
Pe\paren{\PD{q}{t}+\nabla\cdot(\mb{u}q)}=&-\nabla \cdot \mb{J}_q,\\
\mb{J}_q=&-\Sigma(q)\paren{\frac{1}{\sqrt{4S+q^2}}\nabla q+\nabla \phi},\\
\Sigma(q)&=\frac{1}{2}\paren{(D_{\rm C}+D_{\rm A})\sqrt{4S+q^2}+(D_{\rm C}-D_{\rm A})q},\label{Sigmaq}
\end{align}
where we dropped the subscript ``$(0)$" for notational simplicity. It is not difficult to see that the 
effective conductivity coefficient $\Sigma(q)$ is positive.

From \eqref{dlessconcbc} we see that $q$ satisfies:
\begin{equation}\label{qbc}
q=0 \text{ on } \partial \Omega \text{ or } q\to 0 \text{ as } \abs{\mb{x}}\to \infty.
\end{equation}
On the interface $\Gamma$, the (dimensionless version of the) flux boundary condition in \eqref{concbc} 
results in the boundary condition:
\begin{equation}\label{qint1}
\jump{\mb{J}_q\cdot \mb{n}}=0.
\end{equation}
The second boundary condition in \eqref{concbc} produces:
\begin{equation}\label{qint2}
\at{\frac{l_{\rm C}}{2}\paren{q+\sqrt{4+q^2}}}{\Gamma_{\rm e}}=\at{\frac{1}{2}\paren{q+\sqrt{4S_{\rm i}+q^2}}}{\Gamma_{\rm i}}.
\end{equation}
We point out that, thanks to the thermodynamic restriction \eqref{th_restriction}, this condition is mathematically equivalent to
\begin{equation}
\at{\frac{l_{\rm A}}{2}\paren{-q+\sqrt{4+q^2}}}{\Gamma_{\rm e}}=\at{\frac{1}{2}\paren{-q+\sqrt{4S_{\rm i}+q^2}}}{\Gamma_{\rm i}}.
\end{equation}
Relation \eqref{th_restriction}, therefore, ensures that the assumption of small $\alpha$ leads to a 
mathematically consistent limiting problem.

\begin{figure}
\begin{center}
\includegraphics[width=0.35\textwidth]{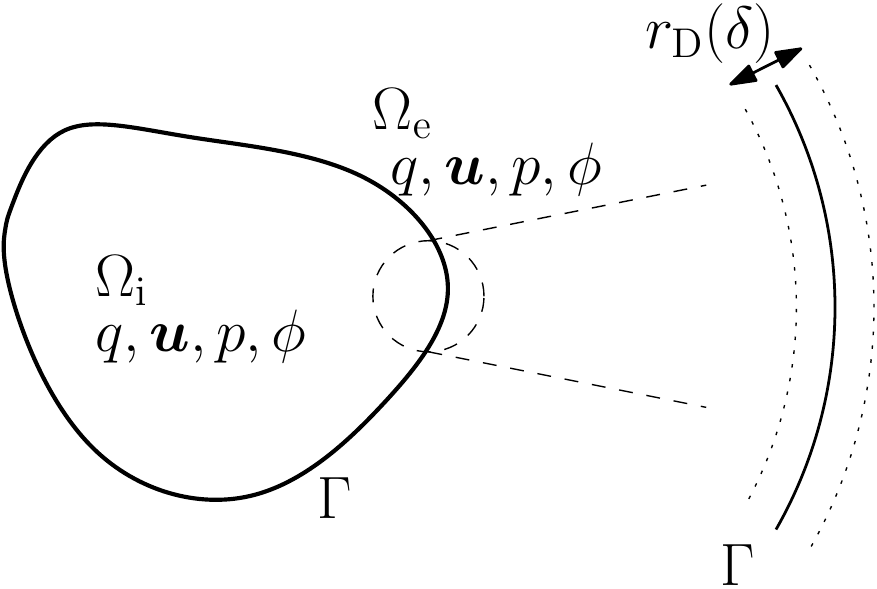}
\end{center}
\caption{\label{charge_diff_fig} Setup for the charge diffusion model. 
In contrast to the modified Saville model (Figure \ref{modified_saville_fig}),
the unknown functions in $\Omega_{\rm i,e}$ are now {the charge $q$, the velocity and pressure fields $\bm{u},p$ and 
the electrostatic potential $\phi$}.
As $\delta\to 0$, a boundary layer of width $r_{\rm D}$ ($\delta$ in dimensionless units) 
develops on both sides of the interface $\Gamma$, as shown on the right.}
\end{figure}

In the resulting reduced model the unknown variables are thus $q,\phi, \mb{u}$ and $p$, where the equations for $q$
were given above (see Figure \ref{charge_diff_fig}). 
The equation for $\phi$ and $\mb{u}$ remain the same as the original model except that we should 
replace $c-a$ in \eqref{dlesscapoisson} and \eqref{dlesscastokes} with $q$:
\begin{align}
\label{dlesspoisson}
-\delta^2 \nabla \cdot \paren{\epsilon\nabla \phi}&=q,\\
\label{dlessstokes}
\delta^2(\mu\Delta \mb{u}-\nabla p)&=q\nabla \phi, \; \nabla \cdot \mb{u}=0.
\end{align}
The interface conditions \eqref{dlesspoissonbc}, \eqref{dlstressbc} and \eqref{dlnoslip} as well as the outer boundary conditions for $\phi$
and $\mb{u}$ remain the same as the modified Saville model.
We shall call this the {\em charge diffusion model}. 
Henceforth, we consider the charge diffusion model instead of the modified Saville model.

We may arrive at the charge diffusion model via a model that is slightly different from the modified Saville model, 
a derivation that may in some cases be more physically relevant.
Suppose that the dielectric itself undergoes ionic dissociation. Then, S in \eqref{dissociation} should be considered 
the dielectric rather than the solute. However, the dielectric in the interior and exterior are different. Let us suppose 
that the the dissociation is such that the cation is the same in both the interior and exterior dielectrics (for example, 
it could be a proton).
The dissociation reaction is
\begin{equation}
{\rm S}_{\rm i,e} \rightleftharpoons {\rm C}^++{\rm A}^-_{\rm i,e}
\end{equation}
in $\Omega_{\rm i}$ and $\Omega_{\rm e}$ respectively where the ${\rm S}_{\rm i,e}$ are the interior/exterior dielectrics
and ${\rm A}_{\rm i,e}$ are the interior/exterior anions. A leaky dielectric is a poor conductor, 
and we thus assume that most of the dielectric is not dissociated. Thus, the ratio between the dissociated and non-dissociated 
dielectric, $\alpha$, is assumed small. The concentrations for $\rm{A}^-_{\rm i,e}$ as well as ${\rm C}^+$ satisfy the dimensionless 
equations \eqref{dlessa} and \eqref{dlessc}, but it would not be appropriate to model the concentration dynamics of $\rm{S}_{\rm i,e}$ with \eqref{dlesss}.
This is because the diffusion equation describes the dynamics of a dilute solute; 
after all, the dielectric is the solvent itself and as such it fills the space. 
{One possibility will be to assume that the concentration of $S$ can be calculated 
from the volume unoccupied by the solutes.  This will lead to small modifications for the electrodiffusion 
equations for the ions as well (see, for example, \citet{zhou2011mean}). An even simpler possibility, valid in the case of dilute solutes, 
will be to assume that $S$ is spatially constant.}
The cation concentration $c$ satisfies the interface condition \eqref{dlessconcic} whereas the anion concentrations 
$a_{\rm i,e}$ satisfy
\begin{equation}
D_{\rm A_{\rm i,e}}\paren{\nabla a_{\rm i,e}-a_{\rm i,e}\nabla \phi}\cdot \mb{n}=0,
\end{equation}
since ${\rm A}_{\rm i,e}$ do not cross the solvent interface $\Gamma$. 
We can then recover 
the charge diffusion model as $\alpha\to 0$. 

An important difference between the first derivation via the modified Saville model 
and second derivation outlined above is that, in the second 
derivation, the constant $S$ of \eqref{Sexp} is a material property that does not depend on any external boundary condition
at $\partial \Omega$ or the far field; in the first derivation, $S$ depended on the value of the solute concentration 
at the boundary $\partial \Omega$, but in the second, $S$ depends on the concentration of the dielectric, a material property. 
As we can see from \eqref{qeqn}, $S$ determines the conductivity; the second derivation supports the view that 
the conductivity should be a material property.
The only constant in the charge diffusion model that depends not on the material 
property but on the property {\em between} materials is the partition coefficient. 

The partition coefficients
determine whether there is a voltage jump ({\em Galvani potential}) 
and thus an electric double layer (EDL) across the interface $\Gamma$,
{as already discussed in Section \ref{EDLGP} using a thermodynamic argument. Below, 
we illustrate this with a slightly different argument. We emphasize that this heuristic picture will later be justified 
through systematic asymptotic calculations in Sections \ref{derivation_of_TM} and \ref{subsec:lc_neq_la}.}

Consider the special problem in which {fluid velocity is $0$} and the boundary voltage at $\partial \Omega$
to be $\phi_{\rm b}=0$. In this case, we see that $q=0$ (with $\phi=0$) satisfies \eqref{qeqn}, \eqref{qbc}, 
\eqref{qint1}, \eqref{qint2} so long as
\begin{equation}\label{lCsqrtS1}
l_{\rm C}=\sqrt{S_{\rm i}}.
\end{equation}
According to \eqref{th_restriction}, this condition is satisfied when 
\begin{equation}\label{lClA}
l_{\rm C}=l_{\rm A}.
\end{equation}
In other words, condition \eqref{lClA} ensures that the 
state in which the system is everywhere electroneutral ($q=0$) is a steady state solution of the system.
Under \eqref{lClA}, \eqref{qint2} can be reduced to the condition:
\begin{equation}\label{qint2eq}
\at{l_{\rm C}q}{\Gamma_{\rm e}}=\at{q}{\Gamma_{\rm i}}.
\end{equation}
If $l_{\rm C}\neq l_{\rm A}$, {a globally electroneutral steady state with no interfacial layer is impossible,}
and we expect an accumulation of charge on the 
interface $\Gamma$ resulting 
in a voltage jump across the interface, even in the absence of an externally imposed electric field.
In fact, it is possible to obtain an expression for this voltage jump by the following heuristic argument,
{which uses some results from Appendix \ref{app:energy}}.
At equilibrium, we expect the chemical potential, 
or energy per unit charge, to be equal on both sides of the interface. Equations 
\eqref{chargediff} and \eqref{CD_energies} suggest that:
\begin{equation}
\at{(\mu_q+E_q)}{\Omega_{\rm e}}=\at{(\mu_q+E_q)}{\Omega_{\rm i}}.
\end{equation}
In the bulk, $q$ should approximately be equal to $0$, and therefore, we have:
\begin{equation}
\at{\phi}{\Omega_{\rm e}}=\ln\sqrt{S_{\rm i}}-\ln l_{\rm C}+\at{\phi}{\Omega_{\rm i}}.
\end{equation}
Thus,
\begin{equation}\label{phijumpheuristic}
\at{\phi}{\Omega_{\rm i}}-\at{\phi}{\Omega_{\rm e}}=\ln\paren{\frac{l_{\rm C}}{\sqrt{S_{\rm i}}}}=\frac{1}{2}\ln\paren{\frac{l_{\rm C}}{l_{\rm A}}}.
\end{equation}
We see that $l_{\rm C}\neq l_{\rm A}$ leads to a voltage jump across the interface 
{ and hence an electric double layer (see Figure \ref{galvani_EDL_fig}).}
This suggests that the cases $l_{\rm C}=l_{\rm A}$ and $l_{\rm C}\neq l_{\rm A}$ are qualitatively different and we will thus
treat these two cases separately.

\section{Taylor-Melcher Limit}\label{leaky}
\subsection{Derivation of the Taylor-Melcher Model}\label{derivation_of_TM}
As discussed above, the cases $l_{\rm C}=l_{\rm A}$ and $l_{\rm C}\neq l_{\rm A}$ are fundamentally different.
In this section, we consider the case $l_{\rm C}=l_{\rm A}$. Under this assumption, 
we consider the limit $\delta\to 0$ in the charge diffusion model to derive the TM model.
In the calculations to follow, we scale $Pe$ and $\gamma$ with respect to $\delta$ as follows:
\begin{equation}\label{scaling}
Pe=\frac{\chi}{\delta^2}, \; \gamma=\wh{\gamma}\delta^2,
\end{equation}
where $\chi$ and $\wh{\gamma}$ are constant. 
As we shall see, this scaling yields the full TM model with surface charge convection for arbitrary interfacial geometry.
{In fact, there are three time scales in the TM model, the Maxwell-Wagner charge relaxation time, the electrohydrodynamic 
time and the capillary time scale \citep{salipante2010electrohydrodynamics}. All terms in the TM model will be important only when 
these three time scales are of the same order, and the above scaling ensures this.}
Different distinguished limits can be obtained depending on how we scale the two dimensionless numbers $Pe$ and $\gamma$.
The scaling of $Pe$ with respect to $\delta$ determines whether surface charge convection will be important. 
This is summarized in Table \ref{dist_limits}. 
$Pe=\mc{O}(\delta^{-1})$ leads to the TM model with surface charge convection as an $\mc{O}(\delta)$ correction.
For smaller $Pe$, surface charge convection is negligible.
We shall not present these calculations here since they are quite similar to (and simpler than) the calculations presented here. 
In Appendix \ref{app:energy}, we shall also see that \eqref{scaling} is the natural one as dictated by the free energy identity.

\begin{table}
\begin{center}
\begin{tabular}{c|c}
$Pe$ or $Pe_{\rm E}$ & surface charge convection \\
\hline
$\mc{O}(\delta^{-2})$ or $\mc{O}(\delta_{\rm E}^{-2})$ & $\mc{O}(1)$ \\
$\mc{O}(\delta^{-1})$ or $\mc{O}(\delta_{\rm E}^{-1})$ & $\mc{O}(\delta)$ or $\mc{O}(\delta_{\rm E})$\\
$\mc{O}(\delta^{k})$ or $\mc{O}(\delta_{\rm E}^k)$, $k\geq 0$ & smaller than $\mc{O}(\delta)$ or $\mc{O}(\delta_{\rm E})$\\
\end{tabular}
\end{center}
\caption{\label{dist_limits} The presence of surface charge convection in the limiting TM model 
depending on the scaling of $Pe$ or $Pe_{\rm E}$ with respect to $\delta$ or $\delta_{\rm E}$. 
The dimensionless constants $Pe_{\rm E}$ and $\delta_{\rm E}$ pertain to asymptotics under strong electric 
fields discussed in Section \ref{large_voltage}, and are defined in \eqref{dless_strong} and \eqref{PeE}.
}
\end{table}

We expand $q$ in powers of $\delta $ as:
\begin{equation}\label{qpower}
q=q_0+\delta q_1+\delta^2 q_2 +\cdots,
\end{equation}
and likewise for other variables $\phi$ and $\mb{u}$.

Let us first consider equations in the bulk or outer layer. We see from \eqref{dlesspoisson} that:
\begin{equation}\label{q01zero}
q_0=q_1=0.
\end{equation}
This is also compatible with \eqref{qeqn}. The leading non-trivial equations we obtain 
from \eqref{dlesspoisson}, \eqref{qeqn} and \eqref{dlessstokes} are:
\begin{align}
\chi\paren{\PD{q_2}{t}+\nabla \cdot (\mb{u}_0q_2)}&=\nabla \cdot (\sigma\nabla \phi_0),\label{bulk_q2}\\
-\nabla \cdot(\epsilon\nabla \phi_0)&=q_2,\label{bulk_poisson}\\
\mu\Delta \mb{u}_0-\nabla p_0&=q_2\nabla \phi_0,\; \nabla \cdot \mb{u}_0=0,\label{bulk_stokes}
\end{align}
where the conductivity $\sigma$ is given by:
\begin{equation}
\sigma=\Sigma(q_0=0)=(D_{\rm C}+D_{\rm A})\sqrt{S}=\begin{cases}
(D_{\rm{C,i}}+D_{\rm{A,i}})\sqrt{l_{\rm C}l_{\rm A}} &\text{ in } \Omega_{\rm i},\\
D_{\rm{C,e}}+D_{\rm{A,e}} &\text{ in } \Omega_{\rm e}.
\end{cases}
\end{equation}
where we used the relation \eqref{Sexp}, \eqref{lCsqrtS1} and \eqref{th_restriction} in the last equality.
We shall sometimes use the notation $\sigma_k, k={\rm i,e}$ to denote the value of $\sigma$ in the two regions.
In this section, $l_{\rm C}=l_{\rm A}$ by assumption, and therefore, $\sqrt{S}=l_{\rm C}$ in $\Omega_{\rm i}$.

\begin{figure}
\begin{center}
\includegraphics[width=0.5\textwidth]{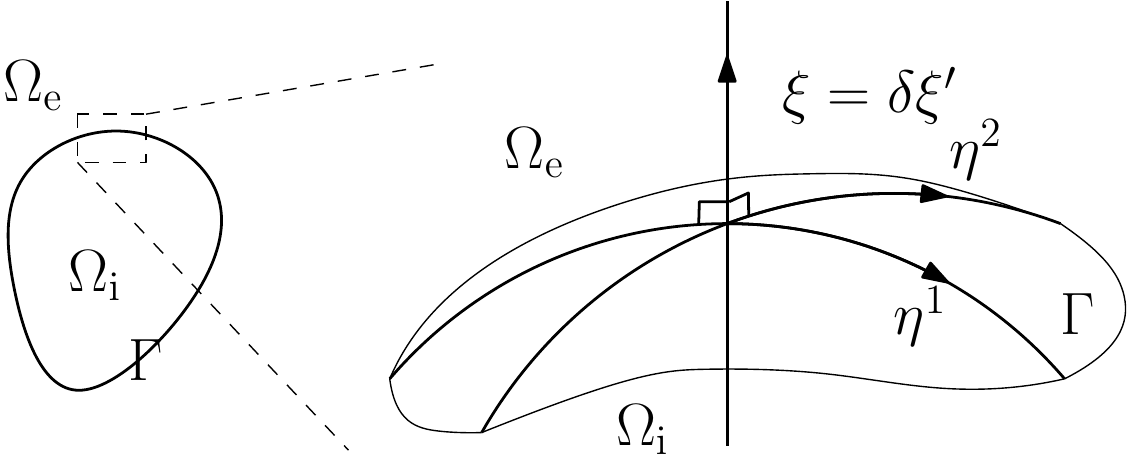}
\end{center}
\caption{\label{curvilinear} The curvilinear coordinate system. $\xi$ is perpendicular to the 
surface $\Gamma$, where $\xi<0$ is on the $\Omega_{\rm i}$ side. The $\xi=0$ surface corresponds 
to $\Gamma$ and $\bm{\eta}=(\eta^1,\eta^2)$ is the coordinate system fitted to $\Gamma$.}
\end{figure}

In order to obtain interface conditions for the above equations in the outer layer, 
we perform a boundary layer analysis near $\Gamma$.
Small $\delta$ in \eqref{dlesspoisson} implies a boundary layer of thickness $\delta $ near $\Gamma$ (see Figure \ref{charge_diff_fig}).
To study this boundary layer, we introduce an curvilinear coordinate system 
$(\xi,\bm{\eta})=(\xi,\eta^1,\eta^2)$ fitted to the interface $\Gamma$ (Figure \ref{curvilinear}).
Let $\bm{\eta}$ be a local coordinate system on the surface $\Gamma$ so that $\mb{x}=\mb{X}(\bm{\eta},t)$ is the 
cartesian coordinate position of the evolving interface. We let points of fixed $\bm{\eta}$ move with the normal 
velocity of the interface:
\begin{equation}
\PD{\mb{X}}{t}=u_\perp(\bm{\eta},t)\mb{n}(\bm{\eta},t), \quad u_\perp(\bm{\eta},t)=\mb{u}(\mb{X}(\bm{\eta},t),t)\cdot \mb{n}(\bm{\eta},t),
\end{equation}
where, $\mb{n}$ is the outward pointing unit normal (pointing from $\Omega_{\rm i}$ to $\Omega_{\rm e}$).
The map $\mc{T}$:
\begin{equation}\label{coordT}
\mc{T}(\xi,\bm{\eta},t)\mapsto \mb{X}(\bm{\eta},t)+\xi \mb{n}(\bm{\eta},t)
\end{equation}
defines the desired local curvilinear coordinate system. The coordinate $\xi$
is thus the signed distance function from the interface, 
where $\xi>0$ is on the $\Omega_{\rm e}$ side and $\xi<0$ on the $\Omega_{\rm i}$ side.
We use an arbitrary point $\mb{x}=\mb{x}_*\in \Gamma$ at 
time $t=t_*$ as the origin in the $(\xi,\bm{\eta})$ coordinate system in the following boundary layer analysis.

%

In an abuse of notation, $q, \phi$ and other scalar functions will
be seen interchangeably as functions of $(\xi,\bm{\eta},t)$ as well as of $(\mb{x},t)$.
For the velocity field $\mb{u}$, we introduce the velocity functions $(u,v^1,v^2)$
adapted to the curvilinear coordinate system:
\begin{equation}\label{vcurvilinear}
\mb{u}(\mb{X},t)=u(\xi,\bm{\eta},t)\PD{\mc{T}}{\xi}+v^i(\xi,\bm{\eta},t)\PD{\mc{T}}{\eta^i}
\end{equation}
where $\mc{T}$ is the local coordinate map \eqref{coordT}. In the above and henceforth, 
we shall use the summation convention for repeated indices. The function $u$ is the fluid velocity 
normal to level sets of the signed distance function of $\Gamma$
and $v^1, v^2$ are the fluid velocities tangent to the same level sets. 

First we introduce the stretched boundary layer coordinate system $\xi' \equiv \xi/\delta$
close to the interface, with 
the inner layer variables denoted by $\wt{\cdot}$:
\begin{equation}
\wt{q}(\xi',\bm{\eta},t)=q(\xi/\delta,\bm{\eta},t),
\end{equation}
and likewise for $\phi, u, v_1$ and $v_2$. 
We then expand each of these variables in powers of $\delta $ as follows:
\begin{equation}
\wt{q}=\wt{q}_0+\delta \wt{q}_1+\delta^2 \wt{q}_2\cdots.
\end{equation}
Expressions resulting from the Stokes equation \eqref{dlessstokes} 
in inner layer coordinates are discussed in Appendix \ref{tensors}.

In the calculations to follow,
the symbol $\jump{\cdot}$, applied to an inner layer variable, is the jump 
in the value across $\xi'=0$
($\jump{\cdot}=\at{\cdot}{\xi'=0-}-\at{\cdot}{\xi'=0+}$).
When applied to an outer layer variable, the symbol $\jump{\cdot}$ denotes the jump
in the value across $\xi=0$: $\jump{\cdot}=\at{\cdot}{\xi=0-}-\at{\cdot}{\xi=0+}$.

From equations \eqref{dlesspoisson} and \eqref{dlessstokes} (see \eqref{stokescurv0-1}) we have:
\begin{equation}
\label{q00}
-\PDD{2}{\wt{\phi}_0}{\xi'}=\wt{q}_0, \quad \wt{q}_0\PD{\wt{\phi}_0}{\xi'}=0 
\rightarrow \wt{q}_0=0,\quad\PD{\wt{\phi}_0}{\xi'}=0.
\end{equation}
%
%
The interface conditions \eqref{dlesspoissonbc} yield
\begin{equation}
\jump{\wt{\phi}_0}=0.
\end{equation}
Given the matching conditions 
\begin{equation}\label{phi0limits}
\lim_{\xi'\to \pm \infty} \wt{\phi}_0=\lim_{\xi\to 0\pm} \phi_0\equiv \phi_\pm,
\end{equation}
we see that
\begin{equation}\label{phi0}
 \wt{\phi}_0\equiv \lim_{\xi\to 0\pm} \phi_0 \text{ for } -\infty<\xi'<\infty.
\end{equation}
In particular, this implies that
\begin{equation}\label{phicont}
\jump{\phi_0}=0.
\end{equation}
This shows that the voltage, to leading order, must be continuous across the boundary layer.
Expressions \eqref{q00} and \eqref{phi0} are compatible with the following additional equations and boundary 
conditions that we obtain as leading order equations from 
\eqref{dlesspoissonbc} and \eqref{qint2eq}:
\begin{equation}
\jump{\epsilon \PD{\wt{\phi}_0}{\xi'}}=0, \; l_{\rm C}\at{\wt{q}_0}{\xi'=0+}=\at{\wt{q}_0}{\xi'=0-}.
\end{equation}
Note in particular that $\wt{q}_0\equiv 0$ is compatible with \eqref{qint2} only because $l_{\rm C}=l_{\rm A}$ (in which 
case \eqref{qint2} reduces to \eqref{qint2eq}). 

We now turn to equations at the next order. From \eqref{dlessstokes} we obtain 
\begin{equation}
\PD{\wt{u}_0}{\xi'}=0, \quad \mu\PDD{2}{\wt{v}^i_0}{\xi'}=\wt{q}_0g^{ij}\PD{\wt{\phi}_0}{\eta^j}=0,\; i=1,2,
\end{equation}
where we used \eqref{q00} in the second expression. 
{This shows that
\begin{equation}\label{UVWconsts}
\wt{u}_0=\begin{cases} 
U_-& \text{ for } \xi'<0,\\
U_+&\text{ for } \xi'>0,
\end{cases}
\quad
\wt{v}^i_0=\begin{cases} 
V_-^i\xi'+W_-^i &\text{ for } \xi'<0,\\
V_+^i\xi'+W_+^i &\text{ for } \xi'>0,
 \end{cases}
\end{equation}
where $U_\pm, V_\pm^i, W_\pm^i$ do not depend on $\xi'$.
Condition \eqref{dlnoslip} 
at the interface $\xi'=0$ gives rise to:
\begin{equation}\label{u0vi0eq0}
\jump{\wt{u}_0}=\jump{\wt{v}^i_0}=0.
\end{equation}
The matching conditions at $\xi'=\pm\infty$ are given by:
\begin{equation}\label{u0vi0}
\lim_{\xi'\to \pm \infty} \wt{u}_0=\lim_{\xi\to 0\pm} u_0, \; \; \lim_{\xi'\to \pm \infty} \wt{v}^i_0=\lim_{\xi\to 0\pm} v^i_0.
\end{equation}
Implicit in the above matching conditions is that all of the above limits exist. In particular, the 
limits of $\wt{v}_0^i$ at $\xi'\to \pm\infty$ must exist, and therefore, $V_\pm^{i}=0$
in \eqref{UVWconsts}. Using \eqref{u0vi0eq0}, we thus conclude that:}
\begin{equation}\label{wtu0}
\wt{u}_0\equiv \lim_{\xi\to 0\pm} u_0 \text{ and } \wt{v}^i_0\equiv \lim_{\xi\to 0\pm} v^i_0, i=1,2, \text{ for } -\infty<\xi'<\infty.
\end{equation}
In particular, this implies that:
\begin{equation}\label{ucont}
\jump{u_0}=\jump{v^i_0}=0,\;\; i=1,2.
\end{equation}
This shows that the velocity field in the outer layer is continuous across the boundary layer to leading order.

We need further interface conditions solve the outer layer equations \eqref{bulk_q2}-\eqref{bulk_stokes}.
We first consider the incompressibility condition in \eqref{dlessstokes} and \eqref{dlnoslip}, from which we obtain (see \eqref{divcurv1}):
\begin{equation}
\PD{\wt{u}_1}{\xi'}+\kappa \wt{u}_0+\frac{1}{\sqrt{\abs{g}}}\PD{}{\eta^i}\paren{\sqrt{\abs{g}}\wt{v}^i_0}=0, \; \jump{\wt{u}_1}=0,
\end{equation}
where $\kappa$ is the sum of the two principal curvatures of the surface $\Gamma$ and 
$\abs{g}$ is the determinant of the metric tensor $g_{ij}$ associated with the interface:
\begin{equation}
g_{ij}=\PD{\mb{X}}{\eta^i}\cdot \PD{\mb{X}}{\eta_j}, \; \abs{g}=\det(g_{ij}).
\end{equation}
By \eqref{u0vi0}, $\wt{u}_0$ and $\wt{v}^i_0$ do not depend on $\xi'$ and are equal to the outer layer 
values $u_0$ and $v^i_0$ respectively. Thus:
\begin{equation}\label{u1exp}
\wt{u}_1=-\paren{\kappa u_0+\frac{1}{\sqrt{\abs{g}}}\PD{}{\eta^i}\paren{\sqrt{\abs{g}}v^i_0}}\xi'+\wt{u}_{\perp,1}
\end{equation}
where $\wt{u}_{\perp, 1}$ is the value of $\wt{u}_1$ at $\xi'=0$.

We first focus on the equations for $q$ and $\phi$.
From \eqref{qeqn} and \eqref{dlesspoisson}, we obtain the following equations:
\begin{align}\label{q1eqn}
\chi\paren{\PD{\wt{q}_1}{t}+(\wt{u}_1-\wt{u}_{\perp,1})\PD{\wt{q}_1}{\xi'}+\wt{v}^i_0\PD{\wt{q}_1}{\eta^i}}
&=-\PD{\wt{J}_{q,1}}{\xi'},\\
\wt{J}_{q,1}&=-\frac{1}{2}(D_{\rm C}+D_{\rm A})\PD{\wt{q}_1}{\xi'}-\sigma\PD{\wt{\phi}_1}{\xi'},\\
\label{phi1poisson}
-\epsilon\PDD{2}{\wt{\phi}_1}{\xi'}&=\wt{q}_1, 
\end{align}
with interface conditions:
\begin{equation}\label{intcond1}
\jump{\wt{\phi}_1}=\jump{\epsilon\PD{\wt{\phi}_1}{\xi'}}=\jump{\wt{J}_{q,1}}=0, \; \;l_{\rm C}\at{\wt{q}_1}{\xi'=0+}=\at{\wt{q}_1}{\xi'=0-}.
\end{equation}
The matching conditions for $\wt{q}_1$ are, by \eqref{q01zero}:
\begin{equation}\label{q1match}
\lim_{\xi'\to \pm\infty} \wt{q}_1=0.
\end{equation}

The above, together with matching conditions for $\wt{\phi}_1$, to be discussed shortly, should be
sufficient to match the outer and inner layer solutions. However, the impossibility of obtaining closed-form 
solutions for the inner layer equations makes the matching procedure difficult without additional assumptions. 
We assume that $\wt{q}_1$ is integrable:
\begin{equation}\label{qintegral}
\int_{-\infty}^\infty \abs{\wt{q}_1}d\xi'<\infty.
\end{equation}
Furthermore, we assume that:
\begin{equation}\label{qlimits}
\lim_{\xi'\to \pm\infty}\PD{\wt{q}_1}{\xi'}=0, \; \lim_{\xi'\to \pm \infty} \xi'\wt{q}_1=0.
\end{equation}
The above two conditions in \eqref{qlimits} are likely to be consequences of \eqref{qintegral}, \eqref{q1match} 
together with the fact that $\wt{q}_1$ satisfies \eqref{q1eqn}-\eqref{phi1poisson}, and hence, redundant.

{To obtain matching conditions for $\wt{\phi}_1$, we invoke Kaplun's matching procedure 
by introducing an intermediate coordinate system 
$\xi_{\rm m}$ which scales as $\xi=\delta^\nu \xi_{\rm m}, 0<\nu<1$. The result of this analysis is that:
\begin{equation}\label{phi1match}
\lim_{\xi'\to \pm\infty}\PD{\wt{\phi}_1}{\xi'}=\lim_{\xi\to 0\pm}\PD{\phi_0}{\xi}.
\end{equation}
The derivation of these conditions relies on \eqref{phi1poisson} and \eqref{qintegral}, 
which ensures that the limits $\lim_{\xi'\to \pm \infty} \partial\wt{\phi}_1/\partial \xi'$ exist.}
An immediate consequence of the above is that we may integrate \eqref{phi1poisson} from $\xi'=-\infty$ to $\infty$ to find that
\begin{equation}\label{dphi0jump}
\jump{\epsilon\PD{\phi_0}{\xi}}=q_\Gamma, \quad q_\Gamma\equiv\int_{-\infty}^\infty \wt{q}_1d\xi'.
\end{equation}

Next, integrate equation \eqref{q1eqn} from $\xi=0$ to $\infty$. Let us first consider the left hand side. We have:
\begin{equation}
\begin{split}
&\int_0^\infty\paren{\PD{\wt{q}_1}{t}+(\wt{u}_1-\wt{u}_{\perp,1})\PD{\wt{q}_1}{\xi'}+\wt{v}^i_0\PD{\wt{q}_1}{\eta^i}}d\xi'\\
=&\PD{q_\Gamma^+}{t}+\paren{\kappa u_0+\frac{1}{\sqrt{\abs{g}}}\PD{}{\eta^i}\paren{\sqrt{\abs{g}}v^i_0}}q_\Gamma^++v^i_0\PD{q_\Gamma^+}{\eta^i}, \quad
q_\Gamma^+\equiv\int_0^\infty \wt{q}_1d\xi'.
\end{split}
\end{equation}
Note that $q_\Gamma^+$ is well-defined thanks to \eqref{qintegral}. The second term in the integrand was integrated 
by parts, where we used \eqref{u1exp} and \eqref{qlimits}. In the last term in the integrand, we replaced $\wt{v}^i_0$
with the outer layer value using \eqref{u0vi0}. Let us now turn to the right hand side of \eqref{q1eqn}.
We integrate to obtain:
\begin{equation}
\int_0^\infty -\PD{\wt{J}_{q,1}}{\xi'}d\xi'=\at{\sigma\PD{\phi_0}{\xi}}{\xi=0+}+\at{\wt{J}_{q,1}}{\xi'=0+}
\end{equation}
where we used \eqref{qlimits} and \eqref{phi1match}. Combining the above two equations, we have:
\begin{equation}
\chi\paren{\PD{q_\Gamma^+}{t}+\kappa u_0q_\Gamma^++\frac{1}{\sqrt{\abs{g}}}\PD{}{\eta^i}\paren{\sqrt{\abs{g}}v^i_0q_\Gamma^+}}
=\at{\sigma\PD{\phi_0}{\xi}}{\xi=0+}+\at{\wt{J}_{q,1}}{\xi'=0+}.
\end{equation}
We may perform a similar calculation on the $\xi'<0$ side and use \eqref{intcond1} to conclude that:
\begin{equation}\label{qgammaeqn}
\chi\paren{\PD{q_\Gamma}{t}+\kappa u_0q_\Gamma+\frac{1}{\sqrt{\abs{g}}}\PD{}{\eta^i}\paren{\sqrt{\abs{g}}v^i_0q_\Gamma}}
=-\jump{\sigma\PD{\phi_0}{\xi}},
\end{equation}
where $q_\Gamma$ was defined in \eqref{dphi0jump}.

To obtain the stress boundary conditions, 
let us first focus on the velocity field component $u$.
Equation \eqref{dlessstokes} yields \eqref{stokescurv01}, which together with \eqref{q00} and \eqref{wtu0} gives:
\begin{equation}
\label{fuinner}
\mu\PDD{2}{\wt{u}_1}{\xi'}-\PD{\wt{p}_0}{\xi'}-\wt{q}_1\PD{\wt{\phi}_1}{\xi'}=0.
\end{equation}
The interface conditions \eqref{dlnoslip} and \eqref{dlstressbc} at $\xi'=0$ yield (see \eqref{Sigmanormal}):
\begin{equation}\label{intinneru}
\jump{2\mu\PD{\wt{u}_1}{\xi'}-\wt{p}_0}=-\jump{\frac{\epsilon}{2}\paren{\paren{\PD{\wt{\phi}_1}{\xi'}}^2-g^{ij}\PD{\wt{\phi}_0}{\eta^i}\PD{\wt{\phi}_0}{\eta^j}}}
-\wh{\gamma}\kappa,
\end{equation}
where $g^{ij}$ are the components of the inverse of the metric tensor $g_{ij}$.
The matching conditions at $\xi'=\pm\infty$ for $p_0$ are
\begin{equation}\label{matchp1}
\lim_{\xi'\to \pm \infty} \wt{p}_0=\lim_{\xi\to 0\pm} p_0.
\end{equation}
Note that the incompressibility condition in the outer layer, at $\xi=0\pm$, can be expressed as:
\begin{equation}
\at{\paren{\PD{u_0}{\xi}+\kappa u_0+\frac{1}{\sqrt{\abs{g}}}\PD{}{\eta^i}\paren{\sqrt{\abs{g}}v^i_0}}}{\xi=0\pm}=0.
\end{equation}
Comparing this expression with \eqref{u1exp} yields:
\begin{equation}\label{matchu1}
\PD{\wt{u}_1}{\xi'}=\at{\PD{u_0}{\xi}}{\xi=0\pm}.
\end{equation}
We see from \eqref{u1exp} that $\wt{u}_1$ is a linear function of $\xi'$. 
This observation may be used together with \eqref{fuinner} to obtain:
\begin{equation}\label{fuinnermod}
2\mu\PDD{2}{\wt{u}_1}{\xi'}-\PD{\wt{p}_0}{\xi'}=\wt{q}_1\PD{\wt{\phi}_1}{\xi'}.
\end{equation}
Let us integrate the above from $\xi'=0$ to $\xi'=\infty$. We have:
\begin{equation}
\at{\paren{2\mu\PD{u_0}{\xi}-p_0}}{\xi=0+}-\at{\paren{2\mu\PD{\wt{u}_1}{\xi'}-\wt{p}_0}}{\xi'=0+}=\int_0^\infty \wt{q}_1\PD{\wt{\phi}_1}{\xi'}d\xi',
\end{equation}
where we used \eqref{matchu1} to obtain the first term in the above. 
Note that the right hand side of the above is indeed integrable thanks to our assumption \eqref{qintegral}.
Indeed, using \eqref{phi1poisson}, we have:
\begin{equation}
\begin{split}
\int_0^\infty \wt{q}_1\PD{\wt{\phi}_1}{\xi'}d\xi'&=-\int_0^\infty \epsilon\PDD{2}{\wt{\phi}_1}{\xi'}\PD{\wt{\phi}_1}{\xi'}d\xi'\\
&=\at{\frac{\epsilon}{2}\paren{\PD{\wt{\phi}_1}{\xi'}}^2}{\xi'=0+}
-\lim_{\xi'\to \infty}{\frac{\epsilon}{2}\paren{\PD{\wt{\phi}_1}{\xi'}}^2}\\
&=\at{\frac{\epsilon}{2}\paren{\PD{\wt{\phi}_1}{\xi'}}^2}{\xi'=0+}
-\at{\frac{\epsilon}{2}\paren{\PD{\phi_1}{\xi}}^2}{\xi=0+}.
\end{split}
\end{equation}
In the last equality, we used \eqref{phi1match}, which in turn relied on condition \eqref{qintegral}.
Integrating \eqref{fuinnermod} with from $\xi'=0$ to $-\infty$
and following the same procedure, we obtain a similar relation on the $\xi'<0$ side. If we combine the calculations 
at the $\xi'>0$ and $\xi'<0$ sides, we find that:
\begin{equation}
{\jump{2\mu\PD{u_0}{\xi}-p_0+\frac{\epsilon}{2}\paren{\PD{\phi_0}{\xi}}^2}}
={\jump{2\mu\PD{\wt{u}_1}{\xi'}-\wt{p}_0+\frac{\epsilon}{2}\paren{\PD{\wt{\phi}_1}{\xi}}^2}}
\end{equation}
We may now combine this with \eqref{intinneru} and \eqref{phi0} to find that:
\begin{equation}\label{stressperp}
\jump{2\mu\PD{u_0}{\xi}-p_0+\frac{\epsilon}{2}\paren{\paren{\PD{\phi_0}{\xi}}^2-g^{ij}\PD{\phi_0}{\eta^i}\PD{\phi_0}{\eta^j}}}=-\wh{\gamma}\kappa
\end{equation}
This is the normal stress balance condition for the outer variables (see \eqref{Sigmanormal}). 

Next we turn to the equations for $v^i$. Equation \eqref{dlessstokes} yields \eqref{stokescurvi1}, which together with \eqref{wtu0} and \eqref{q00} gives:
\begin{equation}\label{fvinner}
\mu\PDD{2}{\wt{v}^i_1}{\xi'}=q_1g^{ij}\PD{\phi_0}{\eta^j}, \; i=1,2.
\end{equation}
We have used \eqref{phi0} to replace the inner layer variable $\wt{\phi}_0$ with the outer layer variable $\phi_0$.
The interface conditions \eqref{dlstressbc} at $\xi'=0$ gives (see \eqref{Sigmanormal}):
\begin{equation}
\jump{\wt{v}^i_1}=0,\; \jump{\mu\paren{\PD{\wt{v}^i_1}{\xi'}+g^{ij}\PD{\wt{u}_0}{\eta^j}}}=0,
\end{equation}
Using Kaplun's matching procedure as we did for $\wt{\phi}_1$, we obtain, with the help of \eqref{qintegral}, 
\begin{equation}\label{matchv1}
\lim_{\xi'\to \pm \infty} \PD{\wt{v}^i_1}{\xi'}=\lim_{\xi\to 0\pm} \PD{v^i_0}{\xi}.
\end{equation}
We may now integrate \eqref{fvinner} over $0<\xi'<\infty$ to obtain
\begin{equation}
\at{\mu\PD{v^i_0}{\xi}}{\xi=0+}-\at{\mu\PD{\wt{v}^i_1}{\xi'}}{\xi'=0+}=\paren{\int_0^\infty \wt{q}_1d\xi'}g^{ij}\PD{\phi_0}{\eta^j}
\end{equation}
where we used \eqref{matchv1}.
We may likewise integrate \eqref{fvinner} over $-\infty<\xi'<0$ and combine this with the above to find that
\begin{equation}
{\jump{\mu\PD{v^i_0}{\xi}}}-{\jump{\mu\PD{\wt{v}^i_1}{\xi'}}}=-q_\Gamma g^{ij}\PD{\phi_0}{\eta^j}, 
\end{equation}
where $q_\Gamma$ was defined in \eqref{dphi0jump}. The above, together with \eqref{matchv1} and \eqref{wtu0}
yields
\begin{equation}\label{stressparallel}
\jump{\mu\paren{\PD{v^i_0}{\xi}+g^{ij}\PD{u_0}{\eta^j}}}=-q_\Gamma g^{ij}\PD{\phi_0}{\eta^j}.
\end{equation}
This is the tangential stress balance condition for the outer variables (see \eqref{Sigmanormal}).
This concludes our discussion of the interfacial boundary conditions.
At the outer boundary $\partial \Omega$ (or at $\abs{\mb{x}}=\infty$), there is no boundary layer, and we simply obtain the 
conditions:  
\begin{equation}\label{outerbcLD}
\mb{u}_0=0, \; q_2=0, \; \phi_0=\phi_{\rm b} \text{ at } \mb{x}\in \partial \Omega \text{ or } \abs{\mb{x}}=\infty. 
\end{equation}\\
%
%

\begin{figure}
\begin{center}
\includegraphics[width=0.3\textwidth]{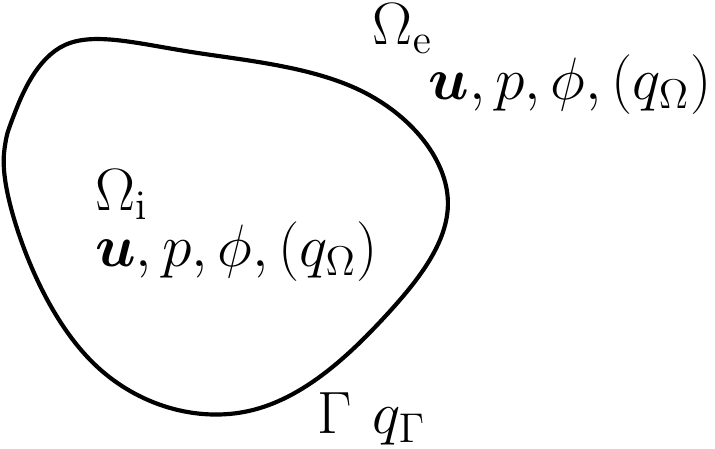}
\end{center}
\caption{\label{taylor_melcher_fig} Setup for the TM model. 
The unknown functions are now {the velocity and pressure fields $\bm{u},p$ and electrostatic potential 
$\phi$ in $\Omega_{\rm i,e}$ and the surface charge density $q_\Gamma$ on $\Gamma$}.
The bulk charge $q_\Omega$, shown in parentheses above, 
can be set identically to $0$ so that we recover the classical
TM model with surface charge advection {(see argument below \eqref{taudef})}.}
\end{figure}

{\bf Summary:}
Let us now collect our results. Dropping the subscript $0$ 
and renaming $q_2$ as $q_{\Omega}$ in \eqref{bulk_q2}-\eqref{bulk_stokes}, we have:
\begin{align}
\PD{q_\Omega}{t}+\nabla \cdot (\mb{u}q_\Omega)&=\nabla \cdot (\wh{\sigma}\nabla \phi),\quad \wh{\sigma}=\frac{\sigma}{\chi},\label{LD_q}\\
-\nabla \cdot(\epsilon\nabla \phi)&=q_\Omega,\label{LD_poisson}\\
\mu\Delta \mb{u}-\nabla p&=\nabla \cdot \Sigma(\bm{u},p)=q_\Omega \nabla \phi,\; \nabla \cdot \mb{u}=0.\label{LD_stokesq}
\end{align}
The boundary conditions for \eqref{LD_poisson} are given by \eqref{phicont} and \eqref{dphi0jump}:
\begin{equation}\label{phiqgamma}
\jump{\phi}=0, \; \jump{\epsilon\PD{\phi}{\mb{n}}}=q_\Gamma,
\end{equation}
where $q_\Gamma$ satisfies equation \eqref{qgammaeqn}, which in the original coordinates, can be written as:
\begin{equation}\label{LD_qGammaeqn}
\partial^\perp_t{q_\Gamma}+\kappa u_\perp q_\Gamma+\nabla_\Gamma \cdot(\mb{u}_\parallel q_\Gamma)
=-\jump{\wh{\sigma}\PD{\phi}{\mb{n}}}, \quad
u_\perp=\mb{u}\cdot \mb{n}, \; \mb{u}_\parallel=\mb{u}-u_\perp\mb{n},
\end{equation}
where $\nabla_\Gamma \cdot$ is the surface divergence operator and $\partial^\perp_t q_\Gamma$ is the time derivative of $q_\Gamma$ 
taken along trajectories that travel with the normal velocity of the interface. 
The boundary conditions for \eqref{LD_stokesq} are \eqref{ucont}, \eqref{stressperp} and \eqref{stressparallel}, which we may rewrite as:
\begin{equation}\label{LD_stokesBC}
\begin{split}
\jump{\mb{u}}&=0, \\
\jump{\Sigma(\mb{u},p)\mb{n}}&=-\jump{\frac{\epsilon}{2}\paren{\paren{\PD{\phi}{\mb{n}}}^2-\abs{\nabla_\Gamma\phi}^2}}\mb{n}
-q_\Gamma \nabla_\Gamma \phi -\wh{\gamma}\kappa\mb{n}.
\end{split}
\end{equation}
With the help of \eqref{phiqgamma}, the last stress boundary condition can also be written as follows:
\begin{equation}
\jump{(\Sigma(\mb{u},p)+\Sigma_{\rm M}(\phi))\mb{n}}=-\wh{\gamma}\kappa \mb{n}.
\end{equation}
The outer boundary conditions are given by \eqref{outerbcLD} (with our modified notation).
The final observation to make is that \eqref{LD_q} can be rewritten as follows using \eqref{LD_poisson} and 
the incompressibility condition in \eqref{LD_stokesq}:
\begin{equation}\label{taudef}
\PD{q_\Omega}{t}+\mb{u}\cdot \nabla q_\Omega=-\frac{1}{\tau}q_\Omega, \; \tau=\frac{\epsilon}{\wh{\sigma}}.
\end{equation}
This makes clear that equation \eqref{LD_q} requires no boundary condition and that $q_\Omega$ decays 
exponentially along fluid particle trajectories. After an initial transient, therefore, bulk charge is absent. 
In other words, the subset of phase space characterized by $q_\Omega\equiv 0$ 
is invariant and exponentially attracting.
We may thus set $q_\Omega\equiv 0$ in \eqref{LD_q}-\eqref{LD_stokesq} to find:
\begin{align}
\Delta \phi&=0,\label{LD_laplace}\\
\mu\Delta \mb{u}-\nabla p&=0,\; \nabla \cdot \mb{u}=0.\label{LD_stokes}
\end{align}
We have recovered the TM model with surface charge convection for arbitrary interface geometry.
The unknown variables are listed in the Figure \ref{taylor_melcher_fig}.

\subsection{Structure of the Boundary Layer}\label{bndry_layer}
The above derivation gives us additional information about the Taylor-Melcher regime
beyond what the TM model provides. Indeed, we now have the equations 
for the charge distribution within the inner space charge layer.
Suppose we have solved the dynamic TM model with suitable initial condition for $q_\Gamma$ and 
initial interface geometry. We may then find the charge distribution $\wt{q}_1$ inside 
the interfacial charge layer by solving the following equation obtained from \eqref{u1exp}-\eqref{phi1poisson}:
\begin{equation}\label{innercharge}
\PD{\wt{q}_1}{t}-\paren{\kappa u +\frac{1}{\sqrt{\abs{g}}}\PD{}{\eta^i}(\sqrt{\abs{g}}v^i)}\xi'\PD{\wt{q}_1}{\xi'}+v^i\PD{\wt{q}_1}{\eta^i}
=D_q\PDD{2}{\wt{q}_1}{\xi'}-\frac{1}{\tau}\wt{q}_1,\; D_q=\frac{D_{\rm C}+D_{\rm A}}{2\chi},
\end{equation}
where $\tau$ is as in \eqref{taudef}. This must be solved under the following conditions obtained from 
\eqref{intcond1}, \eqref{q1match} and \eqref{dphi0jump}:
\begin{equation}\label{innerchargebc}
l_{\rm C}\at{\wt{q}_1}{\xi'=0+}=\at{\wt{q}_1}{\xi'=0-},\quad \lim_{\xi'\to \pm\infty}\wt{q}_1=0, \quad \int_{-\infty}^\infty \wt{q}_1d\xi'=q_\Gamma.
\end{equation}
The functions $u, v^i$ and $q_\Gamma$ are solutions to the TM model and are thus known functions. 
If we take as our initial condition $q_\Gamma=0$, it would be reasonable to take $\wt{q}_1\equiv 0$ as the initial condition 
for \eqref{innercharge}. Once $\wt{q}_1$ is known, we may also obtain $\wt{\phi}_1$ and $\wt{p}_0$ 
using \eqref{phi1poisson}, \eqref{intcond1}, \eqref{phi1match}, \eqref{u1exp}, \eqref{matchp1} and \eqref{fuinnermod}.

To gain insight into the structure of the interfacial charge distribution, let us consider a steady state of the TM model.
The interface is stationary ($u=0)$ and the interfacial charge does not change in time ($\partial \wt{q}_1/\partial t=0$). 
Equation \eqref{innercharge} reduces to:
\begin{equation}
-\paren{\frac{1}{\sqrt{\abs{g}}}\PD{}{\eta^i}(\sqrt{\abs{g}}v^i)}\xi'\PD{\wt{q}_1}{\xi'}+v^i\PD{\wt{q}_1}{\eta^i}
=D_q\PDD{2}{\wt{q}_1}{\xi'}-\frac{1}{\tau}\wt{q}_1
\end{equation}
To simplify further, let us consider this equation at stagnation points, where $v_1=v_2=0$.
\begin{equation}\label{innerchargeODE}
D_q\PDD{2}{\wt{q}_1}{\xi'}-\lambda\xi'\PD{\wt{q}_1}{\xi'}-\frac{1}{\tau}\wt{q}_1=0, \; 
\lambda=-\frac{1}{\sqrt{\abs{g}}}\PD{}{\eta^i}{(\sqrt{\abs{g}}v^i)}=-\nabla_\Gamma\cdot \mb{u}_\parallel.
\end{equation}
It can be shown that such stagnation points always exist when the interface $\Gamma$ is homeomorhpic to a sphere 
for topological reasons.
We have now only to solve the above ODE in $\xi'$ under conditions \eqref{innerchargebc}.
First, define
\begin{equation}\label{taumax}
\tau_{\rm max}=\max(\tau_{\rm i}, \tau_{\rm e}), \; \tau_k=\frac{\epsilon_k}{\wh{\sigma}_k}=\frac{\chi\epsilon_k}{\sigma_k}, \; k={\rm i,e}.
\end{equation}
We let $D_{q,k}, k={\rm i,e}$ denote the value of $D_q$ in the interior ($\xi'<0$) and exterior ($\xi'>0$) fluid respectively.
Note that $\lambda$ is the same on both sides of the interface since the fluid velocity is continuous across the interface.
We have the following result, which is proved in Appendix \ref{kummereqn}.
\begin{proposition}\label{kummer_prop}
Consider equation \eqref{innerchargeODE} under condition \eqref{innerchargebc} and assume $q_\Gamma\neq 0$.
There is a unique solution if and only if $\lambda\tau_{\rm max}<1$. If $\lambda\tau_{\rm max}\geq 1$, there is no solution.
When there is a solution, $\wt{q}_1$ is either positive or negative everywhere (depending of the sign of $q_\Gamma$)
and has the following behavior.
If $\lambda>0$
\begin{equation}
\begin{split}\label{q1asymp+}
\wt{q}_1(\xi')&=q_\Gamma C^+_{\rm i}\abs{\xi'}^{-(\lambda\tau_{\rm i})^{-1}}\paren{1+\mc{O}(\xi'^{-2})} \text{ as } \xi' \to -\infty,\\
\wt{q}_1(\xi')&=q_\Gamma C^+_{\rm e}\xi'^{-(\lambda\tau_{\rm e})^{-1}}\paren{1+\mc{O}(\xi'^{-2})} \text{ as } \xi' \to \infty.
\end{split}
\end{equation}
If $\lambda=0$, 
\begin{equation}
\begin{split}
\wt{q}_1(\xi')&=\frac{q_\Gamma l_{\rm C}}{\sqrt{D_{q, \rm i}\tau_{\rm i}}l_{\rm C}+\sqrt{{D_{q,\rm e}\tau_{\rm e}}}}
\exp\paren{-\frac{\abs{\xi'}}{\sqrt{D_{q,\rm i}\tau_{\rm i}}}} \text{ for } \xi'<0,\\
\wt{q}_1(\xi')&=\frac{q_\Gamma}{\sqrt{D_{q, \rm i}\tau_{\rm i}}l_{\rm C}+\sqrt{{D_{q, \rm e}\tau_{\rm e}}}}
\exp\paren{-\frac{\xi'}{\sqrt{D_{q, \rm e}\tau_{\rm e}}}} \text{ for } \xi'>0.
\end{split}
\end{equation}
If $\lambda<0$, 
\begin{equation}\label{q1asymp-}
\begin{split}
\wt{q}_1(\xi')&=q_\Gamma C^-_{\rm i}\abs{\xi'}^{-(\abs{\lambda}\tau_{\rm i})^{-1}-1}
\exp\paren{-\frac{\abs{\lambda}}{2D_{q,{\rm i}}}\xi'^2}\paren{1+\mc{O}(\xi'^{-2})} \text{ as } \xi'\to -\infty,\\
\wt{q}_1(\xi')&=q_\Gamma C^-_{\rm e}\xi'^{-(\abs{\lambda}\tau_{\rm e})^{-1}-1}
\exp\paren{-\frac{\abs{\lambda}}{2D_{q,{\rm e}}}\xi'^2}\paren{1+\mc{O}(\xi'^{-2})} \text{ as } \xi'\to \infty.
\end{split}
\end{equation}
In the above, $C^\pm_k, k={\rm i,e}$ are positive constants that depend only on $\lambda\tau_k, \lambda/D_{q,k}$ and $l_{\rm C}$.
The above also satisfy the second condition in \eqref{qlimits}.
\end{proposition}

When
\begin{equation}\label{lambdatau}
\lambda=-\nabla_\Gamma\cdot \mb{u}_\parallel<\frac{1}{\tau_{\rm max}}
\end{equation}
at a stagnation point, the above Proposition states that there is a unique inner layer charge distribution 
consistent with the TM model. However, the charge distribution does {\em not}
exhibit the familiar exponential decay with distance from the interface. If $\lambda>0$
the decay is only algebraic. Convective charge accumulation leads to a broader space charge layer.
If $\lambda<0$ at a stagnation point, charge distribution decay is faster than exponential.

At a stagnation point, if $q_\Gamma\neq 0$ and 
\begin{equation}\label{charge_accumulation_overwhelm}
\lambda=-\nabla_\Gamma \cdot \mb{u}_\parallel\geq \frac{1}{\tau_{\max}},
\end{equation}
an inner layer charge distribution consistent with the steady state of the TM model does not exist.
{Indeed, when $\lambda\tau_{\rm max}\geq 1$, 
the asymptotic behavior for $\wt{q}_1$ in \eqref{q1asymp+} 
makes it impossible for $\wt{q}_1$ to satisfy \eqref{qintegral} since:
\begin{equation}
\int_1^\infty \abs{\xi'}^{-(\lambda \tau_{\rm max})^{-1}}d\xi'=\infty.
\end{equation}
The picture that emerges is that}
convective charge accumulation overwhelms bulk charge dissipation when $\lambda \tau_{\rm max}\geq 1$,
and the boundary layer is destroyed {(the reader is referred back to Section \ref{LD_intro} 
and Figure \ref{charge_accumulation_fig} for a heuristic discussion)}. 
Our analysis here is confined to stagnation points, but similar statements may hold for 
points at which $\mb{u}_\parallel\neq 0$. 

\begin{figure}
\begin{center}
\vspace{-0.5cm}
\includegraphics[width=0.5\textwidth]{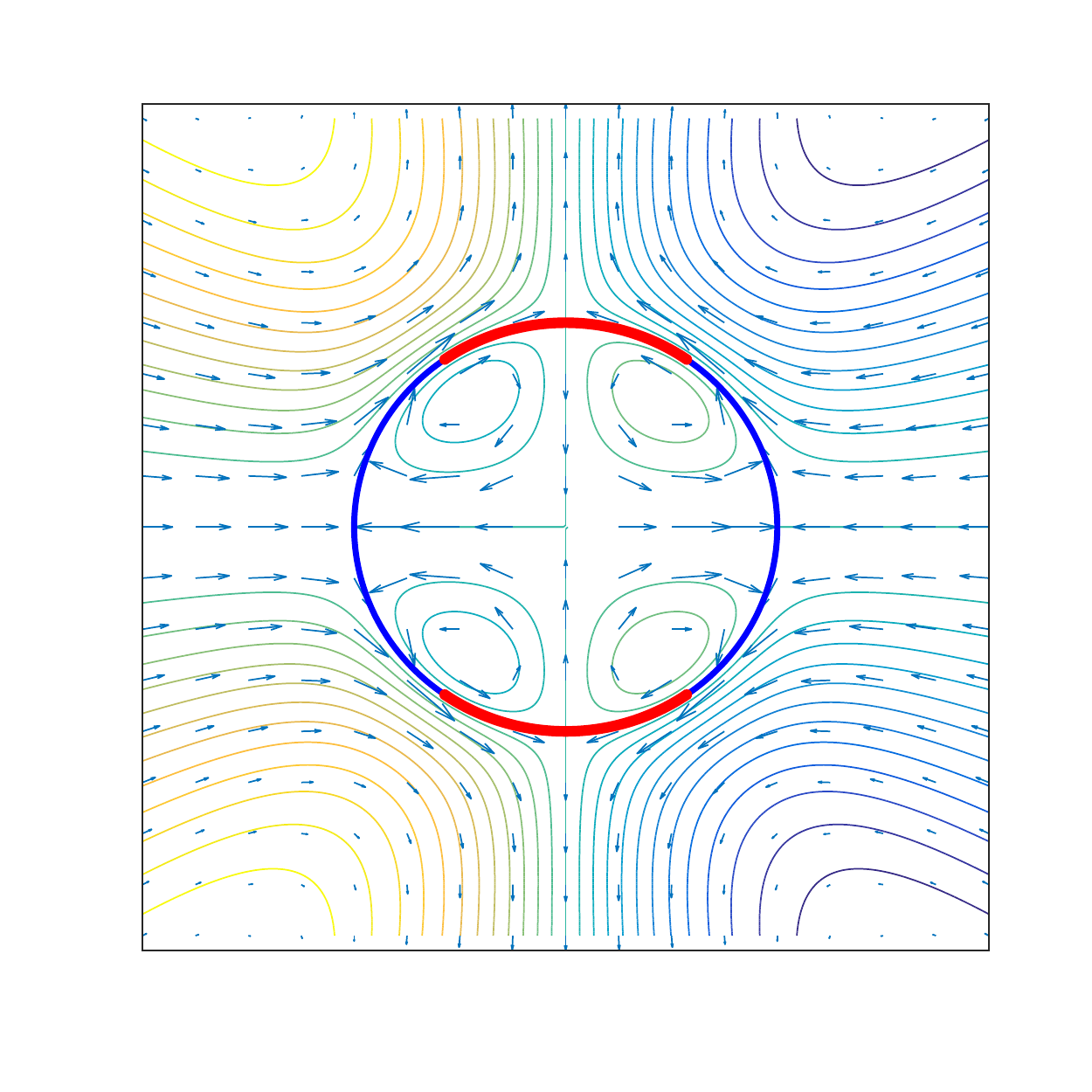}
\vspace{-0.5cm}
\end{center}
\caption{\label{surf_div} A flow field plot indicating regions of positive 
and negative surface divergence $\nabla_\Gamma \cdot \mb{u}_\parallel$. Plotted is the flow field 
around a sphere of the quadrupole vortex of Taylor (horizontal axis is the axis of rotation and is parallel 
to the imposed electric field direction) \citep{Taylor1966_PRSLa}. 
On the sphere, $\nabla_\Gamma \cdot \mb{u}_\parallel<0$ in the region in thick red and $\nabla_\Gamma \cdot \mb{u}_\parallel>0$
in the region in thin blue. 
The charge distribution in the region $\nabla_\Gamma \cdot \mb{u}_\parallel<0$ is expected 
to decay algebraically with distance from the interface.}
\end{figure}

Consider gradually increasing the imposed electric field on a leaky-dielectric droplet.
Suppose the increase in field strength is slow enough that the system is at steady state at each instant. 
As the imposed electric field is increased, surface convection will be stronger and
the space charge layer where $\nabla_\Gamma\cdot \mb{u}_\parallel<0$ will broaden (see Figure \ref{surf_div}).
At a certain threshold field strength, condition \eqref{lambdatau} at a stagnation 
point (or a similar condition at non-stagnation points) may be violated. 
At this threshold, the TM model will cease to be valid. 
The boundary layer will be destroyed possibly resulting in a non-zero charge distribution extending into the bulk.
Recall that the derivation of the stress balance boundary condition also required condition \eqref{qintegral}.
Stress balance across the interface may become impossible as the threshold field strength is reached. 

\subsection{Large Imposed Voltage}\label{large_voltage}

Given that the imposed voltage is large in most experimental setups, it is of interest to ask whether our analysis may be 
extended to this case. Here, we discuss the necessary scaling, state the result and only sketch our derivation, since 
the details are almost exactly the same as our foregoing analysis.

\begin{figure}
\begin{center}
\includegraphics[width=0.5\textwidth]{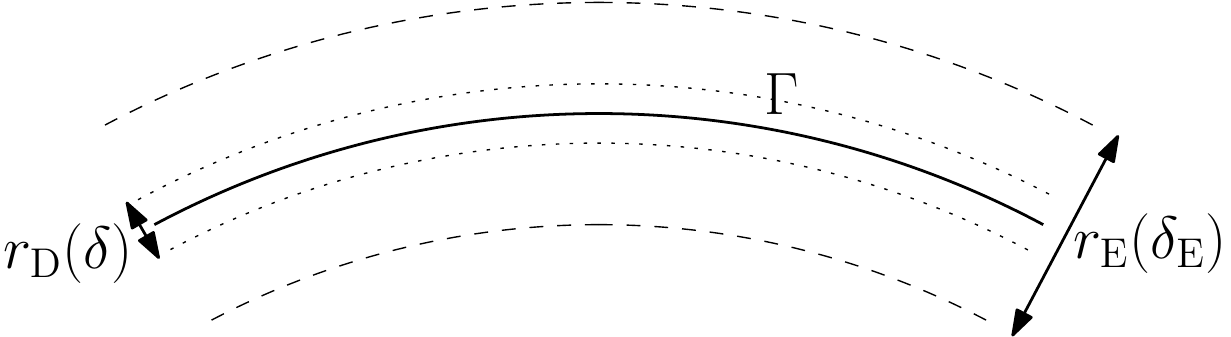}
\end{center}
\caption{\label{two_layers} The boundary layer structure when the externally imposed voltage is large.
There are two layers, the inner layer of thickness $r_{\rm E}$ ($\delta_{\rm E}$ in dimensionless units)
and an inner-inner layer of thickness $r_{\rm D}$ ($\delta$ in dimensionless units).} 
\end{figure}

Let $E_*$ be the representative imposed electric field strength. 
Modify the scaling of $\mb{u},p,\phi$ in \eqref{dless_scaling} by replacing the thermal 
voltage $RT/F$ with the imposed representative voltage $E_*L$.
Define the following constants:
\begin{equation}\label{dless_strong}
\delta_{\rm E}=\frac{\delta}{\omega}=\frac{r_{\rm D}/\omega}{L}=\frac{r_{\rm E}}{L},\; \omega=\sqrt{\frac{RT/F}{E_*L}},\; r_{\rm E}=\sqrt{\frac{\epsilon_*E_*L}{Fc_*}}.
\end{equation}
The length $r_{\rm E}$ is defined by replacing the thermal voltage $RT/F$
in the definition of the Debye length with the externally imposed voltage $E_*L$. 
In the notation of \citet{saville1977electrokinetic,baygents1990circulation,Schnitzer2015_JFM},
$\beta\equiv\omega^{-2}$ is the dimensionless imposed field strength. With $\omega=1$, we are back to the original scaling.
When $\omega\ll 1$, we can perform a similar analysis to that presented above provided
\begin{equation}\label{scaling_large}
\delta_{\rm E}\ll \omega\ll 1.
\end{equation}
This is equivalent to the parametric ordering
\begin{equation}\label{scaling_large2}
1\ll \omega^{-2}=\beta\ll \delta^{-1}
\end{equation} 
assumed in \citet{Schnitzer2015_JFM}.
This leads to an inner layer of thickness $r_{\rm E}$ and an inner-inner layer of thickness $r_{\rm D}$ (Figure \ref{two_layers}).
The presence of two layers is similar to \citet{baygents1990circulation} 
except that there the thickness of the wider layer is $L/\beta$.
Outside of these boundary layers, the TM model is valid. The magnitude of surface charge convection is 
determined as in Table \ref{dist_limits}, depending on the magnitude of $Pe_{\rm E}=Pe\omega^{-2}$ with respect to $\delta_{\rm E}$.

We now briefly discuss the modifications needed to the foregoing analysis.
The dimensionless equations of the modified Saville model in Section \ref{non_dimensionalization} change as follows.
Equation \eqref{dlessc} becomes:
\begin{equation}\label{PeE}
\begin{split}
Pe_{\rm E}\paren{\PD{c}{t}+\nabla \cdot (\mb{u}c)}
&=\nabla \cdot \paren{D_{\rm C}\paren{\omega^2\nabla c+c\nabla \phi}}+\frac{k_{\rm E}}{\alpha}\paren{s-\frac{ca}{K}},\\
Pe_{\rm E}&=Pe\omega^{-2}, \; k_{\rm E}=k\omega^2.
\end{split}
\end{equation}
and the interface conditions for the solutes change accordingly.
In equations \eqref{dlesscapoisson}, \eqref{dlesscastokes} and its attendant interface conditions, $\delta_{\rm E}$ replaces 
every instance of $\delta$.

We first take the limit $\alpha\to 0$ in the modified Saville model to obtain the charge diffusion model as in Section \ref{CD}.
Equation \eqref{qeqn} for $q$ in the new scaling takes the following form.
\begin{equation}
Pe_{\rm E}\paren{\PD{q}{t}+\nabla\cdot(\mb{u}q)}=
-\nabla \cdot \mb{J}_q, \; \mb{J}_q=-\Sigma(q)\paren{\frac{\omega^2}{\sqrt{4S+q^2}}\nabla q+\nabla \phi},
\end{equation}
where $\Sigma(q)$ is the same as in \eqref{Sigmaq}. Equations \eqref{dlesspoisson}, 
\eqref{dlessstokes} and their boundary conditions remain the same except that $\delta$
should be replaced by $\delta_{\rm E}.$

We next take the limit $\delta_{\rm E}\to 0$ scaling 
\begin{equation}\label{scaling_strong}
Pe_{\rm E}\equiv Pe\omega^{-2}=\chi_{\rm E}\delta_{\rm E}^{-2}, \; \gamma=\wh{\gamma}_{\rm E}\delta_{\rm E}^2
\end{equation}
as in \eqref{scaling}. The boundary layer coordinate is scaled 
so that $\xi=\delta_{\rm E}\xi'$.
The resulting TM model is exactly the same as in \eqref{LD_q}-\eqref{LD_stokesBC}
except that the constant $\wh{\sigma}$ should be replaced
by $\wh{\sigma}_{\rm E}=\sigma/\chi_{\rm E}$ and $\wh{\gamma}$ by $\wh{\gamma}_{\rm E}$.
The equation in the interfacial layer \eqref{innercharge} takes the form:
\begin{equation}\label{innercharge_omega}
\begin{split}
\PD{\wt{q}_1}{t}-&\paren{\kappa u +\frac{1}{\sqrt{\abs{g}}}\PD{}{\eta^i}(\sqrt{\abs{g}}v^i)}\xi'\PD{\wt{q}_1}{\xi'}+v^i\PD{\wt{q}_1}{\eta^i}
=\omega^2D_{q,\rm E}\PDD{2}{\wt{q}_1}{\xi'}-\frac{1}{\tau_{\rm E}}\wt{q}_1,\\
D_{q,\rm E}&=\frac{D_{\rm C}+D_{\rm A}}{2\chi_{\rm E}}, \; \tau_{\rm E}=\frac{\epsilon}{\wh{\sigma}_{\rm E}}.
\end{split}
\end{equation}
The boundary conditions at $\xi'=\pm\infty$ and $\xi'=0\pm$ remain the same.

Let us finally let $\omega \to 0$. The constant $\omega$ only appears inside the boundary layer equations, 
so it only affects the behavior in the charge layer of width $\delta_{\rm E}$ (or $r_{\rm E}$ in physical dimensions).
If we let $\omega\to 0$ in \eqref{innercharge_omega}, we lose the second order spatial derivative, making it impossible 
to satisfy the boundary conditions. This indicates the presence of a boundary layer of width 
$\omega\delta_{\rm E}=\delta$ (or $r_{\rm D}$ in physical dimensions). 
Confining our analysis to stagnation points as in Section \ref{bndry_layer}, we obtain, in place of \eqref{innerchargeODE}
\begin{equation}\label{innerchargeODE_omega}
\omega^2D_{q,\rm E}\PDD{2}{\wt{q}_1}{\xi'}-\lambda\xi'\PD{\wt{q}_1}{\xi'}-\frac{1}{\tau_{\rm E}}\wt{q}_1=0, \; 
\lambda=-\frac{1}{\sqrt{\abs{g}}}\PD{}{\eta^i}{(\sqrt{\abs{g}}v^i)}=-\nabla_\Gamma\cdot \mb{u}_\parallel.
\end{equation}
This equation is in fact just a rescaled version of \eqref{innerchargeODE_omega} and can be solved in exactly the same 
way. Taking the limit as $\omega\to 0$ in the resulting expression, one sees that the charge $\wt{q}_1$ will be identically 
equal to $0$ in the inner layer (and not the inner-inner layer) when $\nabla_\Gamma\cdot \mb{u}_\parallel>0$.

\section{Interfacial Double Layer and Electrophoresis}\label{DL}

\subsection{Leading Order Equations for the initial layer with $l_C \neq l_A$}\label{subsec:lc_neq_la}

We now consider the case $l_{\rm C}\neq l_{\rm A}$. 
In this case, there is a double layer at the interface $\Gamma$ as was suggested at the end of Section \ref{CD}.
We consider the limit $\delta\to 0$ as in Section \ref{leaky}.
Here, the dimensionless parameters are scaled so that
\begin{equation}\label{scaling2}
Pe=\mc{O}(\delta^0), \; \gamma=\wh{\gamma}\delta .
\end{equation}
These scalings are different from the scalings \eqref{scaling} used in Section \ref{leaky}.
%
%
{As will shall see,
the scaling of the surface tension strength $\gamma$ is chosen so we can obtain a well-posed limiting problem.
The scaling of $Pe$ is chosen for analytical feasibility; the choice in \eqref{scaling} leads to a difficult
analytical problem. We shall return to this point in Section \ref{theory}.}
%
%
We expand the variables $q,\phi, \mb{u}$ and $p$ in powers of $\delta$ as follows:
\begin{equation}\label{dlexpansion}
\begin{split}
q &= q_0 + {\delta} q_1 + {\delta^2} q_2 + \cdots,\\
\phi &= \phi_0 + {\delta } \phi_1 + {\delta^2} \phi_2 +\cdots,\\
\mb{u}&=\frac{1}{\delta}\mb{u}_{-1}+\mb{u}_0+\delta \mb{u}_1+\cdots,\\
p&=\frac{1}{\delta^2}p_{-2}+\frac{1}{\delta }p_{-1}+p_0+\cdots.
\end{split}
\end{equation}
{As we shall see, the stronger surface tension 
(see scaling of $\gamma$ in \eqref{scaling2} compared with \eqref{scaling})
and the presence of the Galvani potential
necessitates velocity and pressure fields of order $1/\delta$ (and $1/\delta^2)$. 
We shall see, in fact, that $\bm{u}_{-1}$ represents the initial fluid velocity transient.
}

We 
first consider the equations in the outer layer.
{Substituting the expansions in \eqref{dlexpansion} into 
\eqref{dlesspoisson}, we find that \eqref{q01zero} holds exactly as before.
Plugging this into \eqref{qeqn}, we find that
\begin{equation}\label{ohm}
\nabla \cdot (\sigma \nabla \phi_0)=0 \text{ and hence } \Delta \phi_0=0.
\end{equation}
The above expression corresponds to \eqref{bulk_q2} of the Section \ref{derivation_of_TM}, except for the absence 
of the advection term thanks to our assumption that  $Pe=\mc{O}(1)$. In the implication above, we used the 
fact that $D_{\rm C,A}$ and hence $\sigma$ is spatially constant (within $\Omega_{\rm i}$ and $\Omega_{\rm e}$ respectively).
Equation \eqref{q01zero} and \eqref{ohm}, applied to \eqref{dlesspoisson} gives us:
\begin{equation}\label{q2zero}
q_2=0.
\end{equation}
We remark that the derivation of \eqref{ohm} and \eqref{q2zero} 
is different from the way we derived \eqref{LD_laplace} and $q_\Omega=q_2=0$ in Section \ref{derivation_of_TM}, 
where these were derived as a consequence 
of \eqref{bulk_q2} and \eqref{bulk_poisson} (see \eqref{taudef}).}

{Since $q_0=q_1=0$ by \eqref{q01zero},} we conclude that
\begin{align}
\nabla p_{-2}&=0,\\
\mu\Delta \mb{u}_{-1}-\nabla p_{-1}&=0, \; \nabla \cdot \mb{u}_{-1}=0.\label{stokes-1}
\end{align}
From the first equation, we see that $p_{-2}$ is constant within $\Omega_{\rm i}$ and $\Omega_{\rm e}$ respectively. We set
\begin{equation}\label{p-1}
p_{-2}=\begin{cases}
p_- &\text{ in } \Omega_{\rm i},\\ 
p_+&\text{ in } \Omega_{\rm e}.
\end{cases}
\end{equation}

We now turn to equations in the inner layer.
We introduce a curvilinear boundary layer coordinate system as in the previous Section.
From \eqref{qeqn}, we have:
\begin{align}
\PD{\wt{J}_{q,0}}{\xi'}&=0,\label{Jq0const}\\
\wt{J}_{q,0}=&-\Sigma(\wt{q}_0)\paren{\frac{1}{\sqrt{4S+\wt{q}_0^2}}\PD{\wt{q}_0}{\xi'}+\PD{\wt{\phi}_0}{\xi'}}.\label{Jq0}\end{align}
From \eqref{dlesspoisson}, we have:
\begin{equation}\label{poissoninner0}
-\epsilon\PDD{2}{\wt{\phi}_0}{\xi'}=\wt{q}_0.
\end{equation}
The above two equations are satisfied for $\xi'>0$ and $\xi'<0$.
The interface conditions at $\xi'=0$ are given by \eqref{dlesspoissonbc}, \eqref{qint1} and \eqref{qint2}:
\begin{equation}\label{intinner0}
\begin{split}
\jump{\wt{\phi}_0}&=\jump{\epsilon\PD{\wt{\phi}_0}{\xi'}}=\jump{\wt{J}_{q,0}}=0,\\
l_{\rm C}\at{\paren{\wt{q}_0+\sqrt{4+\wt{q}_0^2}}}{\xi'=0+}&=\at{\paren{\wt{q}_0+\sqrt{4S_{\rm i}+\wt{q}_0^2}}}{\xi'=0-}.
\end{split}
\end{equation}
Here, $\at{\cdot}{\xi'=0\pm}$ denotes the limiting value of quantity of interest from the positive and negative sides of $\xi'=0$ respectively
and $\jump{\cdot}=\at{\cdot}{\xi'=0-}-\at{\cdot}{\xi'=0+}$.
The matching conditions must be that, at $\xi\to \pm \infty$, the boundary layer values matches the limiting values in the 
outer layer.
\begin{equation}\label{q0phi0limits}
\lim_{\xi'\to \pm \infty} \wt{\phi}_0=\lim_{\xi\to 0\pm} \phi_0\equiv \phi_\pm, \; \; \lim_{\xi'\to \pm \infty} \wt{q}_0=\lim_{\xi\to 0\pm} q_0=0.
\end{equation}
The last equality follows from \eqref{q01zero}. 
We now solve the above system of equations. 
From \eqref{Jq0const}, we see that $\wt{J}_{q,0}$ is a constant. From \eqref{Jq0}, we see that 
\begin{equation}
\PD{}{\xi'}\paren{\ln\paren{\wt{q}_0+\sqrt{4S+\wt{q}_0^2}}+\wt{\phi}_0}=-\frac{\wt{J}_{q,0}}{\Sigma(\wt{q}_0)}.
\end{equation}
Given \eqref{q0phi0limits}, we conclude that 
\begin{equation}
\abs{\int_0^{\pm\infty}\frac{\wt{J}_{q,0}}{\Sigma(\wt{q}_0)}d\xi'}<\infty.
\end{equation}
Since $\wt{q}_0$ is assumed to go to $0$ as $\xi'\to \infty$, $\Sigma(\wt{q}_0)$ remains bounded (and positive)
for all $\xi'$, and thus the above integral is unbounded unless $\wt{J}_{q,0}=0$. Thus, $\wt{J}_{q,0}=0$ and 
\begin{equation}
\ln\paren{\wt{q}_0+\sqrt{4S+\wt{q}_0^2}}+\wt{\phi}_0=\ln\paren{2\sqrt{S}}+\phi_\pm,
\end{equation}
where the $+$ $(-)$ sign is valid for $\xi'>0$ $(\xi'<0)$.
Solving the above for $\wt{q}_0$ we obtain
\begin{equation}
\wt{q}_0=-2\sqrt{S}\sinh\paren{\wt{\phi}_0-\phi_\pm}.
\end{equation}
Substituting the above into \eqref{poissoninner0} we obtain the following equation for $\wt{\phi}_0$
\begin{equation}
\epsilon\PDD{2}{\wt{\phi}_0}{\xi'}=2\sqrt{S}\sinh\paren{\wt{\phi}_0-\phi_\pm}.
\end{equation}
This equation can be solved analytically to yield:
\begin{equation}\label{wtphiexp}
\begin{split}
\wt{\phi}_0&=\phi_\pm+2\ln\paren{\frac{1+A_\pm\exp(\mp\lambda_\pm\xi')}{1-A_\pm\exp(\mp\lambda_\pm\xi')}}, \\
\lambda_+&=\sqrt{2}\epsilon_{\rm e}^{-1/2}, \; \lambda_-=\sqrt{2}\epsilon_{\rm i}^{-1/2}S_{\rm i}^{1/4},
\end{split}
\end{equation}
where again the $+$ $(-)$ sign corresponds to the expression for $\xi'>0$ $(\xi'<0)$.
{The ratio between the Debye layer thicknesses on the two sides of the interface is thus given by:
\begin{equation}
\frac{\lambda_+}{\lambda_-}=\sqrt{\frac{\epsilon_{\rm i}}{\epsilon_{\rm e}\sqrt{S}}}.
\end{equation}
}
Using condition \eqref{q0phi0limits} to determine the constants $\phi_\pm$ and $A_\pm$, we find
\begin{equation}
\jump{\phi_0}=\phi_--\phi_+=\ln \paren{\frac{l_{\rm C}}{\sqrt{S_{\rm i}}}}=\frac{1}{2}\ln \paren{\frac{l_{\rm C}}{l_{\rm A}}}\equiv \phi^{\Delta},\label{phi0jump}
\end{equation}
and 
\begin{equation}\label{Apm}
\begin{split}
A_+&=\frac{\sqrt{\rho(\rho+\zeta)}-\sqrt{1+\rho\zeta}}{\sqrt{\rho(\rho+\zeta)}+\sqrt{1+\rho\zeta}}, \; 
A_-=\frac{\sqrt{\rho+\zeta}-\sqrt{\rho(1+\rho\zeta})}{\sqrt{\rho+\zeta}+\sqrt{\rho(1+\rho\zeta})},\\
\rho&=\sqrt{\frac{l_{\rm C}}{\sqrt{S_{\rm i}}}}, \; \zeta=\sqrt{\frac{\epsilon_{\rm e}}{\epsilon_{\rm i}\sqrt{S_{\rm i}}}}=\frac{}.
\end{split}
\end{equation}
Note that in \eqref{phi0jump} we recovered \eqref{phijumpheuristic}.
We further note that there is a jump in voltage across the interface $\Gamma$ if $l_{\rm C}\neq l_{\rm A}$.
Define the interior and exterior voltage differences $\phi^\Delta_{\rm i,e}$:
\begin{equation}\label{phiDeltaie}
\begin{split}
\phi^\Delta_{\rm i}&=\wt{\phi}_0(-\infty)-\wt{\phi}_0(0)=-2\ln\paren{\frac{1+A_-}{1-A_-}}, \\
\phi^\Delta_{\rm e}&=\wt{\phi}_0(0)-\wt{\phi}_0(\infty)=2\ln\paren{\frac{1+A_+}{1-A_+}}.
\end{split}
\end{equation}
Clearly $\phi^\Delta_{\rm i}+\phi^\Delta_{\rm e}=\phi^\Delta$.
It is not difficult to see from \eqref{Apm} that $\phi^\Delta_{\rm i,e}$ have the same sign.
Noting that $\ln \rho=\phi^\Delta/2$, we may expand $A_\pm$ and hence $\phi^\Delta_{\rm i,e}$ in terms of $\phi^\Delta$ to find:
\begin{equation}\label{phiDeltaieapprox}
\phi^\Delta_{\rm i}=\frac{\zeta}{1+\zeta}\phi^\Delta+\mc{O}((\phi^\Delta)^2), \quad \phi^\Delta_{\rm e}=\frac{1}{1+\zeta}\phi^\Delta+\mc{O}((\phi^\Delta)^2).
\end{equation}
The parameter $\zeta$ is thus the ratio between $\phi^\Delta_{\rm i}$ and $\phi^\Delta_{\rm e}$ for small $\phi^\Delta$.
{We may also calculate the amount of charge that accumulates on the interface:
\begin{equation}
\int_{-\infty}^0 \wt{q}_0d\xi'=-\int_0^\infty \wt{q}_0d\xi'=
\paren{(\epsilon_{\rm i}\lambda_-)^{-1}+(\epsilon_{\rm e}\lambda_+)^{-1}}^{-1}\phi^\Delta +\mc{O}((\phi^\Delta)^2).
\end{equation}
For small $\phi^\Delta$, therefore, the interface can be viewed as two capacitor in series, with capacitance $\epsilon_{\rm i}\lambda_-$
and $\epsilon_{\rm e}\lambda_+$.
}
We also note that
\begin{equation}\label{phi0eta}
\PD{\wt{\phi}_0}{\eta^i}=\at{\PD{\phi_0}{\eta^i}}{\xi=0+}=\at{\PD{\phi_0}{\eta^i}}{\xi=0-}.
\end{equation}
The latter inequality is true even if $\phi_0$ has a jump, thanks to \eqref{phi0jump} which says 
that the magnitude of the jump $\phi^\Delta$ is independent of $\eta^i$.

We now focus on the fluid equations.
Since the expansions for the velocity and pressure start at $\mc{O}(\delta ^{-1})$ and $\mc{O}(\delta^{-2})$, 
the calculations in Appendix \ref{tensors} do not apply, but the calculations there are easily modified for the case at hand.
For $\wt{u}_{-1}$ we have: 
\begin{equation}\label{intinneruDL0}
-\PD{\wt{p}_{-2}}{\xi'}=\wt{q}_0\PD{\wt{\phi}_0}{\xi'}, \; \PD{\wt{u}_{-1}}{\xi'}=0,
\end{equation}
with interface conditions:
\begin{equation}\label{u-1p-2jump}
\jump{\wt{u}_{-1}}=0, \; \jump{2\mu\PD{\wt{u}_{-1}}{\xi'}-\wt{p}_{-2}}=0.
\end{equation}
The matching conditions are:
\begin{equation}
\lim_{\xi'\to \pm \infty} \wt{u}_{-1}=\lim_{\xi\to 0\pm} u_{-1}, \; \; \lim_{\xi'\to \pm \infty} \wt{p}_{-2}=\lim_{\xi\to 0\pm} p_{-2}\equiv p_\pm.
\end{equation}
We thus see that
\begin{equation}\label{u-1const}
\wt{u}_{-1}=\lim_{\xi\to 0\pm}u_{-1},
\end{equation}
and therefore, 
\begin{equation}\label{u-1cont}
\jump{u_{-1}}=0.
\end{equation}
For $\wt{p}_{-2}$, we see that 
\begin{equation}\label{p-2exp}
\wt{p}_{-2}=\begin{cases}
p_++(\epsilon_{\rm e}/2)(\partial \wt{\phi}_0/\partial\xi')^2 &\text{ for }\xi'>0, \\
p_-+(\epsilon_{\rm i}/2)(\partial \wt{\phi}_0/\partial\xi')^2 &\text{ for }\xi'<0.
\end{cases}
\end{equation}
Since $p_\pm$ does not depend $\eta^i$ (see \eqref{p-1}), $\wt{p}_{-2}$ does not depend on $\bm{\eta}$.
Given \eqref{u-1p-2jump} and \eqref{u-1const}, $p_{-2}$ is continuous at $\xi'=0$
and therefore, 
\begin{equation}
\at{\paren{p_+-\frac{\epsilon_{\rm e}}{2}\paren{\PD{\wt{\phi}_0}{\xi'}}^2}}{\xi'=0+}=\at{\paren{p_--\frac{\epsilon_{\rm i}}{2}\paren{\PD{\wt{\phi}_0}{\xi'}}^2}}{\xi'=0-}.
\end{equation}
After some algebra, we find that:
\begin{equation}
\jump{p_{-2}}=p_{-}-p_+=\frac{\epsilon_{\rm i}}{2}\paren{\frac{4\lambda_-A_-}{1-A_-^2}}^2-\frac{\epsilon_{\rm e}}{2}\paren{\frac{4\lambda_+A_+}{1-A_+^2}}^2,
\end{equation}
where $\lambda_\pm, A_\pm$ are defined in \eqref{wtphiexp} and \eqref{Apm}.
When $l_{\rm C}=l_{\rm A}$, $A_\pm=0$ and thus there is no pressure difference to leading order.
Otherwise, there will in general be a pressure difference.

We may also consider the equations for $\wt{v}^i_{-1}$. This analysis, the details of which we omit, yields:
\begin{equation}
\wt{v}^i_{-1}=\lim_{\xi\to 0\pm}v^i_{-1}, \; i=1,2.\label{v-1const}
\end{equation}
In particular, we have:
\begin{equation}\label{vi-1cont}
\jump{v^i_{-1}}=0, \; i=1,2.
\end{equation}

We now turn to equations at the next order to obtain the boundary conditions for \eqref{ohm}.
From \eqref{qeqn} and \eqref{dlesspoisson} 
we have:
\begin{align}
\PD{\wt{J}_{q,1}}{\xi'}&=0,\label{Jq1const2}\\
\wt{J}_{q,1}&=-\Sigma(\wt{q}_0)\PD{}{\xi'}\paren{\frac{\wt{q}_1}{\sqrt{4S+\wt{q}_0^2}}+\wt{\phi}_1},\label{Jq1exp2}\\
-\wt{q}_1&=\epsilon\paren{\PDD{2}{\wt{\phi}_1}{\xi'}+\kappa \PD{\wt{\phi}_0}{\xi'}}\label{q1exp2}.
\end{align}
with interface conditions from \eqref{dlesspoissonbc}, \eqref{qint1} and \eqref{qint2} given by:
\begin{align}\label{phi1q1bc}
\jump{\wt{\phi}_1}=\jump{\epsilon\PD{\wt{\phi}_1}{\xi'}}=\jump{\wt{J}_{q,1}}=\jump{\frac{\wt{q}_1}{\sqrt{4S+\wt{q}_0^2}}}=0.
\end{align}
We see from \eqref{Jq1const2} and the above that $\wt{J}_{q,1}$ is constant throughout $-\infty<\xi'<\infty$.
Let this constant be equal to $J_0$.
From \eqref{Jq1exp2} and \eqref{q1exp2}, we obtain the equation:
\begin{equation}
-\epsilon\PDD{2}{}{\xi'}\paren{\frac{\wt{q}_1}{\sqrt{4S+\wt{q}_0^2}}}+\wt{q}_1=-\epsilon\kappa\PD{\wt{\phi}_0}{\xi'}
+\epsilon J_0\PD{}{\xi'}\paren{\frac{1}{\Sigma(\wt{q}_0)}},
\end{equation}
with interface conditions:
\begin{equation}
\jump{\epsilon\PD{}{\xi'}\paren{\frac{\wt{q}_1}{\sqrt{4S+\wt{q}_0^2}}}}=\jump{\frac{\wt{q}_1}{\sqrt{4S+\wt{q}_0^2}}}=0.
\end{equation}
It is not difficult to see that this equation for $\wt{q}_1$ has a unique bounded solution that decays exponentially to $0$
as $\xi'\to \pm \infty$. It will be useful later to have a somewhat more explicit form for $\wt{q}_1$ and $\wt{\phi}_1$.
Define the functions $\psi_\kappa$ and $\psi_J$ as being solutions to the following equations:
\begin{align}\label{psikappa}
-\epsilon\PDD{2}{\psi_\kappa}{\xi'}+\sqrt{4S+\wt{q}_0^2}\psi_\kappa&=-\epsilon\PD{\wt{\phi}_0}{\xi'},\\
\label{psiJ}
-\epsilon\PDD{2}{\psi_J}{\xi'}+\sqrt{4S+\wt{q}_0^2}\psi_J&=\epsilon\PD{}{\xi'}\paren{\frac{1}{\Sigma(\wt{q}_0)}},
\end{align}
with interface conditions:
\begin{equation}
\jump{\psi_{\cdot}}=\jump{\epsilon\PD{\psi_\cdot}{\xi'}}=0
\end{equation}
where $\psi_\cdot$ is a place holder for either $\psi_\kappa$  and $\psi_J$. We also require that both $\psi_\kappa$
and $\psi_J$ decay to $0$ at $\xi'\to \pm\infty$. 
Note here that $\psi_\kappa$ and $\psi_J$ depend solely on $\wt{q}_0$ and $\wt{\phi}_0$ and hence, 
like $\wt{q}_0$ and $\wt{\phi}_0$, are universal functions that only depend on the parameters of the system.
Then, we have:
\begin{align}
\wt{q}_1&=\sqrt{4S+\wt{q}_0^2}\paren{\kappa\psi_\kappa+J_0\psi_J}\\
\label{phi1psi}
\PD{\wt{\phi}_1}{\xi'}&=-\kappa\PD{\psi_\kappa}{\xi'}-J_0\PD{\psi_J}{\xi'}-\frac{J_0}{\Sigma(\wt{q}_0)}.
\end{align}
We can now find $\wt{\phi}_1$ the above together with the first of the interface conditions in \eqref{phi1q1bc}.
To relate $J_0$ to the outer layer variables, we use Kaplun's matching procedure to find that:
\begin{equation}\label{sigmaphi0current}
\lim_{\xi\to \pm 0}\sigma\PD{\phi_0}{\xi}=\lim_{\xi'\to\pm \infty}\sigma \PD{\wt{\phi}_1}{\xi'}=-J_0.
\end{equation}
In particular, we have:
\begin{equation}
\jump{\sigma\PD{\phi_0}{\xi}}=0.\label{phi0currcont}
\end{equation}

Let us now turn to the equations for the velocities. For $u$, we have:
\begin{align}
\mu\PDD{2}{\wt{u}_0}{\xi'}-\PD{\wt{p}_{-1}}{\xi'}&=\wt{q_1}\PD{\wt{\phi}_0}{\xi'}+\wt{q}_0\PD{\wt{\phi}_1}{\xi'},\\
\label{incomp1}\PD{\wt{u}_0}{\xi'}+\kappa \wt{u}_{-1}+\frac{1}{\sqrt{\abs{g}}}\PD{}{\eta^i}\paren{\sqrt{\abs{g}}\wt{v}^{i}_{-1}}&=0,
\end{align}
with interface conditions:
\begin{equation}\label{u0gammakappa}
\jump{u_0}=0, \; \jump{2\mu\PD{\wt{u}_0}{\xi'}-\wt{p}_{-1}+\epsilon\PD{\wt{\phi}_1}{\xi'}\PD{\wt{\phi}_0}{\xi'}}=-\wh{\gamma}\kappa.
\end{equation}
Taking the derivative of \eqref{incomp1} with respect to $\xi'$ and using \eqref{u-1const} and \eqref{v-1const}, we see that
\begin{equation}\label{u0eqnmgr}
\PDD{2}{\wt{u}_0}{\xi'}=0.
\end{equation}
Using the above and \eqref{q1exp2} as well as \eqref{poissoninner0}, we have:
\begin{equation}\label{p-1eqn}
\PD{}{\xi'}\paren{2\mu\PD{\wt{u}_0}{\xi'}-\wt{p}_{-1}+\epsilon\PD{\wt{\phi}_1}{\xi'}\PD{\wt{\phi}_0}{\xi'}}=-\kappa\epsilon\paren{\PD{\wt{\phi}_0}{\xi'}}^2.
\end{equation}
We obtain the following from the usual matching procedure. We first have
\begin{equation}\label{u0const2}
\lim_{\xi'\to\pm\infty}\PD{\wt{u}_0}{\xi'}=\lim_{\xi\to0\pm} \PD{u_{-1}}{\xi},
\end{equation}
together with
\begin{equation}\label{u0cont2}
\lim_{\xi'\to\pm\infty}\PD{\wt{u}_0}{\xi'}=\lim_{\xi\to0\pm} \PD{u_{-1}}{\xi}, \quad \jump{u_0}=0.
\end{equation}
We also have:
\begin{equation}
\begin{split}
\label{gammaeff}
\jump{2\mu\PD{u_{-1}}{\xi'}-p_{-1}}&=-\wh{\gamma}_{\rm eff}\kappa,\;\; \wh{\gamma}_{\rm eff}=\wh{\gamma}-\dual{\epsilon\PD{\wt{\phi}_0}{\xi'}}{\PD{\wt{\phi}_0}{\xi'}},\\
\dual{\epsilon\PD{\wt{\phi}_0}{\xi'}}{\PD{\wt{\phi}_0}{\xi'}}&=\frac{8\lambda_+\epsilon_{\rm e}A_+^2}{1-A_+^2}+\frac{8\lambda_-\epsilon_{\rm i}A_-^2}{1-A_-^2},
\end{split}
\end{equation}
where $\dual{\cdot}{\cdot}$ is the standard inner product on $L^2(\mathbb{R})$, 
the space of square integrable functions on the real line
and $\lambda_\pm, A_\pm$ are defined in \eqref{wtphiexp} and \eqref{Apm}.
For $v^i$ we have:
\begin{equation}\label{wtvi02}
\mu\PDD{2}{\wt{v}^i_{0}}{\xi'}-g^{ij}\PD{\wt{p}_{-2}}{\eta^j}=\wt{q}_0g^{ij}\PD{\wt{\phi}_0}{\eta^j}.
\end{equation}
From \eqref{p-2exp}, $\wt{p}_{-2}$ does not depend on $\bm{\eta}$. Thus, 
we have:
\begin{equation}
\mu\PDD{2}{\wt{v}^i_{0}}{\xi'}=\wt{q}_0g^{ij}\PD{\wt{\phi}_0}{\eta^j}.
\end{equation}
This is supplemented with the interface condition:
\begin{equation}\label{intinnervi02}
\jump{v^i_{0}}=0, \; \jump{\mu\paren{\PD{\wt{v}^i_{0}}{\xi'}+g^{ij}\PD{\wt{u}_{-1}}{\eta^j}}}=0.
\end{equation}
Note that \eqref{wtvi02} together with \eqref{poissoninner0} yields:
\begin{equation}\label{muviepsphi}
\PDD{2}{}{\xi'}\paren{\mu \wt{v}^i_{0}+\epsilon \wt{\phi}_0g^{ij}\PD{\phi_0}{\eta^j}}=0,
\end{equation}
where we used \eqref{phi0eta}. Thus, 
\begin{equation}
\wt{v}^i_{0}=-\frac{\epsilon}{\mu}\wt{\phi}_0g^{ij}\PD{\phi_0}{\eta^j}+C_\pm \xi'+B_{\pm}.
\end{equation}
where the $+$ $(-)$ sign is chosen when $\xi'>0$ ($\xi'<0$), where $B_\pm, C_\pm$ are constants to be determined.
The interface conditions \eqref{intinnervi02} yields:
\begin{equation}
\mu_{\rm e}C_+=\mu_{\rm i}C_-, \; B_+-\frac{\epsilon_{\rm e}}{\mu_{\rm e}}\wt{\phi}_0(0)g^{ij}\PD{\phi_0}{\eta^j}
=B_--\frac{\epsilon_{\rm i}}{\mu_{\rm i}}\wt{\phi}_0(0)g^{ij}\PD{\phi_0}{\eta^j}.
\end{equation}
The usual matching procedure yields:
\begin{equation}\label{viApm}
\lim_{\xi\to 0\pm} \PD{v^i_{-1}}{\xi}=\lim_{\xi'\to \pm\infty}\PD{\wt{v}^i_0}{\xi'}=C_{\pm}, 
\; \lim_{\xi \to 0\pm}v^i_{0}=\lim_{\xi'\to\pm\infty}\paren{-\frac{\epsilon}{\mu}\wt{\phi}_0g^{ij}\PD{\phi_0}{\eta^j}+B_{\pm}}
\end{equation}
In particular, we have:
\begin{equation}\label{stressjump-1}
\jump{\mu\paren{\PD{v^i_{-1}}{\xi}+g^{ij}\PD{u_{-1}}{\eta^j}}}=0,
\end{equation}
and 
\begin{equation}\label{vi0slip}
\jump{v^i_{0}}=-\paren{\frac{\epsilon_{\rm i}}{\mu_{\rm i}}\phi^\Delta_{\rm i}+\frac{\epsilon_{\rm e}}{\mu_{\rm e}}\phi^\Delta_{\rm e}}g^{ij}\PD{\phi_0}{\eta^j},
\end{equation}
where $\phi^\Delta_{\rm i,e}$ were defined in \eqref{phiDeltaie}.
As remarked below \eqref{phiDeltaie}, $\phi^\Delta_{\rm i,e}$ have the same sign, and thus, there is a jump in the 
velocity $v^i_{0}$ across the interface $\Gamma$ so long as $l_{\rm C}\neq l_{\rm A}$ and $\partial{\phi_0}/\partial \eta^i\neq 0$.
{This velocity slip is essentially a liquid-liquid version of the Smoluchowski slip formula for liquid-solid interfaces. 
This slip arises due to the fact that there are charges of opposite sign on the two sides of the interface $\Gamma$ (see Figure \ref{velocity_slip_fig}).}

We may now collect our results. 
The governing equations for the leading order electrostatic potential and fluid velocity are:
\begin{align}
\nabla \cdot (\sigma \nabla \phi_0)&=0,\\ 
\jump{\phi_0}&=\ln \paren{\frac{l_{\rm C}}{\sqrt{S_{\rm i}}}}=\frac{1}{2}\ln \paren{\frac{l_{\rm C}}{l_{\rm A}}}, \quad \jump{\sigma\PD{\phi_0}{\xi}}=0,\\ 
\mu\Delta \mb{u}_{-1}-\nabla p_{-1}&=0, \; \nabla \cdot \mb{u}_{-1}=0,\\ 
\jump{\bm{u}_{-1}}&=0,\quad \jump{\Sigma(\bm{u}_{-1},p_{-1})\bm{n}}=-\wh{\gamma}_{\rm eff}\kappa \bm{n}.
\end{align}
The effective tension
$\wh{\gamma}_{\rm eff}$ is defined as
\begin{equation}
\begin{split}
\label{gammaeff0}
\wh{\gamma}_{\rm eff}=\wh{\gamma}-\dual{\epsilon\PD{\wt{\phi}_0}{\xi'}}{\PD{\wt{\phi}_0}{\xi'}},\;
\dual{\epsilon\PD{\wt{\phi}_0}{\xi'}}{\PD{\wt{\phi}_0}{\xi'}}&=\frac{8\lambda_+\epsilon_{\rm e}A_+^2}{1-A_+^2}+\frac{8\lambda_-\epsilon_{\rm i}A_-^2}{1-A_-^2},
\end{split}
\end{equation}
where $\lambda_\pm, A_\pm$ are defined in \eqref{wtphiexp} and \eqref{Apm}.
%
%
The equations for voltage $\phi_0$ and the fluid velocity 
$\mb{u}_{-1}$ are thus completely decoupled, except that the effective surface tension constant is modified to $\wh{\gamma}_{\rm eff}$. 
We point out that the equation for $\mb{u}_{-1}$ is well-posed only if $\wh{\gamma}_{\rm eff}>0$, suggesting that the above asymptotic 
procedure may not be valid if $\wh{\gamma}_{\rm eff}<0$.

Assume that $\wh{\gamma}_{\rm eff}>0$. To leading order, 
the fluid will undergo Stokesian motion without any influence from the electric field. 
An initially deformed droplet will approach a sphere (if initially homeomorphic to a sphere).
The dynamics described by $\mb{u}_{-1}$
thus represents the initial layer, after which the droplet is approximately spherical and $\mb{u}_0$ becomes
the leading order term of the velocity field.
 
\subsection{Perturbation from Sphere}\label{perturb_sphere}

We now consider the dynamics of the droplet after the initial layer and the droplet is approximately spherical. 
This amounts to obtaining equations for $\mb{u}_0$.
We expand the curvature $\kappa$ in terms of $\delta $:
\begin{equation}
\kappa=\kappa_0+\delta \kappa_1+\cdots.
\end{equation}
We note that, strictly speaking, this expansion had to be applied to all preceding calculations. 
It can be checked that this will not have made any difference in our calculations thus far (since this 
would only amount to changing all instances of $\kappa$ to $\kappa_0$ in the foregoing calculations).
Assuming that an equilibrium steady state is reached ($\wh{\gamma}_{\rm eff}>0$)
\begin{equation}\label{assumerest}
u_{-1}=0, \; p_{-1}=\begin{cases}
0 &\text{ in } \Omega_{\rm i}\\
\wh{\gamma}_{\rm eff}\kappa_0 &\text{ in } \Omega_{\rm e}
\end{cases},\; \jump{p_{-1}}= \wh{\gamma}_{\rm eff}\kappa_0,
\end{equation}
where $\kappa_0$ is the constant curvature of the spherical droplet and we have 
normalized the pressure in $\Omega_{\rm e}$ to be $0$. 

{Given \eqref{q2zero} (and \eqref{q01zero}),} the equation for $\mb{u}_0$ in the outer layer is given by:
\begin{equation}\label{stokes02}
\mu\Delta \mb{u}_0-\nabla p_0=0, \; \nabla\cdot \mb{u}_0=0.
\end{equation}
We already have the interfacial conditions \eqref{u0cont2} and \eqref{vi0slip}.
We have only to obtain the stress jump conditions across the interface.
As before, the variables with $\wt{\cdot}$ are for the inner layer.

From \eqref{assumerest} and using \eqref{u-1const}, \eqref{v-1const}, we have:
\begin{equation}\label{u-1v-10}
\wt{u}_{-1}=\wt{v}^i_{-1}=0,
\end{equation}
We caution that $\wt{p}_{-1}$ is {\em not} necessarily constant within the inner layer.
We also have:
\begin{equation}\label{u0vi0div0}
\PD{\wt{u}_0}{\xi'}=0, \; \lim_{\xi'\to \pm\infty}\PD{\wt{v}^i_0}{\xi'}=0,
\end{equation}
where we used \eqref{u0const2} in the first equality and \eqref{viApm} in the second.
Let us now consider the equations for $u$. From \eqref{dlessstokes} using \eqref{u-1v-10} 
and \eqref{u0vi0div0} we have (see \eqref{divcurv1} and \eqref{stokescurv01}):
\begin{align}
\label{stokescurv01mgr}
\mu\PDD{2}{\wt{u}_1}{\xi'}-\PD{\wt{p}_0}{\xi'}&=\wt{q}_0\PD{\wt{\phi}_2}{\xi'}+\wt{q}_1\PD{\wt{\phi}_1}{\xi'}+\wt{q}_2\PD{\wt{\phi}_0}{\xi'},\\
\label{divcurv1mgr}
0&=\PD{\wt{u}_1}{\xi'}+\kappa_0\wt{u}_0+\frac{1}{\sqrt{\abs{g}}}\PD{}{\eta^i}\paren{\sqrt{\abs{g}}\wt{v}^i_0}.
\end{align}
From \eqref{dlstressbc} and \eqref{phi0eta}, we have (see \eqref{Sigmanormal})
\begin{equation}
\jump{2\mu\PD{\wt{u}_1}{\xi'}-p_0+\frac{\epsilon}{2}\paren{\paren{\PD{\wt{\phi}_1}{\xi'}}^2+2\PD{\wt{\phi}_0}{\xi'}\PD{\wt{\phi}_2}{\xi'}-g^{ij}\PD{\phi_0}{\eta^i}\PD{\phi_0}{\eta^j}}}=-\wh{\gamma}\kappa_1.
\end{equation}
Take the derivative of \eqref{divcurv1mgr} with respect to $\xi'$, multiply by $\mu$ and add to \eqref{stokescurv01mgr}. 
Using \eqref{u-1v-10}, we obtain:
\begin{equation}\label{u1p02}
2\mu\PDD{2}{\wt{u}_1}{\xi'}-\PD{\wt{p}_0}{\xi'}=-\frac{\mu}{\sqrt{\abs{g}}}\PD{}{\eta^i}\paren{\sqrt{\abs{g}}\PD{\wt{v}^i_0}{\xi'}}
+\wt{q}_0\PD{\wt{\phi}_2}{\xi'}+\wt{q}_1\PD{\wt{\phi}_1}{\xi'}+\wt{q}_2\PD{\wt{\phi}_0}{\xi'}.
\end{equation}
Note that $\wt{q}_2$ satisfies:
\begin{equation}
-\wt{q}_2=\epsilon\paren{\PDD{2}{\wt{\phi}_2}{\xi'}+\kappa_0\PD{\wt{\phi}_1}{\xi'}+\kappa_1\PD{\wt{\phi}_0}{\xi'}+\Delta_{\bm{\eta}} \wt{\phi}_0},
\end{equation}
where $\Delta_{\bm{\eta}}$ is the Laplace-Beltrami operator on the interface:
\begin{equation}
\Delta_{\bm{\eta}} \phi_0=\frac{1}{\sqrt{\abs{g}}}\PD{}{\eta^i}\paren{\sqrt{\abs{g}}g^{ij}\PD{\phi_0}{\eta^j}}.
\end{equation}
Using \eqref{muviepsphi}, \eqref{viApm} and \eqref{u0vi0div0}, we have:
\begin{equation}\label{muviepsphi2}
\mu \PD{\wt{v}^i_{0}}{\xi'}+\epsilon \PD{\wt{\phi}_0}{\xi'}g^{ij}\PD{\phi_0}{\eta^j}=0.
\end{equation}
From this, we see that
\begin{equation}
\frac{\mu}{\sqrt{\abs{g}}}\PD{}{\eta^i}\paren{\sqrt{\abs{g}}\PD{\wt{v}^i_0}{\xi'}}+\epsilon\PD{\wt{\phi}_0}{\xi'}\Delta_{\bm{\eta}}\phi_0=0,
\end{equation}
Using the above relations, \eqref{poissoninner0} and \eqref{q1exp2}, equation \eqref{u1p02} can be rewritten as:
\begin{equation}
\begin{split}
&\PD{}{\xi'}\paren{2\mu\PD{\wt{u}_1}{\xi'}-p_0+\frac{\epsilon}{2}\paren{\paren{\PD{\wt{\phi}_1}{\xi'}}^2+2\PD{\wt{\phi}_0}{\xi'}\PD{\wt{\phi}_2}{\xi'}}}\\
=&-2\kappa_0\epsilon\PD{\wt{\phi}_0}{\xi'}\PD{\wt{\phi}_1}{\xi'}
-\kappa_1\epsilon\paren{\PD{\wt{\phi}_0}{\xi'}}^2.
\end{split}
\end{equation}
We may now apply the usual matching procedure to obtain:
\begin{equation}\label{stressbcu02}
\begin{split}
&\jump{2\mu\PD{u_0}{\xi}-p_0+\frac{\epsilon}{2}\paren{\paren{\PD{\phi_0}{\xi}}^2-g^{ij}\PD{\phi_0}{\eta^i}\PD{\phi_0}{\eta^j}}}\\
=&-\wh{\gamma}_{\rm eff}\kappa_1+2\kappa_0\paren{\frac{\epsilon_{\rm i}}{\sigma_{\rm i}}\phi_{\rm i}^\Delta+\frac{\epsilon_{\rm e}}{\sigma_{\rm e}}\phi_{\rm e}^\Delta-I_J}J_0-2\kappa_0^2I_\kappa
\end{split}
\end{equation}
where $\wh{\gamma}_{\rm eff}$ was defined in \eqref{gammaeff}, $\phi_{\rm i,e}^\Delta$ were defined in \eqref{phiDeltaie} and
\begin{equation}\label{IJkappa}
\begin{split}
I_\kappa&=\dual{\epsilon\PD{\wt{\phi}_0}{\xi'}}{\PD{\psi_\kappa}{\xi'}}, \; I_J=\dual{\epsilon\PD{\wt{\phi}_0}{\xi'}}{\PD{\psi_J}{\xi'}+R_\Delta},\\
R_\Delta&=\begin{cases}
\Sigma(\wt{q}_0)^{-1}-\sigma_{\rm i}^{-1} &\text{ for } \xi'<0,\\
\Sigma(\wt{q}_0)^{-1}-\sigma_{\rm e}^{-1} &\text{ for } \xi'>0.
\end{cases}
\end{split}
\end{equation}
In the above derivation, we used \eqref{phi1psi}, 
\eqref{psikappa} and \eqref{psiJ}. Note that $I_\kappa$ and $I_J$ depend only on the parameters of the system of equations.

We now turn to the equations for $v^i$. We have (see \eqref{stokescurvi1sphere}):
\begin{equation}\label{vi12}
\mu\paren{\PDD{2}{\wt{v}^i_{1}}{\xi'}+2\kappa_0\PD{\wt{v}^i_{0}}{\xi'}}-g^{ij}\PD{p_{-1}}{\eta^j}=\wt{q}_0g^{ij}\PD{\wt{\phi}_1}{\eta^j}+\wt{q}_1g^{ij}\PD{\phi_0}{\eta^j},
\end{equation}
where we used \eqref{phi0eta} in the last term.
The interface condition is given by:
\begin{equation}\label{wtvi1u0int}
\jump{\mu\paren{\PD{\wt{v}^i_{1}}{\xi'}+g^{ij}\PD{\wt{u}_0}{\eta^j}}}=0.
\end{equation}
From \eqref{u0eqnmgr} and \eqref{p-1eqn}, we have:
\begin{equation}
\PD{}{\xi'}\paren{\PD{\wt{p}_{-1}}{\eta^i}-\epsilon\frac{\partial^2\wt{\phi}_1}{\partial \xi'\partial \eta^i}\PD{\wt{\phi}_0}{\xi'}}=0
\end{equation}
Recalling that $p_{-1}$ is spatially constant in the outer layer by assumption, we see that
\begin{equation}
\PD{\wt{p}_{-1}}{\eta^i}-\epsilon\frac{\partial^2\wt{\phi}_1}{\partial \xi'\partial \eta^i}\PD{\wt{\phi}_0}{\xi'}=0.
\end{equation}
Plugging this back into \eqref{vi12} and using \eqref{poissoninner0}, we obtain:
\begin{equation}
\mu\paren{\PDD{2}{\wt{v}^i_{1}}{\xi'}+2\kappa_0\PD{\wt{v}^i_{0}}{\xi'}}=\PD{}{\xi'}\paren{\epsilon\PD{\wt{\phi}_0}{\xi'}g^{ij}\PD{\wt{\phi}_1}{\eta^j}}+\wt{q}_1g^{ij}\PD{\phi_0}{\eta^j}.
\end{equation}
Further using \eqref{muviepsphi2} and \eqref{q1exp2}, 
\begin{equation}\label{mu2vi1}
\mu\PDD{2}{\wt{v}^i_{1}}{\xi'}=\PD{}{\xi'}\paren{\epsilon\PD{\wt{\phi}_0}{\xi'}g^{ij}\PD{\wt{\phi}_1}{\eta^j}}
+\paren{\epsilon\kappa_0 \PD{\wt{\phi}_0}{\xi'}-\epsilon\PDD{2}{\wt{\phi}_1}{\xi'}}g^{ij}\PD{\phi_0}{\eta^j}.
\end{equation}
Let us integrate both sides from $\xi'=-\infty$ to $\infty$. The first term on the right yields:
\begin{equation}
\int_{-\infty}^\infty\PD{}{\xi'}\paren{\epsilon\PD{\wt{\phi}_0}{\xi'}g^{ij}\PD{\wt{\phi}_1}{\eta^j}}d\xi'=\jump{\epsilon\PD{\wt{\phi}_0}{\xi'}g^{ij}\PD{\wt{\phi}_1}{\eta^j}}=0.
\end{equation}
where we used the continuity of $\epsilon{\partial\wt{\phi}_0}/{\partial\xi'}$ and $\wt{\phi}_1$ across $\xi'=0$ (see \eqref{intinner0} and \eqref{phi1q1bc}) in the second equality. 
Let us turn to the second term on the right hand side of \eqref{mu2vi1}. 
Noting that $g^{ij}\partial\phi_0/\partial\eta^j$ does not depend on $\xi'$, it is sufficient to compute the following.
\begin{equation}
\int_{-\infty}^\infty \paren{\epsilon\kappa_0 \PD{\wt{\phi}_0}{\xi'}-\epsilon\PDD{2}{\wt{\phi}_1}{\xi'}}d\xi'
=-\kappa_0\paren{\epsilon_{\rm i}\phi^\Delta_{\rm i}+\epsilon_{\rm e}\phi^\Delta_{\rm e}}-\paren{\frac{\epsilon_{\rm i}}{\sigma_{\rm i}}-\frac{\epsilon_{\rm e}}{\sigma_{\rm e}}}J_0.\\
\end{equation}
In the above, we used the definitions of $\phi_{\rm i,e}^\Delta$ in \eqref{phiDeltaie}, as well as \eqref{sigmaphi0current} and \eqref{intinner0} for the second integrand.
The matching procedure for \eqref{mu2vi1} can now be completed using the above together with \eqref{wtvi1u0int}, from which we find that
\begin{equation}\label{stressbcvi02}
\jump{\mu\paren{\PD{v^i_{0}}{\xi}+g^{ij}\PD{u_0}{\eta^j}}}
=\paren{\kappa_0\paren{\epsilon_{\rm i}\phi^\Delta_{\rm i}+\epsilon_{\rm e}\phi^\Delta_{\rm e}}+\paren{\frac{\epsilon_{\rm i}}{\sigma_{\rm i}}-\frac{\epsilon_{\rm e}}{\sigma_{\rm e}}}J_0}g^{ij}\PD{\phi_0}{\eta^j}.
\end{equation}
We have thus obtained the stress boundary conditions \eqref{stressbcu02} and \eqref{stressbcvi02} for $u_0$ and $v^i_{0}$
respectively. Together with \eqref{u0cont2} and \eqref{vi0slip}, these conditions constitute the interface conditions for the 
Stokes equation \eqref{stokes02}. It is important to note that, when the voltage jump $\jump{\phi_0}$ is equal to $0$ 
(or $l_{\rm C}=l_{\rm A}$), 
the above interface conditions reduce to those of the TM model without convection. In this sense, the 
above calculation generalizes the TM model (without convection, applied to a sphere) 
in the presence of an interfacial double layer charge density.

\subsection{Spherical Drop under Uniform Electric Field}\label{sph_drop}

We now apply the foregoing calculations to the case of 
a spherical drop under a uniform electric field.
The calculation here parallels that of \citet{Schnitzer2015_JFM}.
Assuming the viscous drop maintains a spherical shape with a dimensionless radius of $1$, 
we use spherical polar coordinates for our calculation.
The voltage is given by:
\begin{equation}
\phi=
\begin{cases}
-Er\cos\theta\frac{3\sigma_{\rm e}}{2\sigma_{\rm e}+\sigma_{\rm i}}+\frac{1}{2}\ln(l_{\rm C}/l_{\rm A}) &\text{ for } r<1,\\
-E\cos\theta\paren{r+\frac{\sigma_{\rm e}-\sigma_{\rm i}}{2\sigma_{\rm e}+\sigma_{\rm i}}r^{-2}} &\text{ for } r\geq 1,
\end{cases}
\end{equation}
where $r$ is the radial coordinate and $\theta$ is the polar angle. The electric field is pointing in the $\theta=0$ direction 
and is of magnitude $E$. In the above and in what follows, we have omitted 
the subscript $0$ indicating the order in the asymptotic expansion. We shall only be interested in calculating the order $0$
quantities. Let $u_r$ and $u_\theta$ be the velocities in the $r$ and $\theta$ directions.
The azimuthal component of the flow will be $0$ by symmetry.
We suppose the velocity field is $0$ in the far field. 
Using \eqref{u0cont2} and \eqref{vi0slip}, at the interface $r=1$, we have the conditions:
\begin{align}
\jump{u_r}&=0,\label{uric}\\
\jump{u_\theta}&=-\paren{\frac{\epsilon_{\rm i}}{\mu_{\rm i}}\phi^\Delta_{\rm i}+\frac{\epsilon_{\rm e}}{\mu_{\rm e}}\phi^\Delta_{\rm e}}
\frac{3\sigma_{\rm e}E}{2\sigma_{\rm e}+\sigma_{\rm i}}\sin\theta.\label{uthetaic}
\end{align}
We now turn to the stress jump conditions. Let $\Sigma_{rr}$ and $\Sigma_{r\theta}$ be the $rr$ and $r\theta$ components of the fluid stress,
which can be expressed using $u_r$ and $u_\theta$ as:
\begin{equation}
\Sigma_{rr}=2\mu\PD{u_r}{r}-p, \; \Sigma_{r\theta}=\mu\paren{r\PD{}{r}\paren{\frac{u_\theta}{r}}+\frac{1}{r}\PD{u_r}{\theta}}.
\end{equation}
The stress jump conditions \eqref{stressbcu02} and \eqref{stressbcvi02} yield:
\begin{align}
\nonumber \jump{\Sigma_{rr}}=&\frac{1}{2}\paren{\frac{3\sigma_{\rm e}E}{2\sigma_{\rm e}+\sigma_{\rm i}}}^2
\paren{\epsilon_{\rm i}-\epsilon_{\rm e}+\paren{\epsilon_{\rm e}\paren{\paren{\frac{\sigma_{\rm i}}{\sigma_{\rm e}}}^2+1}-2\epsilon_{\rm i}}\cos^2\theta}\label{Sigrric}\\
&+\frac{12\sigma_{\rm e}\sigma_{\rm i}E}{2\sigma_{\rm e}+\sigma_{\rm i}}
\paren{\frac{\epsilon_{\rm i}}{\sigma_{\rm i}}\phi_{\rm i}^\Delta+\frac{\epsilon_{\rm e}}{\sigma_{\rm e}}\phi_{\rm e}^\Delta-I_J}\cos\theta
-8I_\kappa,\\
\jump{\Sigma_{r\theta}}=&\paren{\frac{3\sigma_{\rm e}E}{2\sigma_{\rm e}+\sigma_{\rm i}}}^2\paren{\epsilon_{\rm i}-\epsilon_{\rm e}\paren{\frac{\sigma_{\rm i}}{\sigma_{\rm e}}}}\sin\theta\cos\theta+\frac{6\sigma_{\rm e}E}{2\sigma_{\rm e}+\sigma_{\rm i}}\paren{\epsilon_{\rm i}\phi^\Delta_{\rm i}+\epsilon_{\rm e}\phi^\Delta_{\rm e}}\sin\theta.\label{Sigrthic}
\end{align}
We have used the fact that $\kappa_0=2$ for a unit sphere. Since we assume the drop shape is spherical, $\kappa_1=0$. 
This constraint should lead to restrictions on the parameter values. 
We shall later restore the surface tension term $\gamma \kappa_1$. 

We now introduce the Stokes stream function $\psi(r,\theta)$ 
to solve the above interface problem, which is related to $u_r$ and $u_\theta$ via
\begin{equation}
u_r=\frac{1}{r^2\sin\theta}\PD{\psi}{\theta}, \; u_\theta=-\frac{1}{r\sin\theta}\PD{\psi}{r}.
\end{equation}
We use the well-known separation of variables solution (see, for example, \citet{leal2007advanced}).
From the above interface conditions, the Stokes stream function $\psi$ should be written as:
\begin{equation}
\psi(r,\theta)=
\begin{cases}
\sum_{n=1}^2 \paren{A_nr^{n+3}+B_nr^{n+1}}Q_n(\cos\theta) &\text{ if } r<1,\\
\sum_{n=1}^2\paren{C_nr^{2-n}+D_nr^{-n}}Q_n(\cos\theta) &\text{ if } r\geq 1.
\end{cases}
\end{equation}
where
\begin{equation}
Q_1(z)=1-z^2, \; Q_2(z)=z-z^3.
\end{equation}
The corresponding pressure is computed as
\begin{equation}
p(r,\theta)=
\begin{cases}
\mu_i \paren{ \frac{7}{2}A_2 r^2 + 20 A_1 \cos\theta + \frac{21}{2} A_2 r^2\cos2\theta} &\text{ if } r<1,\\
\mu_e\paren{ C_2r^{-3} + 2 C_1r^{-2}\cos\theta + 3 C_2r^{-3}\cos2\theta} &\text{ if } r\geq 1.
\end{cases}
\end{equation}
Let us determine the coefficients in the above expressions. 
Plugging in the above expression into \eqref{uric}, we obtain:
\begin{equation}
\begin{split}
&2(A_1+B_1)\cos\theta-(A_2+B_2)(1-3\cos^2\theta)\\
&=2(C_1+D_1)\cos\theta-(C_2+D_2)(1-3\cos^2\theta)=\at{u_r}{r=1},\label{uricsub} \\
\end{split}
\end{equation}
Since we assume that the drop maintains its spherical shape, $u_r$ at $r=1$ must be proportional 
to $\cos\theta$. Therefore, we have:
\begin{equation}
A_1+B_1=C_1+D_1, \; A_2+B_2=C_2+D_2=0.
\end{equation}
From \eqref{uthetaic}, we obtain:
\begin{equation}
\begin{split}
&(-4A_1-2B_1+C_1-D_1)\sin\theta+(-5A_2-3B_2-2D_2)\sin\theta\cos\theta\\
&=-\paren{\frac{\epsilon_{\rm i}}{\mu_{\rm i}}\phi^\Delta_{\rm i}+\frac{\epsilon_{\rm e}}{\mu_{\rm e}}\phi^\Delta_{\rm e}}
\frac{3E\sigma_{\rm e}}{2\sigma_{\rm i}+\sigma_{\rm e}}\sin\theta.
\end{split}
\end{equation}
Thus, 
\begin{align}
-4A_1-2B_1+C_1-D_1&=-\paren{\frac{\epsilon_{\rm i}}{\mu_{\rm i}}\phi^\Delta_{\rm i}+\frac{\epsilon_{\rm e}}{\mu_{\rm e}}\phi^\Delta_{\rm e}}
\frac{3E\sigma_{\rm e}}{2\sigma_{\rm e}+\sigma_{\rm i}},\\
-5A_2-3B_2-2D_2&=0.
\end{align}
Likewise, from \eqref{Sigrric} and \eqref{Sigrthic}, we obtain the equations:
\begin{align}
-12\mu_{\rm i}A_1+\mu_{\rm e}(6C_1+12D_1)&=\frac{12\sigma_{\rm e}\sigma_{\rm i}E}{2\sigma_{\rm e}+\sigma_{\rm i}}
\paren{\frac{\epsilon_{\rm i}}{\sigma_{\rm i}}\phi_{\rm i}^\Delta+\frac{\epsilon_{\rm e}}{\sigma_{\rm e}}\phi_{\rm e}^\Delta-I_J},\\
-3(\mu_{\rm i}(A_2-2B_2)-\mu_{\rm e}(6C_2+8D_2))&=\frac{1}{2}\paren{\frac{3\sigma_{\rm e}E}{2\sigma_{\rm e}+\sigma_{\rm i}}}^2
\paren{\epsilon_{\rm e}\paren{\paren{\frac{\sigma_{\rm i}}{\sigma_{\rm e}}}^2+1}-2\epsilon_{\rm i}},\\
-6\mu_{\rm i}A_1+6\mu_{\rm e}D_1&
=\frac{6\sigma_{\rm e}E}{2\sigma_{\rm e}+\sigma_{\rm i}}\paren{\epsilon_{\rm i}\phi^\Delta_{\rm i}+\epsilon_{\rm e}\phi^\Delta_{\rm e}}\\
-\mu_{\rm i}(16A_2+6B_2)+\mu_{\rm e}(6C_2+16D_2)&=\paren{\frac{3\sigma_{\rm e}E}{2\sigma_{\rm e}+\sigma_{\rm i}}}^2\paren{\epsilon_{\rm i}-\epsilon_{\rm e}\paren{\frac{\sigma_{\rm i}}{\sigma_{\rm e}}}}.
\end{align}
We may solve the above for the eight constants. The five equations for $A_2, B_2, C_2$ and $D_2$, are exactly the same 
as those discussed in \citet{Taylor1966_PRSLa}, and this results in a
condition on the parameters for solvability. This leads to the discriminating function for prolate/oblate deformation.
We note that $l_{\rm A}\ne l_{\rm C}$ only affects $A_1, B_1, C_1$ and $D_1$. Thus the voltage jump due to 
$l_{\rm A}\ne l_{\rm C}$ does not
affect the discriminating function at leading order. However, we expect the migration speed of the drop $V_{\rm mgr}$
to depend on such voltage jump because, from \eqref{uricsub}, $V_{\rm mgr}=2(A_1+B_1)$. 
Solving for the constants $A_1, B_1, C_1$ and $D_1$, we obtain:
\begin{equation}
\begin{split}
V_{\rm mgr}=&\frac{2\paren{5\epsilon_{\rm i}\sigma_{\rm e}\mu_{\rm e}\phi_{\rm i}^\Delta
+\epsilon_{\rm e}\paren{4\sigma_{\rm i}\paren{\mu_{\rm i}+\mu_{\rm e}}-\sigma_{\rm e}(\mu_{\rm i}+2\mu_{\rm e})}\phi_{\rm e}^\Delta}}
{\mu_{\rm e}(3\mu_{\rm i}+2\mu_{\rm e})(2\sigma_{\rm e}+\sigma_{\rm i})}E\\
&-\frac{8(\mu_{\rm i}+\mu_{\rm e})\sigma_{\rm i}\sigma_{\rm e}I_J}{\mu_{\rm e}(3\mu_{\rm i}+2\mu_{\rm e})(2\sigma_{\rm e}+\sigma_{\rm i})}E
\end{split}
\end{equation}
Using \eqref{phiDeltaieapprox} and noting that $I_J=\mc{O}((\phi^\Delta)^2)$ we have:
\begin{equation}
V_{\rm mgr}=\frac{2\paren{5\epsilon_{\rm i}\sigma_{\rm e}\mu_{\rm e}\zeta
+4\epsilon_{\rm e}\sigma_{\rm i}\paren{\mu_{\rm i}+\mu_{\rm e}}-\epsilon_{\rm e}\sigma_{\rm e}(\mu_{\rm i}+2\mu_{\rm e})}}
{\mu_{\rm e}(3\mu_{\rm i}+2\mu_{\rm e})(2\sigma_{\rm e}+\sigma_{\rm i})(1+\zeta)}\phi^\Delta E+\mc{O}((\phi^\Delta)^2).
\end{equation}
If $\phi^\Delta$ is small, the droplet migrates in the direction of the sign of $\phi^\Delta E$ when
\begin{equation}\label{Qmgr}
Q_{\rm mgr}=\frac{5\epsilon_{\rm r}\zeta+4\sigma_{\rm r}(1+\mu_{\rm r})}{2+\mu_{\rm r}}>1, \quad \epsilon_{\rm r}=\frac{\epsilon_{\rm i}}{\epsilon_{\rm e}},
\sigma_{\rm r}=\frac{\sigma_{\rm i}}{\sigma_{\rm e}}, \; \mu_{\rm r}=\frac{\mu_{\rm i}}{\mu_{\rm e}}.
\end{equation}
The direction of droplet migration, therefore, is not necessarily equal to $\phi^\Delta E$ as might have been expected.
This is analogous to the conclusions of \citet{baygents1991electrophoresis}, in which the electrophoresis of conducting drops was studied using an electrodiffusion model. 

We now numerically evaluate the flow field.
The only non-trivial detail is the evaluation of the term $I_J$ defined in \eqref{IJkappa},
which in turn requires the numerical solution of $\psi_J$ and $\psi_\kappa$ which satisfies \eqref{psiJ} and \eqref{psikappa}
respectively. The technical difficulty here is that these differential equations are posed on the real line, 
and a straight forward discretization does not work. We take a smooth map of the real line onto a finite interval
and discretize the transformed equations to obtain a numerical solution. The details of this computation are omitted.

\begin{figure}
\begin{center}
\includegraphics[width=\textwidth]{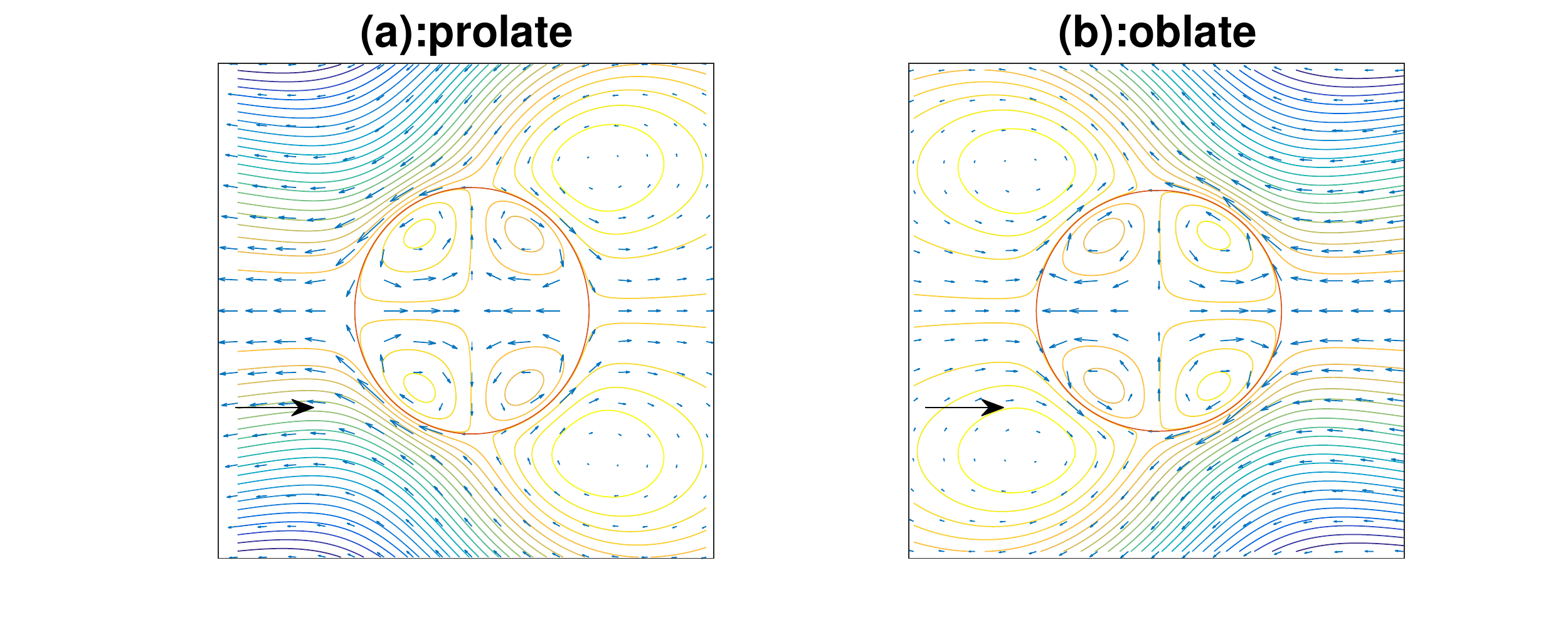}
\end{center}
\caption{\label{electromigration} Flow field around a migrating droplet, in a frame moving with the droplet.
Rotation axis is horizontal.
$D_{\rm C,A}^{\rm i,e}=1$, $\mu^{\rm i,e}=1$, $l_{\rm C}=1.01$, $l_{\rm A}=1/l_{\rm A}$. In Figure (a), 
$\epsilon_{\rm i}=1$, $\epsilon_{\rm e}=2$ and in Figure (b), $\epsilon_{\rm i}=2$, $\epsilon_{\rm e}=1$
so that the flow field is consistent with a prolate/oblate deformation in Figure (a)/(b).
Electric field is in the direction of the arrow.
$\phi^\Delta>0$, $Q_{\rm mgr}>1$ in both figures so that the droplet is moving
in the direction of the arrow.}
\end{figure}

Flow fields around a migrating drop are shown in Figure \ref{electromigration}. 
The flow fields are similar in its appearance to the quadrupole vortex flow field photograph 
reported in \citet{Taylor1966_PRSLa}. In his paper, Taylor reports that the droplet drifts, and the 
flow field shows an asymmetry not explained by the TM model. 

{\section{Discussion and Outlook}\label{discussion}}

\subsection{Scaling and Asymptotics}\label{theory}

First, we note that any asymptotic calculation is only as good as the expansion ansatz 
(\eqref{qpower} for Section \ref{leaky} and \eqref{dlexpansion} for Section \ref{DL}).
It is therefore important future work to validate of our conclusions by computational or analytic means.
{Computational verification will necessarily require numerical methods for the modified Saville or charge diffusion models.
Numerical methods in \citep{tomar2007two,lopez2011charge,berry2013multiphase,hu2015hybrid} may be of particular 
interest in this regard.}

{We have chosen particular scalings for the dimensionless variables to obtain limiting models. 
In Section \ref{CD}, we took the limit $\alpha \to 0$ before taking the limit $\delta \to 0$, as dictated by \eqref{scaling_intro}.
The scaling $\alpha \sim \delta$, for example, may be appropriate in certain situations (see Section \ref{mod_large}). 
In Section \ref{derivation_of_TM}, \eqref{scaling} was chosen to obtain the TM model.
As we saw in Table \ref{dist_limits} other scalings for $Pe$ (when $\phi^\Delta=0$) 
lead to the variants of the TM model. Other distinguished scalings may be worth exploring.
In Section \ref{DL}, scaling \eqref{scaling2} was chosen for analytical feasibility, and 
other scalings may likewise be worthy of further study.
Our analysis there requires that the surface tension coefficient $\gamma$ be greater than twice the electrostatic 
energy stored in the electric double layer (see \eqref{gammaeff} and the discussion at the end of Section \ref{subsec:lc_neq_la}).
When this condition is violated, it does not seem to be possible to take the limit $\delta\to 0$.
This may indicate the presence of an interfacial electrochemical/electrohydrodynamic instability.}

{Scaling of dimensionless variables should 
be guided by their magnitudes in experimental setups,
an important subject to which we now turn.}

%


\subsection{Dimensionless Parameters and Surface Charge Convection}\label{mod_large}

We discuss the size of dimensionless parameters that featured in our analysis.
Let the representative constants be: 
\begin{equation}
T=298K, \; \mu_*=1 {\rm Pa}\cdot{\rm s}, \; L=1{\rm mm}, \; \epsilon_*=4\epsilon_0,\; \sigma_*=10^{-9} {\rm S}/{\rm m}, \; \gamma_*=1{\rm mN/m},
\end{equation}
where $\epsilon_0$ is the permittivity of vacuum and $\sigma_*$ is the representative conductivity. 
All values follow \citet{Saville1997_AnnuRevFluidMech} except $\gamma_*$
which corresponds to the order of magnitude in \citet{salipante2010electrohydrodynamics}.
Following \citet{Saville1997_AnnuRevFluidMech} we set the representative ionic radius $r_{\rm ion}$ to be:
\begin{equation}\label{rion}
r_{\rm ion}=0.25{\rm nm}.
\end{equation}
Assuming the Stokes-Einstein relation, we may set:
\begin{equation}\label{stokes_einstein}
D_*=\frac{k_{\rm B}T}{6\pi \mu_* r_{\rm ion}}, \; \sigma_*=D_*\frac{F^2}{RT}c_*,
\end{equation}
where $k_{\rm B}$ is the Boltzmann constant. The expression for $\sigma_*$ allows for the estimation of $c_*$.
The above yields the following values for the dimensionless parameters.
\begin{equation}\label{dless_constants_values}
\delta=1.8\times 10^{-4}, \; Pe=2.7\times 10^{-2}, \; \gamma=1.3.
\end{equation}
In addition, \citet{Saville1997_AnnuRevFluidMech} uses the above to estimate the parameter $\alpha$ as:
\begin{equation}
\alpha\approx 10^{-4}.
\end{equation}
We note that many of the above parameters may easily vary by a factor of $10$ to $100$ ($\sigma_*$ for example)
in either direction, and thus, should be taken as a rough estimate. 
Since the identity of the conducting ions and their reactions in leaky dielectrics are unknown, 
the estimate for $r_{\rm ion}$ and $\alpha$ above are necessarily uncertain \citep{Saville1997_AnnuRevFluidMech}.

We now discuss the implications of the above values for the applicability of our analysis.
Our most important assumption is the ordering $\alpha\ll\delta\ll 1$ as in \eqref{scaling_intro}.
The above indicates that $\alpha$ and $\delta$ are comparable. 
Note, however, that $\delta$ can be made larger if we assume that $\sigma_*$ is smaller ($\sigma_*=10^{-11}$ to $10^{-12}{\rm S}/{\rm m}$
in \citet{salipante2010electrohydrodynamics}) or $L$ is smaller.
In addition, if the ions present in the leaky dielectrics
are generated by the dissociation of the solvent itself as discussed in Section \ref{CD} (e.g., H$^+$ and OH$^-$ in pure water), the ratio 
$\alpha$ can be significantly smaller (about $10^{-9}$ for pure water). 
The assumption $\alpha\ll\delta\ll 1$ is thus likely to be reasonable for many systems.
It would still be of interest to consider the limit $\alpha\approx \delta\ll 1$, 
possibly a more appropriate scaling in certain systems.

In \eqref{dless_constants_values}, $Pe$ is at most $\mc{O}(1)$ with respect to $\delta$ and $\gamma$ is much larger than $\mc{O}(\delta)$.
According to Table \ref{dist_limits}, this indicates that the TM model without surface charge convection will be valid to leading order.
The large value of $\gamma$ with respect to $\delta^2$ indicates that there will be little deformation from the sphere;
a formal analysis of this situation will be analogous to the calculation in Section \ref{DL}.

Recall from Section \ref{large_voltage} that our analysis can be extended to the case of large imposed voltage provided
\eqref{scaling_large} is satisfied. Given the estimate of $\delta\approx 10^{-4}$ in \eqref{dless_constants_values}, $\beta=\omega^{-2}=10^3$
may be the largest reasonable value for the dimensionless electric field magnitude for our derivation to hold.
In this case, the dimensionless parameters of \eqref{dless_strong} and \eqref{PeE} are:
\begin{equation}
Pe_{\rm E}=Pe\beta=2.7\times 10, \; \delta_{\rm E}=\delta \sqrt{\beta}=5.6\times 10^{-3}.
\end{equation}
Given that these are rough estimates, it is possible that $Pe_{\rm E}=\mc{O}(\delta_{\rm E}^{-1})$ in certain systems.
In this case, Table \ref{dist_limits} indicates that we may obtain surface charge convection as an $\mc{O}(\delta_{\rm E})$ correction.
The surface tension coefficient $\gamma$ is still much larger than $\delta_{\rm E}$, so that deformation from the sphere 
may be small.

The above considerations lead to the question of whether there are any leaky dielectric systems for which surface charge convection 
can be obtained as a leading order term.
According to \eqref{scaling_strong}, we would need:
\begin{equation}
Pe_{\rm E}=Pe \beta \approx \delta_{\rm E}^{-2}=\delta^{-2}\beta^{-1},
\end{equation}
which implies:
\begin{equation}\label{betaPe}
\beta \approx \sqrt{Pe^{-1}} \delta^{-1}.
\end{equation}
This same scaling can be obtained just from the TM model.
Balancing of the convective and conduction terms in the full TM model (see \eqref{phiqgamma} and \eqref{LD_qGammaeqn}) amounts to
equating the {\em electrohydrodynamic time} with the {\em Maxwell-Wagner relaxation time} \citep{salipante2010electrohydrodynamics}:
\begin{equation}
t_{\rm EHD}=\frac{\mu_*}{\epsilon_*E_*^2}\approx \frac{\epsilon_*}{\sigma_*}=t_{\rm MW}.
\end{equation}
This leads to the scaling of $\beta$ in \eqref{betaPe}. The ratio $Re_{\rm E}=t_{\rm MW}/t_{\rm EHD}$ is sometimes referred to as the electric Reynolds number.
We may thus say that the scaling \eqref{betaPe} corresponds to the regime where $Re_{\rm E}=\mc{O}(1)$.

Since we need \eqref{scaling_large2} for the validity of the asymptotic analysis, \eqref{betaPe} implies that
\begin{equation}
Pe\gg 1
\end{equation}
must be satisfied.
Using \eqref{stokes_einstein}, we may compute $Pe$ as:
\begin{equation}
Pe=\frac{r_{\rm ion}}{r_{\rm B}}, \; r_{\rm B}=\frac{q_{\rm el}^2}{6\pi \epsilon_* k_{\rm B}T},
\end{equation}
where $q_{\rm el}$ is the elementary charge. The length $r_{\rm B}$ is the Bjerrum length (up to a factor of $2/3$), 
which is on the order of $0.5$ to $10$nm depending on the dielectric constant. On the other hand, 
ion size is expected to be in the subnanometer range. Thus, barring the use of exotic materials, 
$Pe$ is typically never larger than $1$.

{
Surface charge convection will thus be negligible at thermal voltages, 
but will become appreciable as the imposed voltage becomes larger.
At very large voltages for which the electric Reynolds number is $\mc{O}(1)$, however,
our derivation of the TM model may break down.
Our theory is thus consistent with reports that surface charge convection is important for large imposed voltages
\citep{xu2006settling,salipante2010electrohydrodynamics,vlahovska2016electrohydrodynamic,Lanauze2015_JFM,Das2017_JFM,sengupta2017role},
but experimentally imposed voltages are often higher than the range over which 
our asymptotic calculations are guaranteed to be valid. It is, however, 
quite possible that validity of our asymptotics does not deteriorate too much even at such high voltages.}

{
It is clear that the TM model itself (and thus, any asymptotic derivation of the TM model) fail 
when and after geometric singularities arise.
Recall that the charge diffusion model of Section \ref{CD} depended only on the smallness 
of $\alpha$, and is thus potentially valid at very large field strengths. Direct analysis and simulation of the charge diffusion 
model may help resolve the physics of charge convection and singularity formation under strong fields.}

\subsection{Surface Electrochemistry, Galvani Potential and Electromigration}

It is well-known that many liquid-liquid interfaces have a spontaneous voltage jump known as the Galvani potential (GP) \citep{girault1989electrochemistry,reymond2000electrochemistry}.
In its presence, as shown in Section \ref{DL}, a suspended leaky dielectric droplet will migrate under an imposed electric field.
Our analysis yields an explicit formula for droplet migration velocity. 
{We have, unfortunately, not been able to obtain an estimate for the 
electromigration velocity primarily because of the absence of independent measurements of 
the GP. We may, in turn, be able to use the measured electromigration velocity to obtain an estimate of the GP based on our formula.
We also point out that the electromigration velocity formulae for droplets described in 
\citet{booth1951cataphoresis,baygents1991electrophoresis,pascall2011electrokinetics} are different from those obtained here.
Although this is not surprising given the different modeling assumptions, 
it would be interesting to clarify  the interrelation among these calculations.
}


The properties of the EDL and the resulting GP, which underpins our analysis of droplet migration,
should depend on the details of the surface electrochemistry at the liquid-liquid interface.
Our interfacial boundary conditions for ionic concentrations are the simplest possible.
Unlike \citet{Saville1997_AnnuRevFluidMech,Schnitzer2015_JFM}, the modified Saville model does not incorporate surface ionic concentrations. 
We have assumed that the anions and cations can move across the liquid-liquid interface. 
It is possible that electric current flows across the interface via Faradaic reactions rather than by
simple drift of ions. A more sophisticated model for surface electrochemistry may thus be needed 
for a better understanding of droplet electro-migration and related phenomena \citep{reymond2000electrochemistry}.

{\section{Concluding Remarks}}

In this paper, we have argued that the electrohydrodynamics of leaky dielectrics is the electrohydrodynamics 
of weak electrolyte solutions. The weak electrolyte limit of the modified Saville model leads to the charge diffusion model. 
The charge diffusion model, in turn gives rise to the TM model or droplet electro-migration depending on
the presence of the Galvani potential. We identify droplet migration as an EDL phenomenon, which suggests that 
the electrohydrodynamics of leaky dielectrics is not confined to EML phenomena. Our analysis may have broader 
implications for a unified understanding of EML phenomena, usually associated with the electrohydrodynamics of leaky dielectrics, 
and EDL phenomena, usually associated with electrokinetic behavior of conducting fluids \citep{bazant2015electrokinetics}.

In light of our results, the charge diffusion model may emerge as a suitable model to address problems in electrohydrodynamics
of leaky dielectrics that have so far resisted explanation with the TM model. 
It would thus be of great interest to further study the charge diffusion model, both from analytic and computational standpoints.

More broadly, our analysis highlights the importance of electrochemistry in electrohydrodynamics. 
Despite its importance, there seem to be very few analytical or computational studies on the interplay between 
electrochemistry and electrodiffusion/electrohydrodynamics \citep{bazant2009towards}. 
Our analysis suggests that there is much to explore in this area.\\

Y.M. would like to thank Qiming Wang for introducing him to the leaky dielectric model.
The authors would like to thank Satish Kumar and Howard Stone for discussion and encouragement.
A workshop hosted by the IMA (Institute for Mathematics and its Applications, University of Minnesota) 
led to this project, whose support we gratefully acknowledge. 
Y.M. was supported by NSF DMS-1620316, DMS-1516978 and Y.N.Y. was supported by NSF DMS-1614863, DMS-1412789.

\appendix

{\section{Energy Identities and Model Hierarchy}\label{app:energy}}

{An important feature of the models considered in this paper, the modified Saville (MS), charge diffusion (CD) and 
Taylor-Melcher (TM) models, is that they all satisfy 
free energy identities. In fact, the appropriate scaling of the dimensionless variables in the derivation of the TM model 
can be seen by an examination of the energy identities. We start with the MS model.}


{\subsection{Modified Saville Model}}
{We first discuss the energy identity satisfied by the modified Saville model discussed in Section \ref{sec:modified_saville_model}.
We first introduce a few constants. Set:
\begin{equation}
E_{{\rm X},{\rm i}}-E_{{\rm X},{\rm e}}=RT\ln l_{\rm X}, \; \rm{X}=\rm{C,A,S}
\end{equation}
These are the energy levels already discussed in \eqref{lXE}.}
Suppose $\Omega$ is a bounded domain.
Then, we have:
\begin{equation}\label{energy_full}
\D{\mc{E}}{t}=-\mc{D}+\mc{I},\; \mc{E}=\mc{E}_{\rm chem}+\mc{E}_{\rm elec}+\mc{E}_{\rm surf},\; 
\mc{D}=\mc{D}_{\rm ediff}+\mc{D}_{\rm visc}+\mc{D}_{\rm rct}.
\end{equation}
The components of the free energy $\mc{E}$ are given by:
\begin{equation}
\begin{split}
\mc{E}_{\rm chem}&=\sum_{{\rm X}={\rm C,A,S}}\int_{\Omega}
n_{\rm X}(RT(\ln n_{\rm X}-1)+E_{\rm X})d\mb{x},\\
\mc{E}_{\rm elec}&=\int_\Omega \frac{\epsilon}{2}\abs{\nabla \phi}^2d\mb{x}, \quad \mc{E}_{\rm surf}=\int_{\Gamma} \gamma_* dm_{\Gamma},
\end{split}
\end{equation}
where 
\begin{equation}
E_{\rm X}=\begin{cases}
E_{\rm X,i} &\text{ in } \Omega_{\rm i}\\
E_{\rm X,e}&\text{ in } \Omega_{\rm e}
\end{cases}
\text{ for } {\rm X}={\rm C,A,S}.
\end{equation}
The components of the dissipation $\mc{D}$ are given by: 
\begin{equation}
\begin{split}
\mc{D}_{\rm ediff}&=\sum_{\rm {X}={\rm C,A,S}}\int_\Omega \frac{D_{\rm X}n_{\rm X}}{RT}\abs{\nabla \mu_{\rm X}}^2d\mb{x},\;
 \mu_{\rm X}=RT\ln n_{\rm X}+z_{\rm X}F\phi,\\
\mc{D}_{\rm visc}&=\int_{\Omega} 2\mu\abs{\nabla_S \mb{u}}^2d\mb{x},\\
\mc{D}_{\rm rct}&=\int_\Omega RTk_+s(Q-1)\ln Qd\mb{x},\; Q=\frac{k_-ca}{k_+s}.
\end{split}
\end{equation}
Boundary free energy input $\mc{I}$ is given by:
\begin{equation}
\begin{split}
\mc{I}&=-\int_{\partial \Omega} \paren{\phi j_{\partial \Omega}+
\sum_{\rm {X}={\rm C,A,S}}\mu_{\rm X}f_{\rm X,\partial \Omega}}d{m_{\partial\Omega}},\\
j_{\partial \Omega}&=-\PD{}{t}\paren{\at{\epsilon\PD{\phi}{\mb{n}}}{\partial \Omega}},\; 
f_{{\rm X}, \partial \Omega}=-\at{\frac{D_{\rm X}n_{\rm X}}{RT}\PD{\mu_{\rm X}}{\mb{n}}}{\partial \Omega}.
\end{split}
\end{equation}
In the above, $n_{\rm C,A,S}$ are $c,a,s$ respectively, $z_{\rm X}$ is the valence of each ion 
($z_{\rm C}=-z_{\rm A}=1$ and $z_{\rm S}=0$), $\mb{n}$ is the outward normal on $\partial \Omega$ and  $dm_{\Gamma}$
and $dm_{\partial \Omega}$ denote integration on with the standard surface measures on $\partial \Omega$ and $\partial\Gamma$
respectively. 
The calculations that lead to the above energy identity will be discussed in Section \ref{free_en_deriv}.

The free energy has three components, the chemical and entropic free energy $\mc{E}_{\rm chem}$, the 
electrostatic energy $\mc{E}_{\rm elec}$ and the interfacial surface tension energy $\mc{E}_{\rm surf}$.
The total free energy is dissipated ($\mc{D}\geq 0$) except for possible free 
energy input at the boundary of the domain ($\mc{I}$). The dissipation consists of the viscous dissipation 
$\mc{D}_{\rm visc}$, the electrodiffusive dissipation $\mc{D}_{\rm ediff}$ and the dissipation due 
to the dissociation reaction $\mc{D}_{\rm rct}$. Boundary energy input consists of a term that comes 
from the displacement current and the chemical flux.
We emphasize that such an identity 
is not possible unless \eqref{th_restriction} is satisfied; it can be checked that a violation of this restriction 
will lead to spurious free energy creation at the interface $\Gamma$. 
When $\Omega=\mathbb{R}^3$, the above energy identity still holds if we restrict our integration to 
any bounded smooth domain containing $\Omega_{\rm i}$ and modify the boundary term $\mc{I}$
to include a convective term to the boundary chemical flux and work done by the stress on the boundary.

In the dimensionless variables introduced in Section \ref{non_dimensionalization},
the term in the energy relation \eqref{energy_full} assume the following forms.
For the free energy components, we have:
\begin{equation}\label{energy_dless}
\begin{split}
\mc{E}_{\rm chem}&=\sum_{{\rm X}={\rm C,A}}\mc{E}_{\rm X}+\frac{1}{\alpha}\mc{E}_{\rm S},\\
\mc{E}_{\rm X}&=\int_{\Omega} n_{\rm X}(\ln n_{\rm X}-1+E_{\rm X})d\mb{x}, {\rm X}={\rm C,A,S},\\
\mc{E}_{\rm elec}&=\int_{\Omega} \frac{\delta^2 \epsilon}{2}\abs{\nabla \phi}^2d\mb{x},\quad
\mc{E}_{\rm surf}=\int_{\Gamma}\gamma dm_\Gamma,
\end{split}
\end{equation}
where the dimensionless energy levels are now:
\begin{equation}
E_{\rm X}=\begin{cases}
-\ln l_{\rm X} &\text{ in } \Omega_{\rm i}\\
0 &\text{ in } \Omega_{\rm e}
\end{cases} 
\text{ for } {\rm X}={\rm C,A,S}.
\end{equation}
For the dissipation components and the boundary free energy input, we have:
The components of the dissipation $\mc{D}$ are given by: 
\begin{equation}\label{diss_dless}
\begin{split}
\mc{D}_{\rm ediff}&=\sum_{\rm {X}={\rm C,A}}\mc{D}_{\rm X}+\frac{1}{\alpha}\mc{D}_{\rm S},\\
\mc{D}_{\rm X}&=\int_\Omega \frac{D_{\rm X}}{Pe}n_{\rm X}\abs{\nabla \mu_{\rm X}}^2d\mb{x}, \; 
\mu_{\rm X}=\ln n_{\rm X}+z_{\rm X}\phi,\; {\rm X}={\rm C,A,S},\\
\mc{D}_{\rm visc}&=\int_{\Omega} 2\delta^2\mu\abs{\nabla_S \mb{u}}^2d\mb{x},\; 
\mc{D}_{\rm rct}=\int_\Omega \frac{ks}{\alpha Pe}(Q-1)\ln Qd\mb{x},\; Q=\frac{ca}{Ks}.
\end{split}
\end{equation}
Boundary free energy input $\mc{I}$ is given by:
\begin{equation}
\begin{split}
\mc{I}&=-\int_{\partial \Omega} \paren{\delta^2\phi j_{\partial \Omega}
+\frac{1}{Pe}\sum_{\rm {X}={\rm C,A,S}}\mu_{\rm X}f_{\rm X,\partial \Omega}}d{m_{\partial\Omega}},\\
j_{\partial \Omega}&=-\PD{}{t}\paren{\at{\epsilon\PD{\phi}{\mb{n}}}{\partial \Omega}},\; 
f_{{\rm X}, \partial \Omega}=-\at{D_{\rm X}n_{\rm X}\PD{\mu_{\rm X}}{\mb{n}}}{\partial \Omega}.
\end{split}
\end{equation}

{\subsection{Charge Diffusion Model}}

The charge diffusion model discussed in Section \ref{CD}, like the modified Saville model, satisfies the following energy relation:
\begin{equation}\label{chargediff}
\D{\mc{E}}{t}=-\mc{D}+\mc{I},\; \mc{E}=\mc{E}_{\rm ch}+\mc{E}_{\rm elec}+\mc{E}_{\rm surf}, \; \mc{D}=\mc{D}_{\rm ediff}+\mc{D}_{\rm visc},
\end{equation}
where $\mc{E}_{\rm elec}, \mc{E}_{\rm surf}$ and $\mc{D}_{\rm visc}$ are the same as in \eqref{energy_dless}, \eqref{diss_dless} and 
\begin{equation}\label{CD_energies}
\begin{split}
\mc{E}_{\rm ch}&=\int_{\Omega}\paren{q(\omega_q+E_q)-\sqrt{4S+q^2}}d\mb{x},\; \mc{D}_{\rm ediff}=\int_{\Omega} \frac{1}{Pe}\Sigma(q)\abs{\nabla \mu_q}^2d\mb{x},\\
\omega_q&=\ln\paren{\frac{1}{2}\paren{q+\sqrt{4S+q^2}}}, \; \mu_q=\omega_q+\phi,\; 
E_q=\begin{cases}
-\ln l_{\rm C} &\text{ in } \Omega_{\rm i},\\
0 &\text{ in } \Omega_{\rm e}.
\end{cases}\\
\mc{I}&=-\int_{\partial \Omega} \paren{\delta^2\phi j_{\partial \Omega}+\frac{1}{Pe}\mu_q f_q}d{m_{\partial\Omega}},\\
j_{\partial \Omega}&=-\PD{}{t}\paren{\at{\epsilon\PD{\phi}{\mb{n}}}{\partial \Omega}},\; 
f_{q, \partial \Omega}=-\at{\Sigma(q)\PD{\mu_q}{\mb{n}}}{\partial \Omega}.
\end{split}
\end{equation}
In contrast to the energy identity \eqref{energy_full} of the modified Saville model, 
dissipation due to the dissociation reaction $\mc{D}_{\rm rct}$ as well as 
the terms involving the neutral solute $s$ in $\mc{E}_{\rm chem}$ and $\mc{D}_{\rm ediff}$ are absent.
Indeed, we may obtain the above energy identity in the following fashion. Plug in the expansions in equation~\eqref{alphaexp}
into \eqref{energy_full}, \eqref{energy_dless}, \eqref{diss_dless} where $s_{(0)}, c_{(0)}$ and $a_{(0)}$
are given in \eqref{s0eqn} and \eqref{c0a0q}. As we let $\alpha\to 0$, we see that:
\begin{equation}
\mc{E}^{\rm MS}_{\rm C}+\mc{E}^{\rm MS}_{\rm A}\to \mc{E}^{\rm CD}_{\rm ch}, \; \frac{1}{\alpha}\mc{E}^{\rm MS}_{\rm S}\to -l_{\rm S}\abs{\Omega_{\rm i}},\;
\frac{1}{\alpha}\mc{D}^{\rm MS}_{\rm S}\to 0, \; \mc{D}^{\rm MS}_{\rm rct}\to 0,
\end{equation}
where $\abs{\Omega_{\rm i}}$ is the volume of the region $\Omega_{\rm i}$. 
In the above, the superscripts ${\rm MS}$ and ${\rm CD}$ denote the energies and dissipations in the free energy identities of the modified Saville model 
\eqref{energy_full} and the charge diffusion model \eqref{chargediff} respectively.
Since the flow is incompressible, 
$\abs{\Omega_{\rm i}}$ remains constant in time, and we thus obtain \eqref{chargediff}.

{\subsection{Taylor-Melcher Model and Model Hierarchy}}

The TM model (with or without the assumption that $q_\Omega\equiv 0$, {see discussion between Eq. \eqref{taudef} and \eqref{LD_laplace}})
satisfies the following energy identity:
\begin{equation}\label{LD_energy}
\D{\mc{E}}{t}=-\mc{D}+\mc{I},
\end{equation}
where
\begin{equation}\label{TM_energy}
\begin{split}
\mc{E}&=\mc{E}_{\rm elec}+\mc{E}_{\rm surf}=\int_{\Omega} \frac{\epsilon}{2}\abs{\nabla \phi}^2d\mb{x}+\int_{\Gamma}\wh{\gamma} dm_\Gamma,\\
\mc{D}&=\mc{D}_{\rm cond}+\mc{D}_{\rm visc}=\int_\Omega \paren{\wh{\sigma} \abs{\nabla \phi}^2+2\mu\abs{\nabla_S \mb{u}}^2}d\mb{x},\\
\mc{I}&=-\int_{\partial \Omega} \phi i_{\partial \Omega}d{m_{\partial\Omega}}, \quad
i_{\partial \Omega}=-\paren{\PD{}{t}\paren{\at{\epsilon\PD{\phi}{\mb{n}}}{\partial \Omega}}+\wh{\sigma}\at{\PD{\phi}{\mb{n}}}{\partial \Omega}}.
\end{split}
\end{equation}
This identity can also be obtained by letting $\delta\to 0$ 
in the energy identity \eqref{chargediff} for the charge diffusion model.
Plug in the expressions:
\begin{equation}
\phi=\phi_0+\mc{O}(\delta ), \; q=\delta^2 q_2+\mc{O}(\delta^3), \; \mb{u}=\mb{u}_0+\mc{O}(\delta )
\end{equation}
into \eqref{chargediff}.
The leading order scaling of $q$, is taken to be compatible with our matched asymptotic calculation of Section \ref{derivation_of_TM}.
If we are just interested in obtaining this scaling, however, we have only to look to the Poisson equation \eqref{dlesspoisson}; 
if $\phi_0$ is scaled as $\mc{O}(\delta^0)$, $q$ must be scaled starting at $\mc{O}(\delta^2)$.
We see that 
\begin{equation}
\delta^{-2}(\mc{E}^{\rm CD}_{\rm elec}+\mc{E}^{\rm CD}_{\rm surf})=\mc{E}^{\rm TM}+\mc{O}(\delta ), \quad 
\delta^{-2}\mc{D}^{\rm CD}=\mc{D}^{\rm TM}+\mc{O}(\delta),
\end{equation}
where the superscripts ${\rm CD}$ and ${\rm TM}$ denote the energies and dissipations that appear in the free energy identity for the charge diffusion model 
\eqref{chargediff} and the TM model \eqref{TM_energy} respectively.
Similarly, $\delta^{-2}\mc{I}^{\rm CD}$ reduces to $\mc{I}^{TM}$ to leading order. 
We see that \eqref{scaling} is precisely the scaling of dimensionless parameters that allows each of the energy and dissipation terms in 
\eqref{chargediff} to have the same order of magnitude as $\delta \to 0$.
We have only to consider $\mc{E}^{\rm CD}_{\rm ch}$ in \eqref{chargediff}. We have:
\begin{equation}
\delta^{-2}\mc{E}^{\rm CD}_{\rm ch}=-\delta^{-2}\int_\Omega 2\sqrt{S}d\mb{x}+\int_{\Omega_{\rm i}} q_2\ln(\sqrt{S}_{\rm i}/l_{\rm C})d\mb{x}+\mc{O}(\delta ).
\end{equation}
In order to recover the energy law of the TM model, the time derivative of the above must be $\mc{O}(\delta )$.
The first term is a constant since $\sqrt{S}$ is constant within $\Omega_{\rm i,e}$ respectively and the flow is incompressible.
The second term is $0$ thanks to \eqref{lCsqrtS1}, which is a consequence of our assumption that $l_{\rm C}=l_{\rm A}$.
We see that the appropriate scaling for the dimensionless variables and the parametric constraint $l_{\rm C}=l_{\rm A}$
can both be gleaned by simply taking limits in energy identities.

\begin{figure}
\begin{center}
\includegraphics{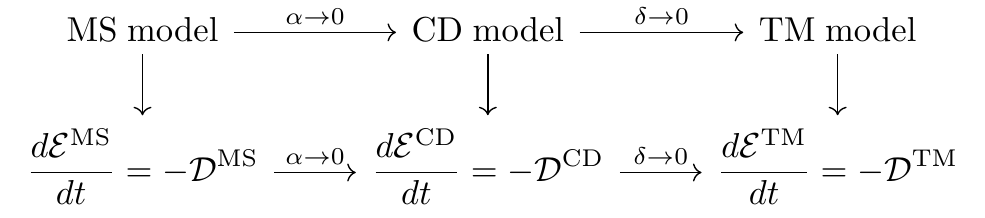}
\end{center}
\caption{\label{comm_diag}Model hierarchy and energy relations. 
The top row lists the modified Saville, charge diffusion and TM models and the bottom row lists the energy relations 
satisfied by the respective models, \eqref{energy_full}, \eqref{chargediff} and \eqref{TM_energy} 
(we have omitted the boundary energy input term $\mc{I}$ to avoid a cluttered diagram). The horizontal arrows correspond 
to taking (asymptotic) limits $\alpha\to 0$ or $\delta\to 0$. The vertical arrows correspond to the ``operation" of forming the free 
energy identity. These arrows commute in the following sense. Taking the asymptotic limit in the complex model and then forming the energy identity, and forming the energy identity of the complex model and then taking the asymptotic limit in the energy identity, 
lead to the same energy identity for the simpler model.}
\end{figure}

{The modified Saville model, the charge diffusion model and the TM model have now been placed 
in a hierarchy of energy relations, the TM model being the simplest and the modified Saville model 
being the most complex. This model hierarchy is expressed in the form of a ``commutative diagram" in Figure \ref{comm_diag}.}


{\subsection{Derivation of Free Energy Relations}\label{free_en_deriv}}

We prove \eqref{LD_energy} of the TM model. The energy identities of the modified Saville and charge diffusion 
models can be proved in a similar fashion.
The calculation to follow parallels that of \citet{Mori2011_PhysicaD}.

Let us calculate the left hand side of \eqref{LD_energy}.
\begin{equation}\label{dEdt}
\begin{split}
&\D{}{t}\int_{\Omega}
\frac{\epsilon}{2}\abs{\nabla \phi}^2d\mb{x}+\int_\Gamma \wh{\gamma} dm_\Gamma\\
=&\int_{\Omega}
\paren{\epsilon\nabla \phi \cdot \nabla\paren{\PD{\phi}{t}}}d\mb{x}
+\int_{\Gamma} 
\paren{\paren{\jump{\frac{\epsilon}{2}\abs{\nabla\phi}^2}+\wh{\gamma}\kappa}u_\perp}
dm_\Gamma\\
=&\int_{\Gamma}\paren{
\jump{\phi\PD{}{\mb{n}}\paren{\epsilon \PD{\phi}{t}}}+\paren{\jump{\frac{\epsilon}{2}\abs{\nabla\phi}^2}+\wh{\gamma}\kappa}u_\perp}
dm_\Gamma\\
&+\int_\Omega \phi \PD{q_\Omega}{t} d\bm{x}+\int_{\partial \Omega}\phi\PD{}{\mb{n}}\paren{\epsilon \PD{\phi}{t}}dm_{\partial \Omega}.
\end{split}
\end{equation}
In the first equality, we used the fact that the interface $\Gamma$
moves with the local flow velocity, where $\bm{u}\cdot \bm{n}=u_\perp$. In the second equality, we integrated 
by parts and used \eqref{LD_poisson}.

In order to proceed further, we must simplify the term $\PD{}{\mb{n}}\paren{\epsilon \PD{\phi}{t}}$ that appears in the integral 
over $\Gamma$ in \eqref{dEdt}. We have:
\begin{equation}
\at{\PD{}{\bm{n}}\paren{\PD{\phi}{t}}}{\Gamma_{k}}=\at{\paren{\partial_t^\perp \paren{\PD{\phi}{\bm{n}}}
+\nabla_\Gamma u_\perp \cdot \nabla_\Gamma \phi-u_\perp\PDD{2}{\phi}{\bm{n}}}}{\Gamma_k}, \; k={\rm i,e}.
\end{equation}
This technical result is proved in Appendix B of \citet{Mori2011_PhysicaD}. Using this, we obtain:
\begin{equation}
\begin{split}
&\int_{\Gamma}\jump{\phi\PD{}{\mb{n}}\paren{\epsilon \PD{\phi}{t}}}dm_\Gamma\\
=&\int_\Gamma\jump{\phi\paren{\partial_t^\perp \paren{\epsilon\PD{\phi}{\bm{n}}}
+\nabla_\Gamma u_\perp \cdot (\epsilon \nabla_\Gamma \phi)-u_\perp\epsilon\PDD{2}{\phi}{\bm{n}}}}dm_\Gamma\\
=&\int_\Gamma\paren{\phi\jump{\partial_t^\perp \paren{\epsilon\PD{\phi}{\bm{n}}}}
-u_\perp\jump{\nabla_\Gamma\cdot (\epsilon \phi \nabla_\Gamma \phi)+\epsilon\PDD{2}{\phi}{\bm{n}}}}dm_\Gamma,
\end{split}
\end{equation}
where we integrated by parts over $\Gamma$ in the second identity and used the fact that $\phi$ and $u_\perp$ are continuous across $\Gamma$
(see \eqref{phiqgamma} and \eqref{LD_stokesBC}). 
Note that the surface $\Gamma$ is closed and thus integration by parts does not result in boundary terms. 
Let us simplify the above integrands. 
First we have:
\begin{equation}
\jump{\partial_t^\perp \paren{\epsilon\PD{\phi}{\bm{n}}}}=\partial_t^\perp\jump{\paren{\epsilon\PD{\phi}{\bm{n}}}}=\partial_t^\perp q_\Gamma.
\end{equation}
Next, we have:
\begin{equation}
\begin{split}
&\jump{\nabla_\Gamma\cdot (\epsilon \phi \nabla_\Gamma \phi)+\epsilon\phi\PDD{2}{\phi}{\bm{n}}}=
\jump{\epsilon \abs{\nabla_\Gamma\phi}^2+\epsilon\phi\paren{\Delta_\Gamma \phi+\PDD{2}{\phi}{\bm{n}}}}\\
&=\jump{\epsilon \abs{\nabla_\Gamma\phi}^2-\epsilon \phi\kappa \PD{\phi}{\bm{n}}-q_\Omega \phi}
=\jump{\epsilon \abs{\nabla_\Gamma\phi}^2}-\kappa q_\Gamma\phi -\jump{q_\Omega}\phi, 
\end{split}
\end{equation}
where $\Delta_\Gamma$ is the Laplace-Beltrami operator on $\Gamma$.
In the second equality above, we used the expression below, which is a consequence of \eqref{LD_poisson}:
\begin{equation}
\at{\epsilon\Delta \phi}{\Gamma_k}=\at{\epsilon \paren{\Delta_\Gamma \phi +\kappa \PD{\phi}{\bm{n}}+\PDD{2}{\phi}{\bm{n}}}}{\Gamma_k}
=-\at{q_\Omega}{\Gamma_k},\; k={\rm i,e}.
\end{equation}
Collecting the above calculations and plugging this back into \eqref{dEdt}, we have:
\begin{equation}
\begin{split}
&\D{}{t}\int_{\Omega}
\frac{\epsilon}{2}\abs{\nabla \phi}^2d\mb{x}+\int_\Gamma \wh{\gamma} dm_\Gamma=I_1+I_2+I_3, \\
I_1=&\int_\Gamma \phi\paren{\partial^\perp_t q_\Gamma+u_\perp\kappa q_\Gamma}dm_\Gamma, \quad I_2=\int_\Omega \phi\PD{q_\Omega}{t}+\int_\Gamma u_\perp\phi \jump{q_\Omega}dm_\Gamma,\\
I_3=&\int_\Gamma\paren{\jump{\frac{\epsilon}{2}\paren{\abs{\PD{\phi}{\bm{n}}}^2-\abs{\nabla_\Gamma\phi}^2}}+\wh{\gamma}\kappa}u_\perp dm_\Gamma.
\end{split}
\end{equation}
Using \eqref{LD_stokesBC}, we see that the integrand of $I_1+I_3$ is equal to:
\begin{equation}
\begin{split}
&\phi (\partial_t^\perp q_\Gamma+\kappa q_\Gamma u_\perp)-[(\Sigma(\mb{u},p))\mb{n}]\cdot\mb{u}-q_\Gamma\mb{u}_{\parallel} \nabla_\Gamma\phi\\
=&-[(\Sigma(\mb{u},p))\mb{n}]\cdot\mb{u}-\nabla_\Gamma\cdot(q_\Gamma\phi\mb{u}_\parallel)-\jump{\wh{\sigma}\PD{\phi}{\mb{n}}}\phi,
\end{split}
\end{equation}
where we used \eqref{LD_qGammaeqn}. Therefore,
\begin{equation}
\begin{split}
I_1+I_3
=&-\int_\Gamma\paren{[(\Sigma(\mb{u},p))\mb{n}]\cdot\mb{u}+\jump{\sigma\PD{\phi}{\mb{n}}}\phi} dm_\Gamma\\
=&-\int_{\Omega}\paren{\wh{\sigma}\abs{\nabla\phi}^2+2\mu\abs{\nabla_{S} \mb{u}}^2}d\mb{x}
+\int_{\partial \Omega} \phi\wh{\sigma}\PD{\phi}{\bm{n}}dm_{\partial \Omega}+I_4,\\
I_4=&-\int_\Omega \paren{\mb{u}\cdot q_\Omega \nabla \phi +\nabla \cdot (\wh{\sigma}\nabla \phi)}d\bm{x}
\end{split}
\end{equation}
where we integrated by parts on the surface $\Gamma$ in the first equality, and integrated by parts over $\Omega$
in the second equality and used \eqref{LD_stokesq}.
Finally, using \eqref{LD_q}, we have:
\begin{equation}
I_2+I_4=-\int_\Omega \nabla \cdot (\bm{u}q_\Omega \phi)d\mb{x}+\int_\Gamma u_\perp\phi \jump{q_\Omega}dm_\Gamma=0.
\end{equation}
This concludes the proof of identity \eqref{LD_energy}.

\section{Stokes Equation in the Inner Layer}\label{tensors}

In the boundary layer analysis, we introduced curvilinear coordinates fitted to the liquid-liquid interface.
Here, we collect vector calculus expressions associated with this coordinate system. 
See \citet{aris1990vectors} for an extensive treatment of the equations of continuum mechanics in curvilinear coordinates.
Recall that the coordinates $(\xi,\eta^1,\eta^2)=(\eta^0,\eta^1,\eta^2)$ 
(we shall use $\eta_0$ or $\xi$ interchangeably)
are given as in \eqref{coordT}. The metric tensor associated with this
coordinate system is given by:
\begin{equation}
G=\begin{pmatrix}
1 & \mb{0}^{\rm T}\\
\mb{0} & \wh{g}
\end{pmatrix}
\end{equation}
where $G$ is the $3\times 3$ metric tensor, $\mb{0}\in\mathbb{R}^2$ is the zero (column) vector
and $\wh{g}$ is the $2\times 2$ metric tensor associated with the coordinates $\eta_1,\eta_2$.
$\wh{g}$ is given by:
\begin{equation}
\wh{g}=g+2\xi h+\xi^2 \wt{h}
\end{equation}
where $g, h$ and $\wt{h}$ are the first, second and third fundamental forms of the surface respectively, 
and are thus functions of $(\eta^1,\eta^2)$ and do not depend on $\xi$:
\begin{equation}
g_{ij}=\PD{\mb{X}}{\eta^i}\cdot \PD{\mb{X}}{\eta_j}, \; 
h_{ij}=\PD{\mb{n}}{\eta^i}\cdot \PD{\mb{X}}{\eta_j}, \; 
\wt{h}_{ij}=\PD{\mb{n}}{\eta^i}\cdot \PD{\mb{n}}{\eta_j}.
\end{equation}
Here, the subscript $i,j$ denote the components of the tensors and $i,j=1,2$. 

We now write the Stokes equation \eqref{dlessstokes} in these coordinates.
The components of the vector field $\mb{u}$ in curvilinear coordinates 
will be denoted $(u,v^1,v^2)=(v^0,v^1,v^2)$ (see \eqref{vcurvilinear}).
Consider first the incompressibility condition. We have:
\begin{equation}\label{divfreecurv}
\nabla \cdot \mb{u}=\frac{1}{\sqrt{\abs{G}}}\PD{}{\eta^\rho}\paren{\sqrt{\abs{G}}v^\rho}
=\frac{1}{\sqrt{\abs{\wh{g}}}}\PD{}{\xi}\paren{\sqrt{\abs{\wh{g}}}u}+\frac{1}{\sqrt{\abs{\wh{g}}}}\PD{}{\eta^i}\paren{\sqrt{\abs{\wh{g}}}v^i}=0
\end{equation}
where $\abs{\cdot}$ is the determinant. In this appendix, Greek indices 
run through $0,1,2$ and whereas Roman indices run through $1,2$.
Note that:
\begin{equation}\label{kappag}
\at{\frac{1}{\sqrt{\abs{\wh{g}}}}\PD{}{\xi}\sqrt{\abs{\wh{g}}}}{\xi=0}=\frac{1}{2\abs{\wh{g}}}\at{\PD{\abs{\wh{g}}}{\xi}}{\xi=0}=g^{ij}h_{ij}=\kappa
\end{equation}
where $g^{ij}$ denotes the components of the inverse of $g$ (following usual convention, and likewise for $\wh{g}$ and $G$)
and $\kappa$ is the sum of the principal curvatures of the interface (twice the mean curvature). 
Introduce the inner layer coordinate $\xi'=\xi/\delta^2$ and postulate a boundary layer expansion in terms of $\delta $ 
starting from:
\begin{equation}
\wt{u}=\wt{u}_0+\delta \wt{u}_1+\cdots,
\end{equation}
and similarly for $v^i$.
Condition \eqref{divfreecurv} gives:
\begin{align}
\label{divcurv0}
\PD{\wt{u}_0}{\xi'}&=0,\\
\label{divcurv1}
\PD{\wt{u}_1}{\xi'}+\kappa \wt{u}_0 +\frac{1}{\sqrt{\abs{g}}}\PD{}{\eta^i}\paren{\sqrt{\abs{g}}\wt{v}^i_0}&=0.
\end{align}

Let us now turn to the momentum balance equation in \eqref{dlessstokes}, written in stress-divergence form:
\begin{equation}\label{stokescurv}
\delta^2\nabla \cdot \Sigma=q\nabla \phi, \; \Sigma=\mu\paren{\nabla \mb{u}+(\nabla \mb{u})^{\rm T}}-pI.
\end{equation}
In the curvilinear coordinate system, the above may be written as:
\begin{equation}
\delta^2\Sigma^{\rho\nu}_{,\nu}=qG^{\rho\nu}\PD{\phi}{\eta^\nu}, \; \Sigma^{\rho\nu}=\mu D^{\rho\nu}-pG^{\rho\nu},
\end{equation}
where the subscripts $,\nu$ denote the covariant derivative and $D^{\rho\nu}$ is given by
\begin{equation}
D^{\rho\nu}=G^{\rho\lambda}v^\nu_{,\lambda}+G^{\nu\lambda}v^\rho_{,\lambda}=G^{\rho\lambda}\paren{\PD{v^\nu}{\eta^\lambda}+\Gamma^\nu_{\lambda\sigma}v^\sigma}
+G^{\nu\lambda}\paren{\PD{v^\rho}{\eta^\lambda}+\Gamma^\rho_{\lambda\sigma}v^\sigma}
\end{equation}
where $\Gamma$ are the Christoffel symbols associated with the metric $G$. Let $(n_0,n_1,n_2)=(1,0,0)$ be the covariant unit
vector in the $\xi=\eta^0$ direction. This is also the unit normal vector on the interface. We have: 
\begin{equation}
D^{\rho\nu}n_\nu=D^{\rho 0}=\PD{v^\rho}{\xi}+\Gamma^\rho_{0\lambda}v^\lambda+G^{\rho\lambda}\paren{\PD{v^0}{\eta^\lambda}+\Gamma^0_{\lambda\sigma}v^\sigma}.
\end{equation}
Noting that:
\begin{equation}\label{chrisrels}
\Gamma^0_{00}=\Gamma^0_{0l}=\Gamma^0_{0l}=\Gamma^l_{00}=0,\; \Gamma^i_{0l}=\Gamma^i_{l0}=\frac{1}{2}\wh{g}^{ik}\PD{\wh{g}_{kl}}{\xi}=-\wh{g}^{ik}\Gamma^0_{kl},
\end{equation}
we have:
\begin{equation}
D^{0\nu}n_\nu=D^{00}=2\PD{u}{\xi}, \quad  
D^{i\nu}n_\nu=D^{i0}=D^{0i}=\PD{v^i}{\xi}+\wh{g}^{ik}\PD{u}{\eta^k}.
\end{equation}
In particular, at $\xi=0$ we have:
\begin{equation}\label{Sigmanormal}
\at{\Sigma^{0\nu}n_\nu}{\xi=0}=2\mu\PD{u}{\xi}-p, \; \at{\Sigma^{i\nu}n_\nu}{\xi=0}=\mu\paren{\PD{v^i}{\xi}+g^{ik}\PD{u}{\eta^k}}.
\end{equation}

Next, we consider the divergence of $D$. We have:
\begin{equation}
D^{\rho\nu}_{,\nu}=\frac{1}{\sqrt{\abs{G}}}\PD{}{\eta^\nu}\paren{\sqrt{\abs{G}}D^{\rho\nu}}+\Gamma^{\rho}_{\nu\lambda}D^{\nu\lambda}.
\end{equation}
For the  $\rho=0$ component, we obtain:
\begin{equation}
\begin{split}
D^{0\nu}_{,\nu}&=\frac{2}{\sqrt{\abs{\wh{g}}}}\PD{}{\xi}\paren{\sqrt{\abs{\wh{g}}}\PD{u}{\xi}}
+\frac{1}{\sqrt{\abs{\wh{g}}}}\PD{}{\eta^i}\paren{\sqrt{\abs{\wh{g}}}\paren{\PD{v^i}{\xi}+\wh{g}^{ik}\PD{u}{\eta^k}}}\\
&-\frac{1}{2}\PD{\wh{g}_{jk}}{\xi}D^{jk}.
\end{split}
\end{equation}
Expanding the above in the inner layer, we have:
\begin{equation}
\begin{split}\label{D0nu}
\delta^2 D^{0\nu}_{,\nu}&=2\PDD{2}{\wt{u}_0}{\xi'}+
\delta \paren{2\PDD{2}{\wt{u}_1}{\xi'}+2\kappa\PD{\wt{u}_0}{\xi'}+\frac{1}{\sqrt{\abs{g}}}\PD{}{\eta^i}\paren{\sqrt{\abs{g}}\PD{\wt{v}^i_0}{\xi}}}+\mc{O}(\delta^2)\\
&=\delta \PDD{2}{\wt{u}_1}{\xi'}+\mc{O}(\delta^2)
\end{split}
\end{equation}
where where we used \eqref{divcurv0} and \eqref{divcurv1} in the second equality. Assume an expansion of $q$ of the form
\begin{equation}
\wt{q}=\wt{q}_0+\delta \wt{q}_1+\delta^2\wt{q}_2+\cdots
\end{equation}
and similarly for $\phi$.
For $\rho=0$, \eqref{stokescurv} thus gives:
\begin{align}
\label{stokescurv0-1}
0&=\wt{q}_0\PD{\wt{\phi}_0}{\xi'},\\
\label{stokescurv00}
0&=\wt{q}_1\PD{\wt{\phi}_0}{\xi'}+\wt{q}_0\PD{\wt{\phi}_1}{\xi'},\\
\label{stokescurv01}
\mu\PDD{2}{\wt{u}_1}{\xi'}-\PD{\wt{p}_0}{\xi'}&=\wt{q}_2\PD{\wt{\phi}_0}{\xi'}+\wt{q}_1\PD{\wt{\phi}_1}{\xi'}+\wt{q}_0\PD{\wt{\phi}_2}{\xi'}.
\end{align}

For $\rho=i=1,2$ we have:
\begin{equation}
\begin{split}
D^{i\nu}_{,\nu}&=\frac{1}{\sqrt{\abs{\wh{g}}}}\PD{}{\xi}\paren{\sqrt{\abs{\wh{g}}}\paren{\PD{v^i}{\xi}+\wh{g}^{ik}\PD{u}{\eta^k}}}
+\wh{g}^{ik}\PD{\wh{g}_{kj}}{\xi}\paren{\PD{v^j}{\xi}+\wh{g}^{jl}\PD{u}{\eta^l}}\\
&+\frac{1}{\sqrt{\abs{\wh{g}}}}\PD{}{\eta^j}\paren{\sqrt{\abs{\wh{g}}}D^{ij}}+\Gamma^i_{jk}D^{jk},
\end{split}
\end{equation}
where we used \eqref{chrisrels} in the above. Expanding the above in the inner layer, we obtain:
\begin{equation}\label{Dinu}
\delta^2 D^{i\nu}_{,\nu}=\PDD{2}{\wt{v}^i_0}{\xi'}
+\delta \paren{\PD{}{\xi'}\paren{\PD{\wt{v}^i_1}{\xi'}+g^{ik}\PD{\wt{u}_0}{\eta^k}}+\kappa\PD{\wt{v}^i_0}{\xi'}+2g^{ik}h_{kj}\PD{\wt{v}^j_0}{\xi'}}
+\mc{O}(\delta^2).
\end{equation}
For $\rho=i=1,2$, \eqref{stokescurv} thus gives:
\begin{align}
\label{stokescurvi0}
\mu\PDD{2}{\wt{v}^i_0}{\xi'}&=\wt{q}_0g^{ij}\PD{\wt{\phi}_0}{\eta^j},\\
\label{stokescurvi1}
\mu\paren{\PDD{2}{\wt{v}^i_1}{\xi'}+\kappa\PD{\wt{v}^i_0}{\xi'}+2g^{ik}h_{kj}\PD{\wt{v}^j_0}{\xi'}}&
=\wt{q}_0g^{ij}\PD{\wt{\phi}_1}{\eta^j}+\wt{q}_1g^{ij}\PD{\wt{\phi}_0}{\eta^j},
\end{align}
where, in the second relation, we used \eqref{divcurv0} and the fact that $g$ does not depend on $\xi'$.
When the interface is a sphere, $2g^{ik}h_{kj}=\kappa \delta^i_j$ where $\delta^i_j$ is the Kronecker delta, 
in which case \eqref{stokescurvi1} reduces to:
\begin{equation}\label{stokescurvi1sphere}
\mu\paren{\PDD{2}{\wt{v}^i_1}{\xi'}+2\kappa\PD{\wt{v}^i_0}{\xi'}}
=\wt{q}_0g^{ij}\PD{\wt{\phi}_1}{\eta^j}+\wt{q}_1g^{ij}\PD{\wt{\phi}_0}{\eta^j}.
\end{equation}

\section{Proof of Proposition \ref{kummer_prop}}\label{kummereqn}
{Proposition \ref{kummer_prop} concerns the solution of equation \eqref{innerchargeODE} 
under the condition \eqref{innerchargebc}. 
When $\lambda\neq 0$, the solution can be written in terms of Kummer functions, 
whose properties we now discuss. The proof of Proposition \ref{kummer_prop} will follow 
after this general discussion.}
For properties of Kummer functions, 
we refer the reader to \citet{olver2010nist}.

{Consider the following differential equation for $x>0$:}
\begin{equation}\label{kummer+}
\DD{2}{y}{x}-x\D{y}{x}-ay=0, \; a>0.
\end{equation}
We are interested in solutions that satisfy:
\begin{equation}\label{liminfty}
\lim_{x\to \infty} y(x)=0.
\end{equation}
Consider the function:
\begin{equation}
f(z)=y(x), \; z=\frac{1}{2}x^2.
\end{equation}
Note that this coordinate transformation is well-defined since $x>0$. After some calculation, we obtain:
\begin{equation}
z\DD{2}{f}{z}+\paren{\frac{1}{2}-z}\D{f}{z}-\frac{a}{2}f=0, z>0.
\end{equation}
This is known as the Kummer differential equation, and its general solution is given by:
\begin{equation}
f(z)=AM\paren{\frac{a}{2},\frac{1}{2},z}+BU\paren{\frac{a}{2},\frac{1}{2},z},
\end{equation}
where $M$ and $U$ are the Kummer functions of the first and second kind, 
and $A, B$ are arbitrary constants. The general solution to \eqref{kummer+} is thus given by:
\begin{equation}
y(x)=AM\paren{\frac{a}{2},\frac{1}{2},\frac{1}{2}x^2}+BU\paren{\frac{a}{2},\frac{1}{2},\frac{1}{2}x^2}.
\end{equation}
From known properties of the functions $M$ and $U$,  we know that, as $x\to \infty$:
\begin{equation}\label{MUinfty}
\begin{split}
M\paren{\frac{a}{2},\frac{1}{2},\frac{1}{2}x^2}&=\frac{2^{(1-a)/2}}{\Gamma\paren{\frac{a}{2}}}x^{a-1}\exp(x^2/2)\paren{1+\mc{O}(x^{-2})},\\
U\paren{\frac{a}{2},\frac{1}{2},\frac{1}{2}x^2}&=2^{a/2}x^{-a}\paren{1+\mc{O}(x^{-2})},
\end{split}
\end{equation}
where $\Gamma(\cdot)$ is the Gamma function. Since we are interested in solutions that satisfy \eqref{liminfty}, 
we have:
\begin{equation}\label{psia}
y(x)=C\psi_a(x), \; \psi_a(x)=\frac{\Gamma\paren{\frac{a+1}{2}}}{\sqrt{\pi}}U\paren{\frac{a}{2},\frac{1}{2},\frac{1}{2}x^2},
\end{equation}
where $C$ is an arbitrary constant.
The function $\psi_a$ has been normalized so that:
\begin{equation}\label{psia1}
\lim_{x\to 0}\psi_a(x)=1.
\end{equation}
We record its behavior as $x\to \infty$:
\begin{equation}\label{psiaasymp}
\psi_a(x)=\frac{\Gamma\paren{\frac{a+1}{2}}}{\sqrt{\pi}}2^{a/2}x^{-a}\paren{1+\mc{O}(x^{-2})}.
\end{equation}
We point out that in fact $\psi_a$ can be extended to an entire function in the complex plane (although $U$ in general cannot)
so that, in particular, $\psi_a$ (and all its derivatives) are defined at $x=0$.

{Next, consider the following differential equation:}
\begin{equation}\label{kummer-}
\DD{2}{y}{x}+x\D{y}{x}-ay=0,\; a>0.
\end{equation}
This is the same as \eqref{kummer+} except that the sign of the second term has changed.
We are again interested in solutions that satisfy \eqref{liminfty}. In this case, 
\begin{equation}
w=\exp(x^2/2)y
\end{equation}
satisfies the equation:
\begin{equation}
\DD{2}{w}{x}-x\D{w}{x}-(a+1)w=0.
\end{equation}
This is just \eqref{kummer+} where $a$ is replaced by $a+1$, and therefore, the general solution to
\eqref{kummer-} is given by:
\begin{equation}
y(x)=\paren{AM\paren{\frac{a+1}{2},\frac{1}{2},\frac{1}{2}x^2}+BU\paren{\frac{a+1}{2},\frac{1}{2},\frac{1}{2}x^2}}\exp(-x^2/2),
\end{equation}
where $A,B$ are arbitrary constants.
Using \eqref{MUinfty}, we see that the solutions satisfying \eqref{liminfty} are given by:
\begin{equation}\label{varphia}
y(x)=C\varphi_a(x), \; \varphi_a(x)=\frac{\Gamma\paren{\frac{a}{2}+1}}{\sqrt{\pi}}U\paren{\frac{a+1}{2},\frac{1}{2},\frac{1}{2}x^2}\exp(-x^2/2),
\end{equation}
where $C$ is an arbitrary constant.
We have normalized $\varphi_a(x)$ so that
\begin{equation}
\lim_{x\to 0}\varphi_a(x)=1,
\end{equation}
and its behavior as $x\to \infty$ is given by:
\begin{equation}
\varphi_a(x)=\frac{\Gamma\paren{\frac{a}{2}+1}}{\sqrt{\pi}}2^{(a+1)/2}x^{-(a+1)}\exp(-x^2/2)\paren{1+\mc{O}(x^{-2})}.
\end{equation}
Like $\psi_a$, $\varphi_a$ also has an analytic continuation as an entire function.

We finally note that
\begin{equation}\label{psiapositive}
\psi_a(x)>0 \text{ and } \varphi_a(x)>0 \text{ for all } 0\leq x<\infty.
\end{equation}
This can be seen as follows.
Note first that $\psi_a(0)=1>0$ and $\psi_a(x)>0$ as $x\to \infty$. If $\psi_a(x)$ is non-positive, it must 
have a local minimum that is either negative or $0$.
By the maximum principle applied to \eqref{kummer+}, $\psi_a(x)$ cannot have a negative minimum.
The function $\psi_a(x)$ cannot attain $0$ as its minimum either, because of uniqueness of 
the ODE initial value problem; if $\psi_a=\psi^{\prime}_a=0$ at one point, then $\psi_a$ would be identically equal to $0$.
The same argument applies to $\varphi_a$. It is also not difficult to see that both $\psi_a$ and $\varphi_a$ are monotone decreasing by a 
similar argument (or using phase plane methods). We omit the proof.

\begin{proof}[{Proof of Proposition \ref{kummer_prop}}]
For $\lambda=0$, equation~\eqref{innerchargeODE} is a linear ODE with constant coefficients that is straightforward to solve.
Suppose $\lambda>0$ and focusing on $\xi'>0$, we may rewrite \eqref{innerchargeODE} as:
\begin{equation}\label{kummera+}
\DD{2}{y}{x}-x\D{y}{x}-ay=0, \; a=\frac{1}{\lambda\tau_{\rm e}}, \; y(x)=\wt{q}_1\paren{\sqrt{\frac{D_{q,\rm e}}{\lambda}}x}.
\end{equation}
{This is nothing other than \eqref{kummer+}. Using \eqref{psia}, we see that 
solutions that decay to $0$ as $\xi'\to \infty$ can be written as}:
\begin{equation}\label{q1+}
\wt{q}_1(\xi')=C\psi_{(\lambda\tau_{\rm e})^{-1}}\paren{\sqrt{\frac{\lambda}{D_{q,\rm e}}}\xi'}, \; \xi'>0.
\end{equation}
where $\psi_a$ is defined in \eqref{psia} and $C$ is a constant to be determined.
Likewise,
\begin{equation}\label{q1-}
\wt{q}_1(\xi')=l_{\rm C}C\psi_{(\lambda\tau_{\rm i})^{-1}}\paren{\sqrt{\frac{\lambda}{D_{q,\rm i}}}\abs{\xi'}}, \; \xi'<0.
\end{equation}
Here, we have used the first condition in \eqref{innerchargebc} and \eqref{psia1}.
Finally, we must determine $C$ using the integral constraint in \eqref{innerchargebc}.
\begin{equation}\label{Ceqn}
\begin{split}
\int_{-\infty}^\infty \wt{q}_1d\xi'&=
\int_{-\infty}^0 l_{\rm C}C\psi_{(\lambda\tau_{\rm i})^{-1}}\paren{\sqrt{\frac{\lambda}{D_{q,\rm i}}}\abs{\xi'}}d\xi'
+\int_0^\infty C\psi_{(\lambda\tau_{\rm e})^{-1}}\paren{\sqrt{\frac{\lambda}{D_{q,\rm e}}}\xi'}d\xi'\\
&=C\paren{l_{\rm C}\sqrt{\frac{D_{q,\rm i}}{\lambda}}\int_0^\infty \psi_{(\lambda\tau_{\rm i})^{-1}}(x)dx+
\sqrt{\frac{D_{q,\rm e}}{\lambda}}\int_0^\infty \psi_{(\lambda\tau_{\rm e})^{-1}}(x)dx}.
\end{split}
\end{equation}
This must be equal to $q_\Gamma$. We can thus solve for $C$ if the sum of integrals in the parentheses 
in the above is nonzero and finite. This is always nonzero since $\psi_a$ is positive by \eqref{psiapositive}.
Noting that $\psi_a(x)$ is bounded, \eqref{psiaasymp} shows that $\psi_a(x)$ is integrable if and only if $a>1$. 
Thus, both $(\lambda\tau_{\rm i})^{-1}$
and $(\lambda\tau_{\rm e})^{-1}$ must be greater than $1$. We can thus solve for $C$ and obtain $\wt{q}_1$
if $\lambda\tau_{\rm max}<1$. If $\lambda\tau_{\rm max}\geq 1$, at least one of the integrals in \eqref{Ceqn} is infinite 
and a solution does not exist unless $q_\Gamma=0$ (in which case $\wt{q}_1$ is identically equal to $0$).
We may obtain \eqref{q1asymp+} from \eqref{q1+}, \eqref{q1-} and \eqref{psiaasymp}.

When $\lambda<0$, for $\xi'>0$, we may rewrite \eqref{innerchargeODE} as:
\begin{equation}\label{kummera-}
\DD{2}{y}{x}+x\D{y}{x}-ay=0, \; a=\frac{1}{\abs{\lambda}\tau_{\rm e}}, \; y(x)=\wt{q}_1\paren{\sqrt{\frac{D_{q,\rm e}}{\abs{\lambda}}}x}.
\end{equation}
{This is equation \eqref{kummer-}. We may thus proceed as in the case $\lambda>0$.}
The function $\varphi_a$ defined in \eqref{varphia} is integrable for any $a$, and thus there always is a unique solution.
\end{proof}


\bibliographystyle{plain}
\bibliography{weakelectrolyte}

\end{document}